\documentclass[
aps,            
prx,            
reprint,        
superscriptaddress,
longbibliography,
noeprint,
]{revtex4-2}

\usepackage{color}
\usepackage{graphicx}
\usepackage{comment}
\usepackage{booktabs}
\graphicspath{{C:/Users/scott/OneDrive/document/kosakaLab/paper/distributed_quantum_computer_architecture/Figure/}}
\graphicspath{{./Figure/}}

\usepackage{fontspec}

\usepackage{amsmath,amssymb,amsfonts}
\usepackage{bm}            
\usepackage{siunitx}       
\usepackage{braket}        
\usepackage{physics}       
\usepackage{graphicx}      
\usepackage{microtype}     
\usepackage[dvipsnames]{xcolor}
\usepackage{hyperref}      
\hypersetup{
	colorlinks=true,
	linkcolor=MidnightBlue,
	citecolor=MidnightBlue,
	urlcolor=MidnightBlue
}

\usepackage{makecell}
\usepackage{listings}

\newif\ifshowJP
\showJPtrue

\usepackage{lineno}

\begin{document}

	\title{Feasibility of Distributed Fault-Tolerant Quantum Computing:\\ From Quantum Links to Algorithms
	}
	
	\author{Takumi Kobayashi}
	\thanks{These three authors contributed equally.}
	\address{Department of Physics, Graduate School of Engineering Science, Yokohama National University, 79-5 Tokiwadai, Hodogaya, Yokohama, Kanagawa, 240-8501, Japan}

	\author{Hodaka Kurokawa}
	\thanks{These three authors contributed equally.}
	\email{kurokawa-hodaka-hm@ynu.ac.jp}
	\address{Institute of Industrial Science (IIS),
	The University of Tokyo,
	4-6-1 Komaba, Meguro-ku, Tokyo 153-8505, Japan}
	
	\address{Quantum Information Research Center, Institute of Advanced Sciences, Yokohama National University, 79-5 Tokiwadai, Hodogaya, Yokohama, Kanagawa, 240-8501, Japan}

	\author{Yasunari Suzuki}
	\thanks{These three authors contributed equally.}
	\address{RIKEN Center for Quantum Computing, 2-1 Hirosawa, Wako, Saitama 351-0198, Japan}

	\author{Hideo Kosaka}
	\email{kosaka-hideo-yp@ynu.ac.jp}
	\address{Department of Physics, Graduate School of Engineering Science, Yokohama National University, 79-5 Tokiwadai, Hodogaya, Yokohama, Kanagawa, 240-8501, Japan}
	\address{Quantum Information Research Center, Institute of Advanced Sciences, Yokohama National University, 79-5 Tokiwadai, Hodogaya, Yokohama, Kanagawa, 240-8501, Japan}

	
\begin{abstract}

Practical fault-tolerant quantum computation requires a large
number of physical qubits, posing a major challenge for
scaling up individual quantum processors.
Distributed architectures offer a promising solution by
interconnecting multiple processors through quantum networks.
However, remote Bell-pair generation remains technically
demanding, making it essential to assess computational
performance under limited communication resources.
In this work, we present a comprehensive resource analysis
of distributed fault-tolerant quantum computation using
superconducting processors connected by microwave or optical
links.
We show that even a single physical communication channel
per node can substantially reduce the physical-qubit
requirements per node under experimentally motivated,
forward-looking microwave-link assumptions of a 1 MHz
physical Bell-pair generation rate and 90$\%$ fidelity,
although at the cost of a roughly tenfold increase in
execution time.
Conversely, reducing the execution-time overhead requires
substantially greater communication resources, including
higher Bell-pair fidelities and extensive multiplexing. 
These results demonstrate the potential feasibility of distributed fault-tolerant quantum computation with limited communication resources and provide a quantitative basis for balancing quantum-link performance, processor size, and execution time.

\end{abstract}

	\maketitle
	
\section{Introduction}

Realizing a practical quantum computer is a major challenge in modern quantum science and technology.
For many practically relevant fault-tolerant applications, 0.1--10 million physical qubits are required \cite{Fowler2012,Preskill2018,Gidney2025}, assuming a fixed qubit arrangement and nearest-neighbor interactions.
However, even in current state-of-the-art devices, the maximum number of physical qubits remains on the order of one hundred to one thousand \cite{AbuGhanem2024,Chiu2025,Manetsch2025}, and reaching the break-even condition \cite{Zobrist2025,Bluvstein2025} at such scales is even more challenging.
Moreover, across leading physical platforms, including superconducting circuits \cite{Kim2023c,AbuGhanem2024,Zobrist2025}, semiconductor devices \cite{Philips2022,Xue2022,Burkard2023}, trapped ions \cite{Schafer2018,Liu2025}, and neutral atoms \cite{Bluvstein2025,Chiu2025,Manetsch2025}, various constraints such as the number of control lines, cooling power, laser power, and available physical space are expected to limit the practical size of a single processor unit.
To address these scalability constraints, architectures interconnecting multiple quantum processor modules through quantum networks are being actively explored \cite{Kimble2008,Jones2012,Nickerson2013,Wehner2018a}.
This architecture is referred to as a distributed quantum computer or a modular quantum computer.

\begin{figure*}[t]
	\centering
	\includegraphics[width=0.95\linewidth]{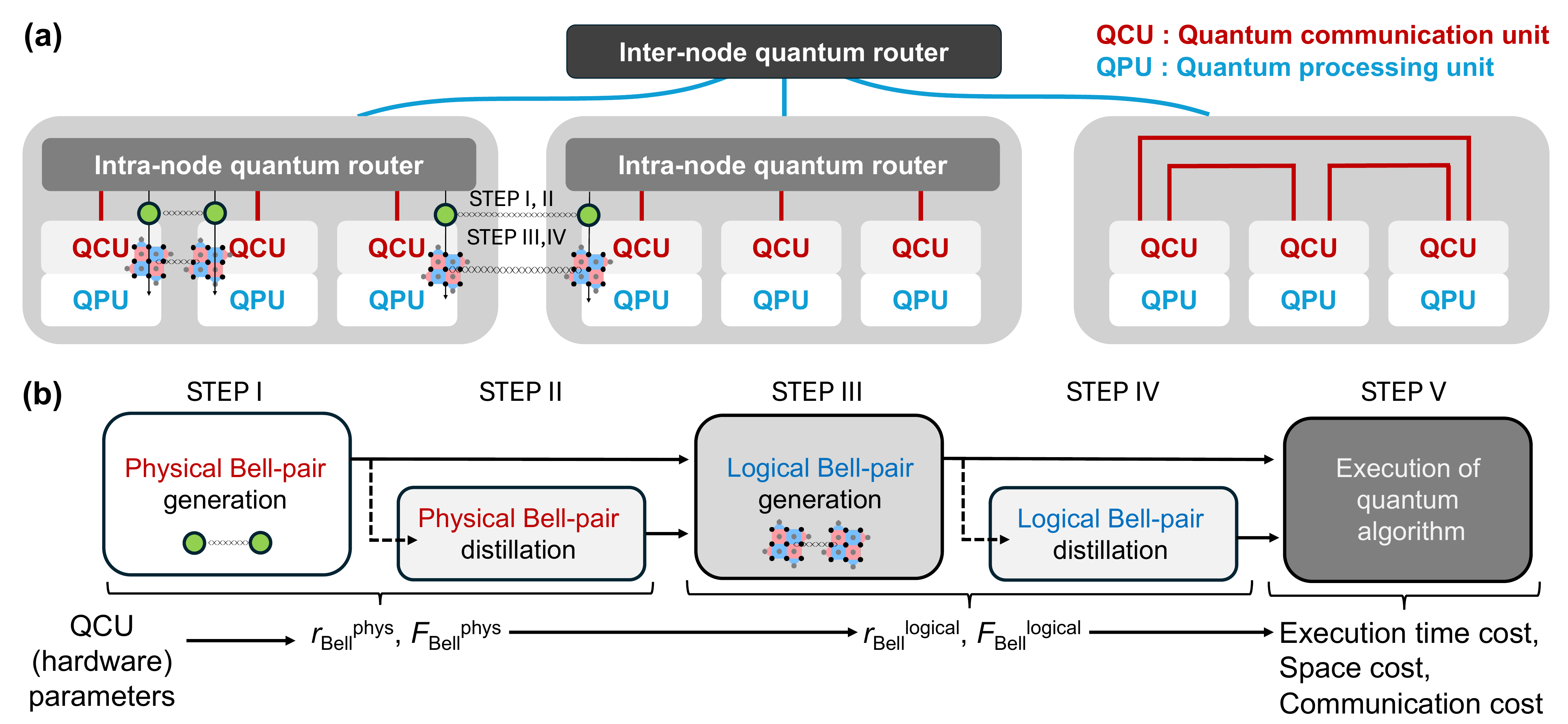}
	\caption{
		Architecture of a distributed fault-tolerant quantum computer and the resource-estimation procedure. (a) Schematic of a distributed fault-tolerant quantum computer. Each node contains quantum processing units (QPUs) and quantum communication units (QCUs). The QPUs within a node are connected via an intra-node router, whereas QPUs in different nodes are linked by an inter-node router. Direct connection between nodes is also possible as shown in the rightmost node. (b) Resource-estimation procedure for distributed quantum computation. The physical Bell-pair generation rate $r_\mathrm{Bell}^\mathrm{phys}$ and its fidelity  $F_\mathrm{Bell}^\mathrm{phys}$ are specified as parameters determined by the quantum communication hardware. Then, the distillation rate and fidelity of physical Bell pairs are calculated, if necessary. Based on the (distilled) physical Bell-pair generation rate and fidelity, the logical Bell-pair generation rate $r_\mathrm{Bell}^\mathrm{logical}$ and fidelity $F_\mathrm{Bell}^\mathrm{logical}$ can be estimated. The logical Bell pairs are also distilled, if needed. Finally, by combining these results with the analysis of quantum-algorithm execution, we obtain the space, time, and communication overheads.
	}
	\label{fig:overview}
\end{figure*}

Progress toward distributed fault-tolerant quantum computation
has been made in quantum-link hardware, methods for remote
logical operations, and system-level resource analysis.
Remote entanglement, a fundamental building block of
distributed quantum computing, has been demonstrated
experimentally using superconducting qubits \cite{Kurpiers2018},
solid-state defects \cite{Bernien2013a,Wei2025},
trapped ions \cite{Nadlinger2022,Saha2025},
and neutral atoms \cite{Covey2023,VanLeent2022}.
For superconducting processors, microwave interconnects
have been demonstrated, while microwave--optical transducers
are being developed for optical links
\cite{Axline2018a,Kurpiers2018,Almanakly2025,Storz2023,
Higginbotham2018,Mirhosseini2020,Sahu2022,Zhao2024}.
Nevertheless, achieving higher Bell-pair generation rates
and fidelities remains technically challenging, while
extensive multiplexing incurs substantial hardware overhead.
To enable fault-tolerant remote operations with imperfect
quantum links, several approaches have been developed,
including remote lattice surgery, direct encoding
of physical Bell pairs, and entanglement distillation
\cite{Ramette2024,Pattison2025,Marqversen2025,Sinclair2025,
Maeda2025}.
Beyond individual remote operations, studies have examined
how distributed architectures, network configurations,
compilation, and scheduling affect the resource requirements
of quantum algorithms
\cite{Wu2023QuComm,Saadatmand2026Modularity,
Sakuma2024QFly,Jacinto2026NetworkRequirements,Naito2026NBQC,
Filippov2025}.

Despite these advances, the connection between realistically
achievable quantum-link performance and fault-tolerant
algorithm-level resource requirements remains insufficiently
quantified.
Existing studies often rely on fixed or simplified
assumptions about other layers, even when multiple stages
are considered.
Consequently, it remains unclear how many physical qubits
per node are required and how long computations take
with realistically achievable quantum links, or what
Bell-pair generation rates, fidelities, and channel counts
are needed to meet target computational requirements.

To address these questions, we develop a unified framework
connecting physical and logical Bell-pair generation and
distillation to fault-tolerant quantum-algorithm execution
for superconducting processors connected by microwave or
optical links.
By optimizing Bell-pair distillation protocols and
network configurations for algorithm execution, we estimate
space, time, and communication overheads for given quantum-link
parameters and, conversely, determine the link performance
required to meet target computational requirements.

Our analysis shows that distributed fault-tolerant quantum
computation can substantially reduce the physical-qubit
requirements per node, even with a single physical
communication channel per node.
Using experimentally motivated, forward-looking microwave-link
parameters of a 1 MHz physical Bell-pair generation rate
and 90$\%$ fidelity,
distributing a computation across $\sim10$ nodes reduces
the required physical-qubit count per node by approximately
an order of magnitude, at the cost of a roughly tenfold
increase in execution time.
Conversely, reducing the execution-time overhead to a factor
of a few requires logical Bell-pair generation rates of
several kHz, demanding higher physical Bell-pair fidelities
and/or multiplexing over tens to hundreds of communication
channels.

These results highlight the potential feasibility of
distributed fault-tolerant quantum computing with limited
communication resources, while providing quantitative
guidance for improving communication efficiency and
advancing its practical realization.

\begin{figure*}[thb]
	\centering
	\includegraphics[width=0.97\linewidth]{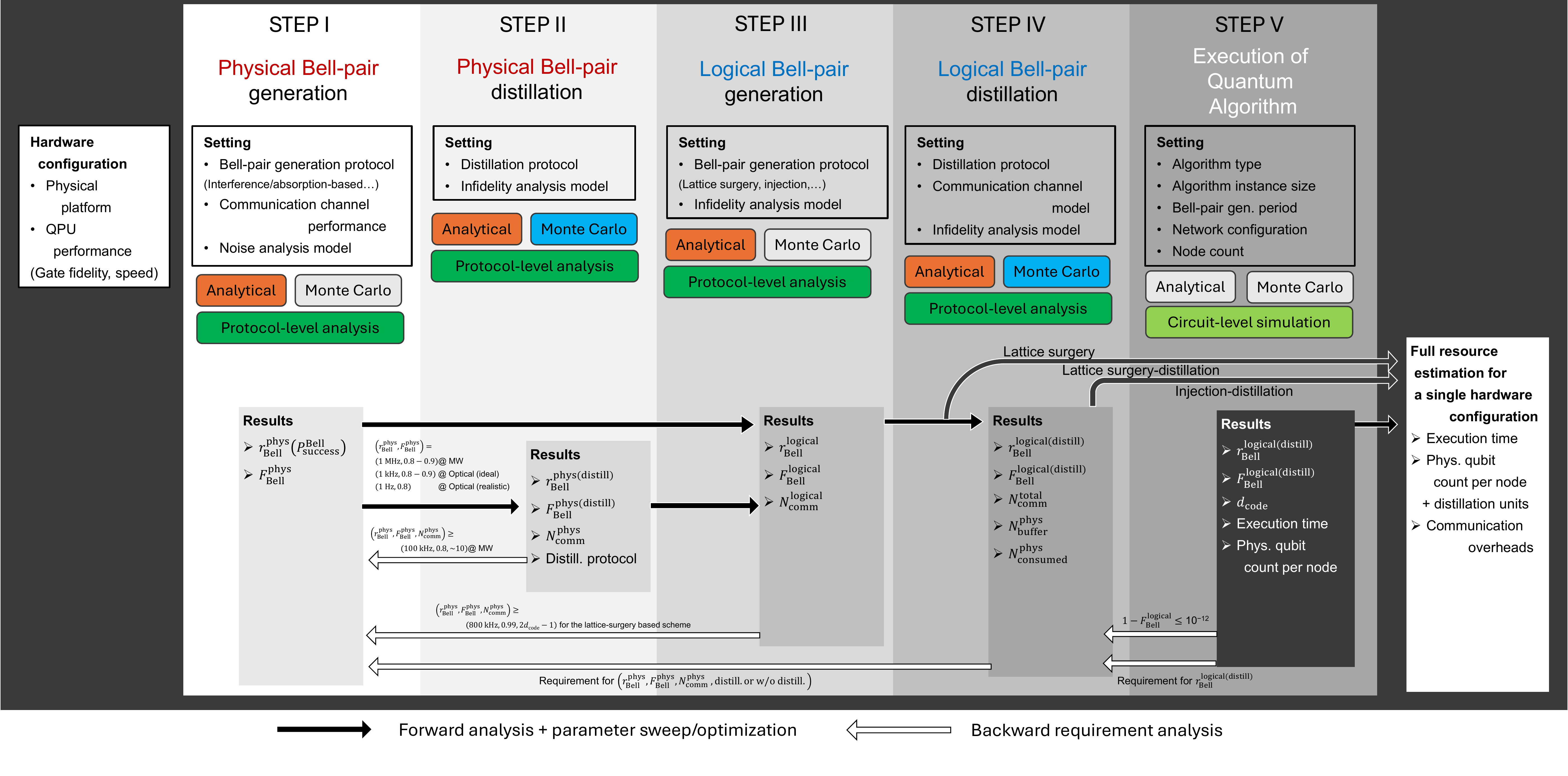}
\caption{
Detailed workflow of the resource analysis outlined in Fig.~\ref{fig:overview}.
Steps I--V connect physical Bell-pair generation and distillation, logical Bell-pair generation and distillation, and quantum-algorithm execution.
Each step combines independently specified settings, including protocols, noise models, and algorithm or network configurations, with parameters obtained from preceding steps, such as Bell-pair generation rates, fidelities, and communication-channel requirements.
Black arrows indicate forward analysis and parameter sweeps, whereas white arrows indicate backward analysis to estimate the communication performance required to satisfy target error rates and computational requirements.
Steps I--IV employ analytical calculations and, where applicable, Monte Carlo simulations of stochastic processes for protocol-level analysis.
Step V employs deterministic circuit-level simulations for algorithm-level resource estimation.
Colored boxes indicate the methods and analysis levels employed in each step, whereas unfilled boxes indicate methods not employed.
}
	\label{fig:calc_flow_summary}
\end{figure*}

\section{Overview of the architecture and its performance analysis}

In this section, we provide an overview of the architecture of a distributed fault-tolerant  quantum computer and outline the procedure for estimating the performance of a distributed system. Distributed  fault-tolerant quantum computation is implemented through quantum operations between logical qubits located in different nodes. First, quantum communication channels are used to share physical Bell pairs between remote nodes. Logical Bell pairs are then generated from physical Bell pairs in a fault-tolerant manner. Remote logical gate operations can be performed by consuming logical Bell pairs shared between nodes, i.e., through logical gate teleportation. Consequently, quantum algorithms, which are compiled into sequences of local and remote logical gate operations, are executed on the distributed quantum system. In other words, if a quantum algorithm can be decomposed into a combination of local logical operations on small quantum nodes and remote logical operations between them, and if logical Bell pairs with sufficiently low error rates can be prepared from physical Bell pairs, the quantum algorithm can be performed on the distributed architecture.

Notably, instead of consuming logical Bell pairs, direct remote two-qubit logical operations, and more generally remote $n$-qubit logical operations, are also possible in a regime in which the quantum links between remote nodes are sufficiently fast and high-fidelity that the nodes can effectively be treated as a single large node. However, for simplicity, we do not consider this case. This is because such a regime demands extremely high performance from the quantum communication hardware, and there is little need to model the communication explicitly in this limit.  Throughout this paper, a physical qubit or physical Bell pair denotes an unencoded qubit or an entangled pair of physical qubits, respectively, whereas a logical qubit is encoded in a surface-code patch and a logical Bell pair denotes an entangled pair of such encoded logical qubits.

An illustrative example of an architecture for a distributed fault-tolerant quantum computer is shown in Fig.~\ref{fig:overview}(a). Quantum processing units (QPUs) in each node communicate with each other via quantum communication units (QCUs). Each QPU hosts logical qubits. A QCU consists of communication qubits that enable remote physical Bell-pair generation and memory/buffer qubits for physical and logical Bell-pair distillation. QCUs are linked by direct connection between them or an intra-node quantum router, which can switch the communication target arbitrarily. Furthermore, the intra-node quantum routers in the nodes are connected to an inter-node quantum router, which controls the communication channels between nodes. For superconducting qubits, intra-node communication can be realized using microwave links, whereas inter-node communication can be based on microwave links, optical links, or their hybridization, depending on the target architecture. In the following analysis, rather than restricting ourselves to the specific architecture illustrated in Fig.~\ref{fig:overview}(a), we consider circle, mesh, tree, and full connectivity between nodes.

\begin{figure*}[t]
	\centering
	\includegraphics[width=180 mm]{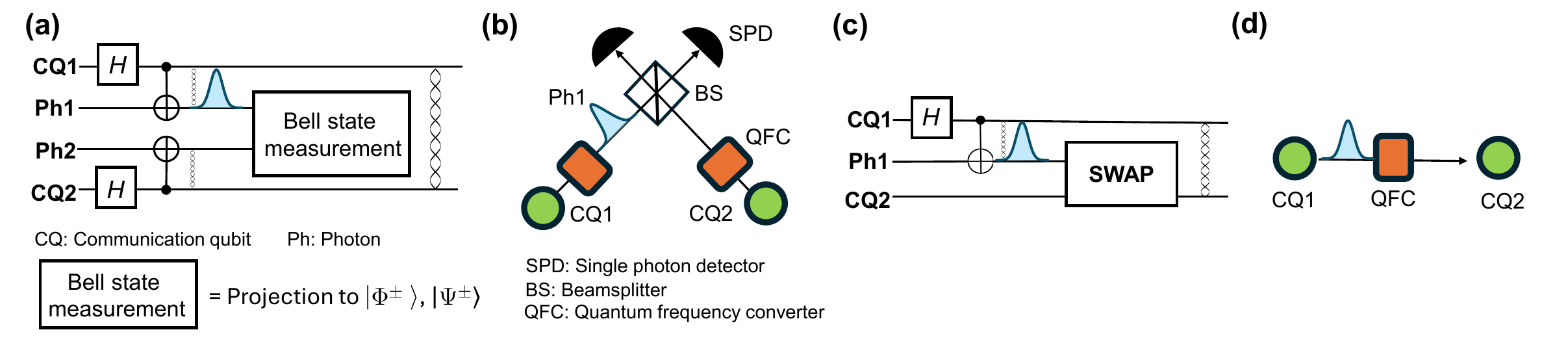}
	\caption{
		Schematic of remote physical Bell-pair generation schemes. (a) Quantum circuit for generating a physical Bell pair between communication qubits (CQs). In each node, entanglement between a communication qubit and a photon (Ph) is generated. A Bell-state measurement on the photons from the two nodes generates a physical Bell pair between the communication qubits. (b) Typical physical implementation of (a). A communication qubit in each node emits an entangled photon. A quantum frequency converter (QFC) adjusts the photon frequency so that the photons can interfere. A beamsplitter at the midpoint erases which-path information, and the photons are detected using single-photon detectors (SPDs). The use of a beamsplitter and SPDs corresponds to an incomplete Bell-state measurement, which projects the photonic state onto two of the four Bell states. Common photonic encodings for this protocol include Fock-state qubits and time-bin qubits. (c) Quantum circuit and (d) schematic for the absorption-based physical Bell-pair generation scheme. The photon entangled with CQ1 is absorbed by CQ2 such that the photonic state is swapped to CQ2. If necessary, a QFC can be used to adjust the photon frequency to match that of CQ2.}
	\label{fig:phys_bell_pair_gen}
\end{figure*}

The end-to-end performance of a distributed fault-tolerant quantum computer, from the hardware to the software levels, is analyzed through five steps, as outlined in Fig.~\ref{fig:overview}(b) and detailed in Fig.~\ref{fig:calc_flow_summary}. First, the physical Bell-pair generation rate, $r_\mathrm{Bell}^\mathrm{phys}$, and fidelity, $F_\mathrm{Bell}^\mathrm{phys}$, are calculated from quantum communication hardware parameters (Step~I). In analyses that focus on the subsequent steps, these quantities can also be treated as effective link parameters and varied independently. If needed, physical Bell-pair distillation is performed to improve fidelity at the cost of generation rate (Step~II). Using the resulting $r_\mathrm{Bell}^\mathrm{phys}$ and $F_\mathrm{Bell}^\mathrm{phys}$, we estimate the logical Bell-pair generation rate, $r_\mathrm{Bell}^\mathrm{logical}$, and fidelity, $F_\mathrm{Bell}^\mathrm{logical}$ (Step~III). If necessary, additional logical Bell-pair distillation is performed to achieve the required fidelity (Step~IV). Finally, by combining the resulting $r_\mathrm{Bell}^\mathrm{logical}$ and $F_\mathrm{Bell}^\mathrm{logical}$ with an analysis of the resource costs of executing quantum algorithms, we estimate the space, time, and communication overheads of distributed fault-tolerant quantum computation (Step~V). In this analysis, we vary the quantum-link parameters, QPU sizes, problem sizes, and network configurations.

In addition to this forward analysis, we perform parameter
sweeps, optimize Bell-pair distillation protocols and network
configurations for algorithm execution, and, where relevant,
trace requirements backward from later steps to determine
the necessary parameter ranges in earlier steps.

In the following sections, we explain the operations associated with Steps~I--V and quantitatively analyze their performance using physically motivated and idealized device parameters.

\section{Step I: Physical Bell-pair generation}

\subsection{Physical Bell-pair Generation Protocols}

In this paper, we consider two representative schemes for remote physical Bell-pair generation: a photon-interference-based scheme and a photon-absorption-based scheme. Although many variants of these schemes exist, including schemes using entangled-photon sources, we consider these two schemes to capture the basic features, namely the rate and fidelity, of remote physical Bell-pair generation protocols.  Although we use a simplified noise model to estimate an upper bound on the fidelity, more detailed noise models are required for precise quantitative predictions.

\subsubsection{Interference-based protocol}

Interference-based physical Bell-pair generation typically consists of the following two steps: entanglement generation between a communication qubit and a photon (flying qubit), followed by a Bell-state measurement on the photons (Fig.~\ref{fig:phys_bell_pair_gen}(a,b)). A Bell-state measurement projects a two-qubit state onto one of the Bell states. The interference-based Bell-state measurement is typically realized using a beamsplitter and single-photon detectors. During photon transmission, quantum frequency conversion can be performed, if necessary, to match the photon frequencies. A feed-forward operation conditioned on the Bell-state measurement outcome can prepare the desired Bell state. Because the four Bell states ($\ket{\Phi^{\pm}},\ket{\Psi^{\pm}}$) can be interconverted using a local single-qubit gate, we specify the type of Bell state only when necessary.

	\begin{figure*}[t]
	\centering
	\includegraphics[width=160 mm]{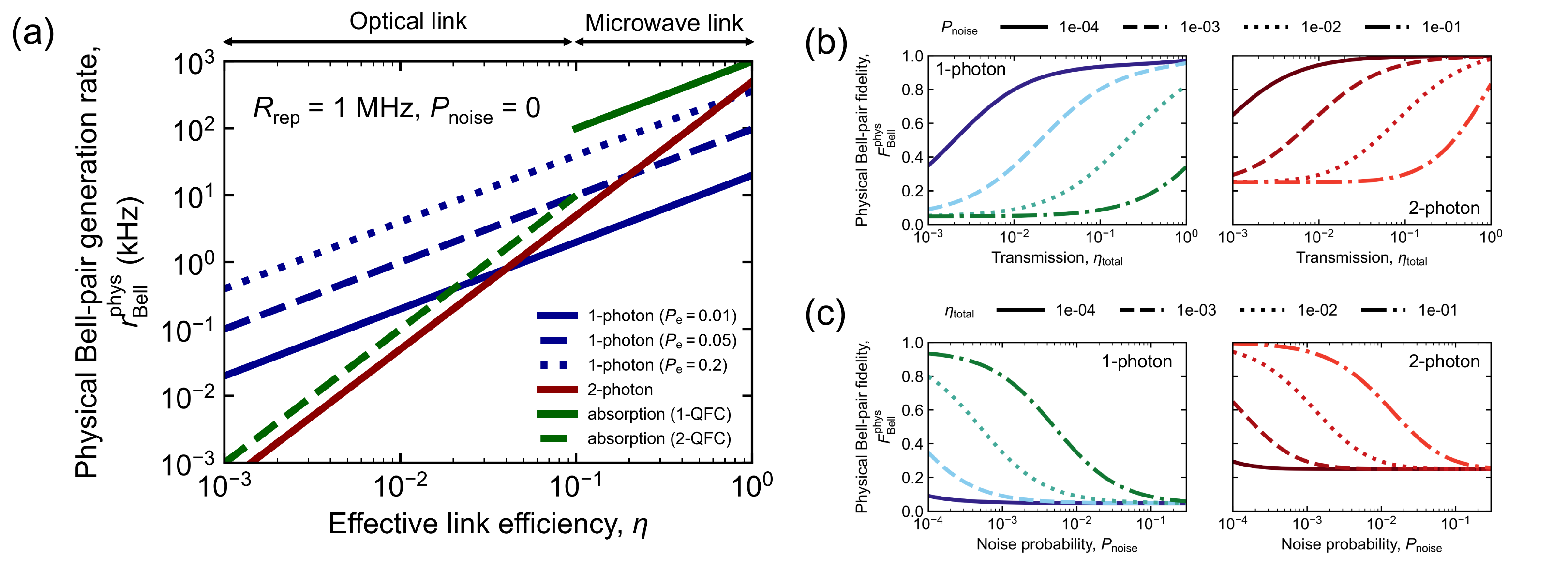}
	\caption
	{Physical Bell-pair generation rate and fidelity. (a) Physical Bell-pair generation rate
$r_\mathrm{Bell}^\mathrm{phys}$ as a function of
the effective single-link efficiency $\eta$
($P_\mathrm{noise}=0$). (b) Physical Bell-pair fidelity  $F_\mathrm{Bell}^\mathrm{phys}$ as a function of $\eta_\mathrm{total}$ for different noise-click probabilities $P_\mathrm{noise}$ for the single-photon protocol ($P_\mathrm{e}=0.05$) (left) and the two-photon protocol (right). (c) $F_\mathrm{Bell}^\mathrm{phys}$ as a function of $P_\mathrm{noise}$ for the single-photon protocol (left) and the two-photon protocol (right).}
	\label{fig:phys_bell_pair_gen_rate_fid}
\end{figure*}

The physical Bell-pair generation rate, $r_\mathrm{Bell}^\mathrm{phys}$, and fidelity, $F_\mathrm{Bell}^\mathrm{phys}$, for the interference-based scheme depend on the details of the protocol. There are two major protocols: the single-photon protocol, or single-click scheme, \cite{Cabrillo1999} and the two-photon protocol, or Barrett--Kok protocol/two-click scheme, \cite{Barrett2005a}. Here, we consider both protocols.

For the single-photon protocol, in the low-excitation and low-noise regime where double-excitation events are rare, the heralded physical Bell-pair generation rate and the conditional fidelity can be expressed as \cite{Zeuthen2020}
\begin{equation}\label{eq:rate_fidelity_1q}
	\begin{aligned}
		r_\mathrm{Bell}^\mathrm{phys}
		&\equiv r_\mathrm{click}
		=
		p_\mathrm{click}R_\mathrm{rep},\\
		p_\mathrm{click}
		&=
		P_\mathrm{e}\eta_\mathrm{total}
		(1-2P_\mathrm{noise})
		(2-P_\mathrm{e}\eta_\mathrm{total})
		+
		2P_\mathrm{noise},\\
		F_\mathrm{Bell}^\mathrm{phys}
		&=
		\\
		&\hspace{-15pt}
		\frac{
			2 P_\mathrm{e} (1 - P_\mathrm{e})\,\eta_\mathrm{total}C
			+ 2 P_\mathrm{e} (1 - P_\mathrm{e})(1 - \eta_\mathrm{total})\,P_\mathrm{noise}
		}{
			p_\mathrm{click}
		},
	\end{aligned}
\end{equation}
where $R_\mathrm{rep}$ is the repetition rate of the qubit--photon entanglement-generation attempt, $P_\mathrm{noise}$ ($\le 1$) is the noise-click probability within the detection frequency and time windows, $P_\mathrm{e}$ ($\le 1$) is the photon-emission probability of a communication qubit, and $\eta_\mathrm{total}$ is the photon transmission efficiency from the qubit to the detector, including the detector quantum efficiency. Here, $r_\mathrm{Bell}^\mathrm{phys}$ is identified with the heralded click rate $r_\mathrm{click}$, because true photon-induced clicks and noise clicks cannot be distinguished experimentally. False heralding events are therefore included in the conditional physical Bell-pair fidelity $F_\mathrm{Bell}^\mathrm{phys}$. The factor $C=(1+V\cos\theta)/2$ accounts for the infidelity caused by the finite coherence of the propagating photon and its phase shift, where $\theta$ is the photon phase shift and $V$ is the photon visibility. The noise-click probability $P_\mathrm{noise}$ originates from detector dark counts, thermal or Raman noise from the quantum frequency converter, and other experimental imperfections. The total detection efficiency, $\eta_\mathrm{total}=\eta_\mathrm{wg}\eta_\mathrm{emi}\eta_\mathrm{qfc}\eta_\mathrm{det}$, is determined by the waveguide propagation efficiency, $\eta_\mathrm{wg}$, the photon-emission coupling efficiency, $\eta_\mathrm{emi}$, the quantum-frequency-conversion efficiency, $\eta_\mathrm{qfc}$, and the detector quantum efficiency, $\eta_\mathrm{det}$. For simplicity, we assume the same $\eta_\mathrm{total}$ for both photon paths.

For the two-photon protocol (\(P_\mathrm{e}=0.5\)), which is based on interference between time-bin qubits and an appropriate bit-flip operation on the communication qubit \cite{Zeuthen2020}, the heralded physical Bell-pair generation rate and conditional fidelity are given by
\begin{equation}\label{eq:rate_fidelity_2q}
	\begin{aligned}
		r_\mathrm{Bell}^\mathrm{phys}
		&\equiv r_\mathrm{click}
		=
		p_\mathrm{click}^{(2)}R_\mathrm{rep},\\
		p_\mathrm{click}^{(2)}
		&=	\frac{1}{2}\\
		&\hspace{-20pt}
		\times
		\left[
		8P_\mathrm{noise}^{2}(1-\eta_\mathrm{total})^{2}
		+
		2P_\mathrm{noise}(4-3\eta_\mathrm{total})\eta_\mathrm{total}
		+
		\eta_\mathrm{total}^{2}
		\right],\\
		F_\mathrm{Bell}^\mathrm{phys}
		&=\\
		&\hspace{-15pt}
		\frac{
			2 P_\mathrm{noise}^{2} (1 - \eta_\mathrm{total})^{2}
			+ 2 P_\mathrm{noise} (1 - \eta_\mathrm{total})\,\eta_\mathrm{total}
			+ \eta_\mathrm{total}^{2}D
		}{
			8 P_\mathrm{noise}^{2} (1 - \eta_\mathrm{total})^{2}
			+ 2 P_\mathrm{noise} (4 - 3\eta_\mathrm{total})\,\eta_\mathrm{total}
			+ \eta_\mathrm{total}^{2}
		},
	\end{aligned}
\end{equation}
where \(p_\mathrm{click}^{(2)}\) is the probability of obtaining a heralding two-click event. The factor \(1/2\) reflects that only two of the four Bell states can be heralded using a beamsplitter and single-photon detectors; in the absence of noise clicks, \(p_\mathrm{click}^{(2)}=\eta_\mathrm{total}^{2}/2\). The factor \(D=1/2+V^2/2\) accounts for the finite coherence of the photons. In contrast to the single-photon protocol, the optical phase shift during the experiment does not reduce the fidelity in this scheme, provided that phase fluctuations are slow compared with the separation between the two time bins. Thus, active phase compensation for optical path-length fluctuations is not required to maintain \(\theta=0\), which is an advantage of the two-photon protocol over the single-photon protocol. In the low-noise limit, the heralded physical Bell-pair generation rate in the single-photon protocol is proportional to \(\eta_\mathrm{total}\), whereas that in the two-photon protocol scales quadratically with \(\eta_\mathrm{total}\). This difference becomes pronounced in the high-loss regime, \(\eta_\mathrm{total}\ll 1\).

\subsubsection{Absorption-based protocol}

The absorption-based physical Bell-pair generation scheme (Fig.~\ref{fig:phys_bell_pair_gen}(c,d)) is realized by emitting a photon entangled with a communication qubit, followed by absorption of the photon by another communication qubit. Photon absorption can be regarded as an effective SWAP operation between the entangled photon and the second communication qubit. This scheme is applicable only to physical platforms with high photon-absorption efficiency and a suitable energy-level structure, such as superconducting qubits \cite{Kurpiers2018}.

For a heralded absorption-based scheme, transmission and absorption losses primarily reduce the usable physical Bell-pair generation rate, which can be approximated as \cite{Kurpiers2019}
\begin{equation}\label{eq:abs_ent_gen_rate}
	r_\mathrm{Bell}^\mathrm{phys}=\eta_\mathrm{total}'R_\mathrm{rep},
\end{equation}
where $\eta_\mathrm{total}'=\eta_\mathrm{emi}\eta_\mathrm{wg}\eta_\mathrm{qfc}\eta_\mathrm{abs}$ is the total transmission efficiency, including the photon-emission efficiency, $\eta_\mathrm{emi}$, and photon-absorption efficiency, $\eta_\mathrm{abs}$. In an unheralded implementation, by contrast, unsuccessful transfer events cannot be discarded and therefore contribute to the error of the generated state rather than simply reducing the generation rate. Thus, the resulting physical Bell-pair fidelity depends strongly on the physical platform, the detailed protocol, and whether transfer failures can be detected. Therefore, we do not provide a general fidelity formula here but instead use experimentally reported generation rates and fidelities.

According to Ref.~\cite{Kurpiers2018}, the upper bounds on
$\eta_\mathrm{emi}$ and $\eta_\mathrm{abs}$ exceeded 0.9, and
the successful transfer probability corresponding to
$\eta_\mathrm{total}'$ was 0.6--0.7.
Early experiments demonstrated physical Bell-pair fidelities
of approximately 0.8--0.9 \cite{Kurpiers2018,Kurpiers2019},
while a more recent flying-microwave-photon experiment achieved
a readout-corrected Bell-pair fidelity of $0.942\pm0.006$
\cite{Qiu2025}.

\subsection{Numerical analysis}

Next, we consider applying these schemes to superconducting qubits. Superconducting qubits can communicate over a microwave link with no quantum frequency conversion or, at most, one quantum-frequency-conversion step. In contrast, quantum communication using an optical link always requires two microwave--optical frequency conversions. Thus, although an optical link provides flexible, reconfigurable, room-temperature, broadband communication channels, it suffers from loss and noise introduced by microwave-to-optical conversion. In the following analysis, we use
$\eta_\mathrm{total}(\eta_\mathrm{total}')\le0.1$
as a representative regime for optical links and
$0.1\le\eta_\mathrm{total}(\eta_\mathrm{total}')\le1$
as a representative regime achievable with microwave links. This assumption is based on recent experimental demonstrations in microwave-optical quantum transducers and superconducting qubit-based quantum communication  \cite{Higginbotham2018,Mirhosseini2020,Sahu2022,Zhao2024,Kurpiers2018,Kurpiers2019}. This classification is not an absolute rule but a working definition used in the following discussion. 

For the fixed parameters used in the calculation, we assume ideal photon
coherence, visibility, and phase stability, corresponding to $C=D=1$.
We set $R_\mathrm{rep}=1$ MHz as a representative repetition rate,
consistent with the sub-microsecond protocol timescale reported for
superconducting microwave quantum communication \cite{Kurpiers2019}.

For the numerical comparison, we use a nominal repetition
rate of $R_\mathrm{rep}=1$ MHz for both protocols.
If the time-bin repetition rate is instead fixed at 1 MHz,
the two-photon generation rate is reduced by approximately
a factor of two because each attempt requires two time bins.

To compare the protocols on a common horizontal axis in
Fig.~\ref{fig:phys_bell_pair_gen_rate_fid}(a), we introduce an
effective single-link efficiency $\eta$. For the interference-based
schemes, $\eta$ corresponds to $\eta_\mathrm{total}$ for each photon
path. For a microwave absorption-based link, we identify
$\eta\simeq\eta_\mathrm{total}'$. For an optical absorption-based
link, two microwave--optical conversion processes are required in
series. Assuming comparable efficiencies for the two conversion
stages and that frequency conversion is the dominant loss channel,
the end-to-end success probability scales approximately as $\eta^2$,
and we therefore use
$r_\mathrm{Bell}^\mathrm{phys}\simeq\eta^2R_\mathrm{rep}$.
We set \(P_\mathrm{noise}=0\) in Fig.~\ref{fig:phys_bell_pair_gen_rate_fid}(a)
to isolate the dependence of the physical Bell-pair generation rate
on transmission loss and to compare the intrinsic scaling of the protocols. 
In this limit, the single-photon protocol gives
\begin{equation}
	r_\mathrm{Bell}^{\mathrm{phys(1ph)}}\simeq
	2P_\mathrm{e}\eta R_\mathrm{rep},
\end{equation}
whereas the two-photon protocol gives
\begin{equation}
	r_\mathrm{Bell}^{\mathrm{phys(2ph)}}=
	\frac{1}{2}\eta^{2}R_\mathrm{rep}.
\end{equation}

Based on the analysis in Fig.~\ref{fig:phys_bell_pair_gen_rate_fid}(a), the resulting physical Bell-pair generation rate ranges from approximately
$1~\mathrm{Hz}$ to several tens of kHz for an optical link, depending on the protocol and parameters, whereas a microwave link can reach $100$--$1000~\mathrm{kHz}$. 
In the high-loss regime, the single-photon protocol becomes increasingly advantageous because its generation rate scales approximately linearly with $\eta$, whereas the two-photon and optical absorption-based schemes scale quadratically with $\eta$. The crossover points depend on $P_\mathrm{e}$ and on the protocol.
For example, for $P_\mathrm{e}=0.05$, the single-photon protocol
becomes faster than the optical absorption-based scheme at approximately
$\eta=0.1$. 
Regarding the generated physical Bell-pair fidelity for an optical link (Fig.~\ref{fig:phys_bell_pair_gen_rate_fid}(b,c)), at $\eta_\mathrm{total}=0.1$, achieving a modest physical Bell-pair fidelity of $0.8$ requires $P_\mathrm{noise}\sim1\times10^{-3}$ for the single-photon protocol and $P_\mathrm{noise}\sim5\times10^{-3}$ for the two-photon protocol. These requirements indicate that substantial simultaneous improvements
in microwave--optical conversion efficiency and added-noise performance
are still required compared with current demonstrations
\cite{Mirhosseini2020,Sahu2022,Zhao2024}.
In contrast, although we do not calculate the physical Bell-pair fidelity for a microwave link here, a fidelity of $0.8$--$0.9$ is feasible based on experimental demonstrations \cite{Kurpiers2018,Kurpiers2019}.

Based on these results, in the following analysis we use representative benchmark parameters for the physical Bell-pair generation rate and fidelity, rather than assigning each case to a single point in Fig.~\ref{fig:phys_bell_pair_gen_rate_fid}. We choose physical Bell-pair generation rates separated by roughly three orders of magnitude to capture the transition from microwave links to optimistic and realistic optical links. For microwave links, we assume \(r_\mathrm{Bell}^\mathrm{phys}=1~\mathrm{MHz}\) and \(F_\mathrm{Bell}^\mathrm{phys}=0.8\)--0.9. For optimistic optical links, we assume \(r_\mathrm{Bell}^\mathrm{phys}=1~\mathrm{kHz}\) and \(F_\mathrm{Bell}^\mathrm{phys}=0.8\)--0.9, whereas for realistic optical links, we assume \(r_\mathrm{Bell}^\mathrm{phys}=1~\mathrm{Hz}\) and \(F_\mathrm{Bell}^\mathrm{phys}=0.8\).

\section{Step II: Physical Bell-pair distillation}
\subsection{Physical Bell-pair Distillation Scheme}

\begin{figure}[t]
	\centering
	\includegraphics[width=85 mm]{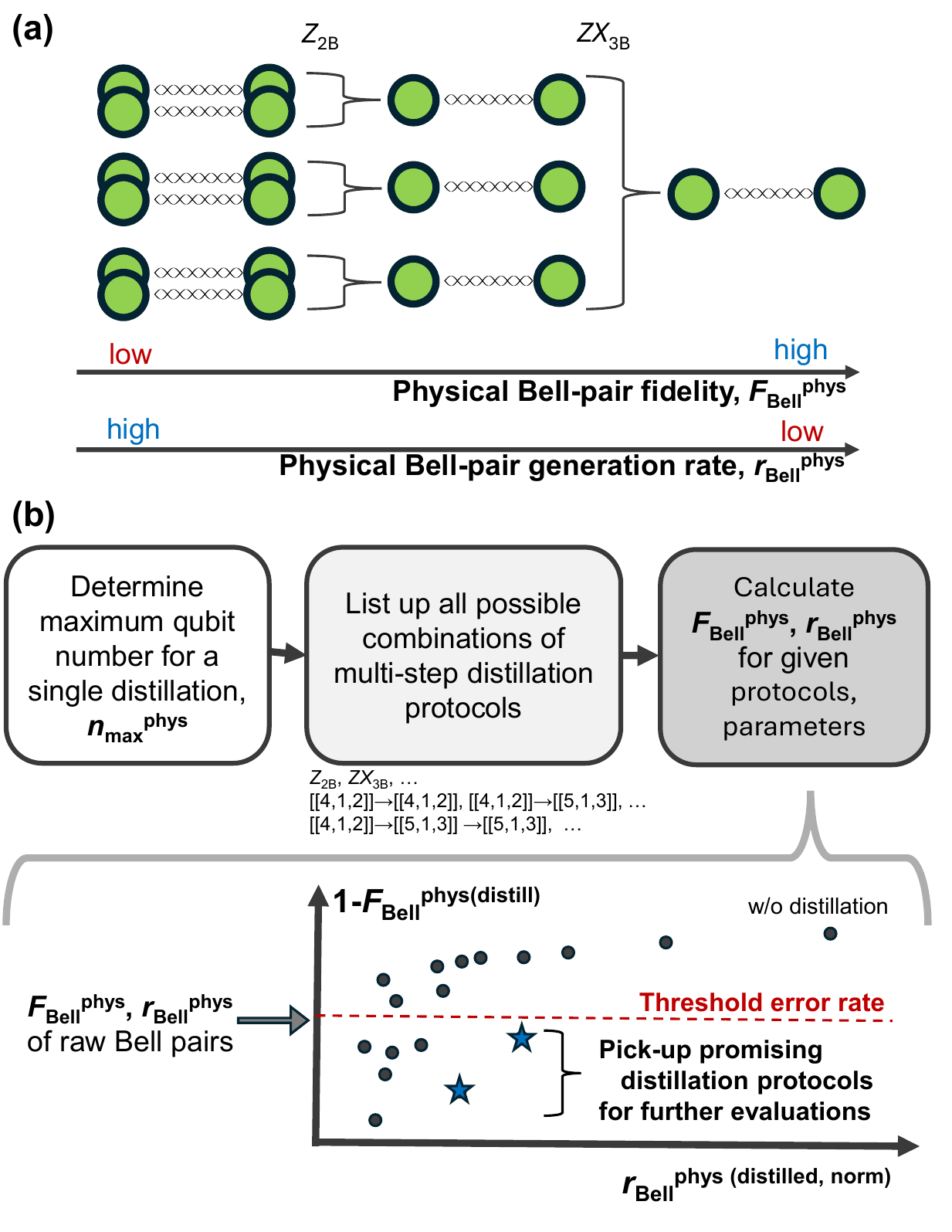}
	\caption
	{Distillation of physical Bell pairs. (a) Schematic of a cascaded physical Bell-pair distillation process, $Z_\mathrm{2B}\rightarrow ZX_\mathrm{3B}$. Physical Bell-pair distillation generates $k$ high-fidelity physical Bell pairs from $n$ low-fidelity physical Bell pairs at the expense of the total physical Bell-pair generation rate. (b) Procedure for selecting higher-performing distillation protocols from various combinations of distillation protocols. Each dot in the bottom graph corresponds to a distillation protocol.}
	\label{fig:phys_distill_rate_fid}
\end{figure}

After generating physical Bell pairs, if necessary, physical Bell-pair distillation can be performed to increase the physical Bell-pair fidelity at the expense of the physical Bell-pair generation rate. The purpose of this section is to estimate the space and time overheads of physical Bell-pair distillation by calculating $r_\mathrm{Bell}^\mathrm{phys}$ and $F_\mathrm{Bell}^\mathrm{phys}$ after distillation, together with the required additional physical qubits, for various distillation protocols and input parameters.

First, we introduce the general concepts and notation for physical Bell-pair distillation used in this paper. In general, physical Bell-pair distillation protocols generate $k$ distilled physical Bell pairs with higher fidelity from $n$ input physical Bell pairs with lower fidelity, with a finite success probability $p_\mathrm{success}^\mathrm{phys,distill}$. This process can be implemented using a combination of single-qubit gates, CNOT gates, and measurements on physical qubits. In the language of error correction, the code distance $d$ characterizes the error-correction capability. 

Distillation protocols can be classified into three types: classical error detection, which can detect either $X$ or $Z$ errors; imperfect quantum error detection, which detects a limited set of $X$ and $Z$ errors; and quantum error detection, which can detect both $X$ and $Z$ errors. We distinguish these three cases using the notation $[n,k,d]$ for the classical case and $[[n,k,d]]$ for the quantum case. For imperfect quantum error detection, we use notation such as $Z_\mathrm{2B}$ and $ZX_\mathrm{3B}$ following Ref.~\cite{Siddhu2025}, where $Z$ ($X$) denotes the stabilizer used and the subscript indicates the number of input physical Bell pairs. Here, we use these codes only for error detection: a code of distance
$d$ can detect errors of weight up to $d-1$, whereas error correction
can correct errors of weight up to
$\lfloor(d-1)/2\rfloor$. In error detection, failed outcomes are discarded, and the same distillation process is repeated until success. Moreover, distillation protocols can be cascaded, e.g., $Z_\mathrm{2B}\rightarrow ZX_\mathrm{3B}\rightarrow\cdots$, to achieve higher fidelity (Fig.~\ref{fig:phys_distill_rate_fid}(a)). In our analysis, we assume isotropic Pauli errors by applying a twirling
operation before each distillation step, such that the $X$, $Y$, and
$Z$ error probabilities are equal.

To estimate the performance of distillation protocols, we select the best distillation protocols from a large set of protocol combinations, e.g., A, B, C, \dots; A$\rightarrow$B, A$\rightarrow$C, \dots; and A$\rightarrow$B$\rightarrow$C, \dots (Fig.~\ref{fig:phys_distill_rate_fid}(b)). The first step is to set the maximum number of input physical Bell pairs
allowed in each elementary distillation protocol,
$n_\mathrm{max}^\mathrm{physical}$.
Here, we set $n_\mathrm{max}^\mathrm{physical}=5$ for simplicity. We identify distillation protocols based on a library of quantum codes \cite{codetable}. For $n_\mathrm{max}^\mathrm{physical}=5$, we consider $Z_\mathrm{2B}$ ($[n,k]=[2,1]$), $ZX_\mathrm{3B}$ ($[n,k]=[3,1]$), $[[4,1,2]]$, and $[[5,1,3]]$, excluding $[[4,2,2]]$ because of the complexity of the error-propagation analysis; for details and implementations of these protocols, see the Appendix.

\begin{figure}[t]
	\centering
	\includegraphics[width=80 mm]{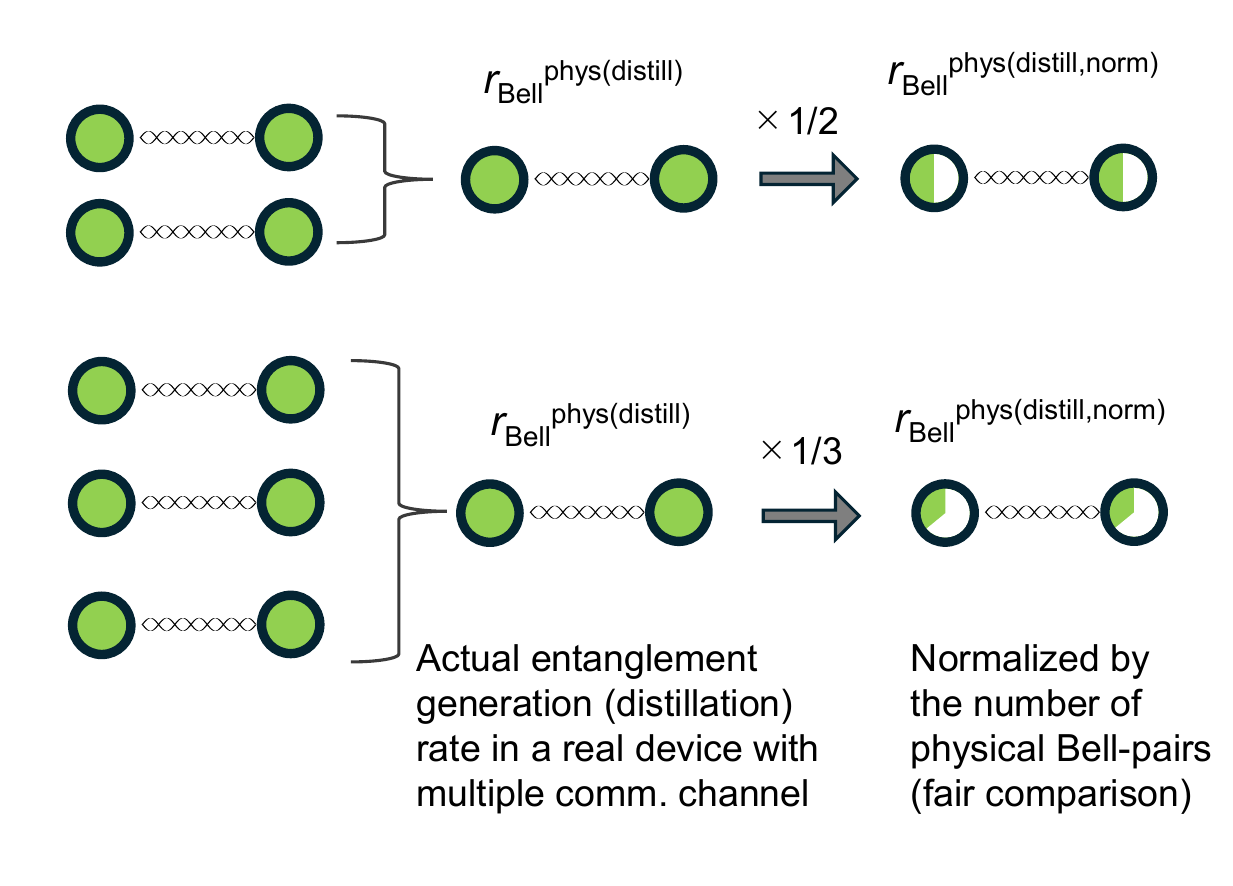}
	\caption
	{Definition of the physical Bell-pair distillation rate. In the situation shown schematically, two (three) parallel  communication channels are required to supply physical Bell pairs for the $Z_\mathrm{2B}$ ($ZX_\mathrm{3B}$) protocol. The resulting rate is the physical Bell-pair distillation rate  $r_\mathrm{Bell}^\mathrm{phys(distill)}$ which corresponds to the actual physical Bell-pair distillation rate in a real device. However, differences in the number of communication channels make comparisons among different distillation protocols unfair. Thus, we define a normalized physical Bell-pair distillation rate $r_\mathrm{Bell}^\mathrm{phys(distill,norm)}$ by dividing $r_\mathrm{Bell}^\mathrm{phys(distill)}$ by the minimum required number of communication channels $N_\mathrm{comm}^\mathrm{phys}$ i.e., the number of parallel communication channels.}
	\label{fig:explanation_of_rate_difinition}
\end{figure}

From these elementary distillation protocols, we construct cascaded distillation protocols such as $Z_\mathrm{2B}$, $Z_\mathrm{2B}\rightarrow ZX_\mathrm{3B}$, and so on. Assuming parallel communication channels, we then estimate $r_\mathrm{Bell}^{\mathrm{phys(distill},i)}$ and $F_\mathrm{Bell}^{\mathrm{phys(distill},i)}$ for each distillation process, which correspond to the physical Bell-pair generation rate and physical Bell-pair fidelity after the $i$th distillation step, as
\begin{equation}\label{eq:phys_rate_analytical}
	\begin{aligned}
		&\frac{1}{r_\mathrm{Bell}^{\mathrm{phys(distill},i=1)}}=\frac{1}{p_\mathrm{success}^{\mathrm{phys(distill},i=1)}}
		\\
		&\times\qty(\frac{\mathbb{E}[\mathcal{N}(p_\mathrm{success}^\mathrm{Bell},n_1)]}{R_\mathrm{rep}}
		+\tau^{\mathrm{phys(distill},i=1)})
		,
		\\
		&\frac{1}{r_\mathrm{Bell}^{\mathrm{phys(distill},i)}}=\frac{1}{p_\mathrm{success}^{\mathrm{phys(distill},i)}}
		\\
		&\times\qty(
		\frac{\mathbb{E}[\mathcal{N}(p_\mathrm{success}^{\mathrm{phys(distill},i-1)},n_i)]}
		{\frac{r_\mathrm{Bell}^{\mathrm{phys(distill},i-1)}}{p_\mathrm{success}^{\mathrm{phys(distill},i-1)}}}
		+\tau^{\mathrm{phys(distill},i)}
		)
		,
	\end{aligned}
\end{equation}
\begin{equation}\label{eq:phys_fidelity_analytical}
	\begin{aligned}
		&1-F_\mathrm{Bell}^{\mathrm{phys(distill},i=1)}
		\\
		&= f_1^\mathrm{phys}\qty(1-F_\mathrm{1q}^{\alpha_{i=1}} F_\mathrm{2q}^{\beta_{i=1}} F_\mathrm{meas}^{\gamma_{i=1}} F_\mathrm{Bell}^{\mathrm{phys}}),
		\\
		&1-F_\mathrm{Bell}^{\mathrm{phys(distill},i)}
		\\
		&= f_i^\mathrm{phys}\qty(1-F_\mathrm{1q}^{\alpha_i} F_\mathrm{2q}^{\beta_i} F_\mathrm{meas}^{\gamma_i}
		F_\mathrm{Bell}^{\mathrm{phys(distill},i-1)}),
	\end{aligned}
\end{equation}
where $\tau^{\mathrm{phys(distill},i)}$ is the execution time of the $i$th distillation step, estimated from the maximum number of qubit gate operations and measurements on a single qubit throughout the distillation circuit. Here, $p_\mathrm{success}^{\mathrm{phys(distill},i)}$ is the success probability of the $i$th distillation step; $\mathbb{E}[\mathcal{N}(p,n)]$ is the expected number of repetition cycles required until all $n$ Bell pairs are successfully prepared,
when the $n$ preparation processes are attempted in parallel with
success probability $p$ per cycle; and $f_i^\mathrm{phys}(1-F^{i-1})$ is the fidelity-improvement function of a distillation protocol as a function of the input physical Bell-pair infidelity. To account for finite operation errors during distillation, we multiply the input Bell-pair fidelity at each distillation step by $F_\mathrm{1q}^{\alpha_i} F_\mathrm{2q}^{\beta_i} F_\mathrm{meas}^{\gamma_i}$, 
where $F_\mathrm{1q}$, $F_\mathrm{2q}$, and $F_\mathrm{meas}$ are the fidelities of single-qubit operations, two-qubit operations, and measurements during distillation, respectively. The exponents $\alpha_i$, $\beta_i$, and $\gamma_i$ are twice the respective operation counts on the most heavily operated qubit in the distillation circuit (Fig.~B1), accounting for operations at both nodes. We assume $F_\mathrm{1q}=F_\mathrm{2q}=F_\mathrm{meas}=99.9\%$. Note that this treatment is an approximation to a full error-propagation analysis, assuming independent errors and  sufficiently low operation error rates ($\ll 1$). 

\begin{figure}[t]
	\centering
	\includegraphics[width=80 mm]{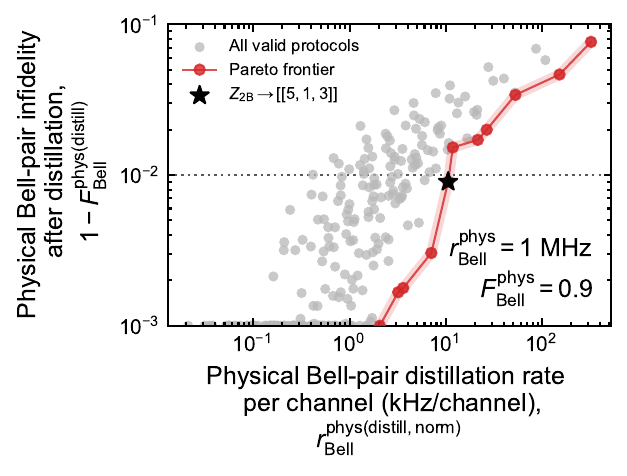}
	\caption
	{Comparison of physical Bell-pair distillation protocols for the input
		of $r_\mathrm{Bell}^\mathrm{phys}=1$ MHz and
		$F_\mathrm{Bell}^\mathrm{phys}=0.9$.
		The red circles indicate the performance frontier among all valid
		protocols, whereas the gray circles indicate the other protocols.
		The star corresponds to the combination
		$Z_\mathrm{2B}\rightarrow[[5,1,3]]$, which is used in the following
		analysis.}
	\label{fig:phys_distill_frontier}
\end{figure}

\begin{table}[]
	\centering
	\caption{Distillation protocols with the highest normalized physical Bell-pair distillation rate
		$r_\mathrm{Bell}^\mathrm{phys(distill,norm)}$ that satisfy the threshold infidelity
		$1-F_\mathrm{Bell}^\mathrm{phys(distill)}<0.01$ for various input raw physical Bell-pair fidelities
		$F_\mathrm{Bell}^\mathrm{phys}$. The underlined protocols are used in
		Fig.~\ref{fig:phys_rate_fid_after_distillation}.}
	\label{tab:set_of_phys_distillation}
	\begin{tabular}{c c c c c c}
		\hline\hline
		$F_{\mathrm{Bell}}^{\mathrm{phys}}$
		& $i=1$ & $i=2$ & $i=3$ & $i=4$
		& $N_{\mathrm{comm}}^{\mathrm{phys}}=\prod_i n_i$ \\
		\hline
		
		0.95
		& [[5,1,3]] &  &  &  & 5 \\
		\cline{2-2}
		& $ZX_{3\mathrm{B}}$ & $ZX_{3\mathrm{B}}$ &  &  & 9 \\
		
		0.9
		& $Z_{2\mathrm{B}}$ & [[5,1,3]] &  &  & 10 \\
		\cline{2-3}
		& $ZX_{3\mathrm{B}}$ & [[5,1,3]] &  &  & 15 \\
		
		0.85
		& $ZX_{3\mathrm{B}}$ & [[5,1,3]] &  &  & 15 \\
		\cline{2-3}
		& $ZX_{3\mathrm{B}}$ & $Z_{2\mathrm{B}}$ & [[5,1,3]] &  & 30 \\
		
		0.8
		& $ZX_{3\mathrm{B}}$ & $ZX_{3\mathrm{B}}$ & [[5,1,3]] &  & 45 \\
		\cline{2-4}
		& $ZX_{3\mathrm{B}}$ & $Z_{2\mathrm{B}}$ & $Z_{2\mathrm{B}}$ & [[5,1,3]] & 60 \\
		
		\hline\hline
	\end{tabular}
\end{table}

\begin{figure*}[t]
	\centering
	\includegraphics[width=150 mm]{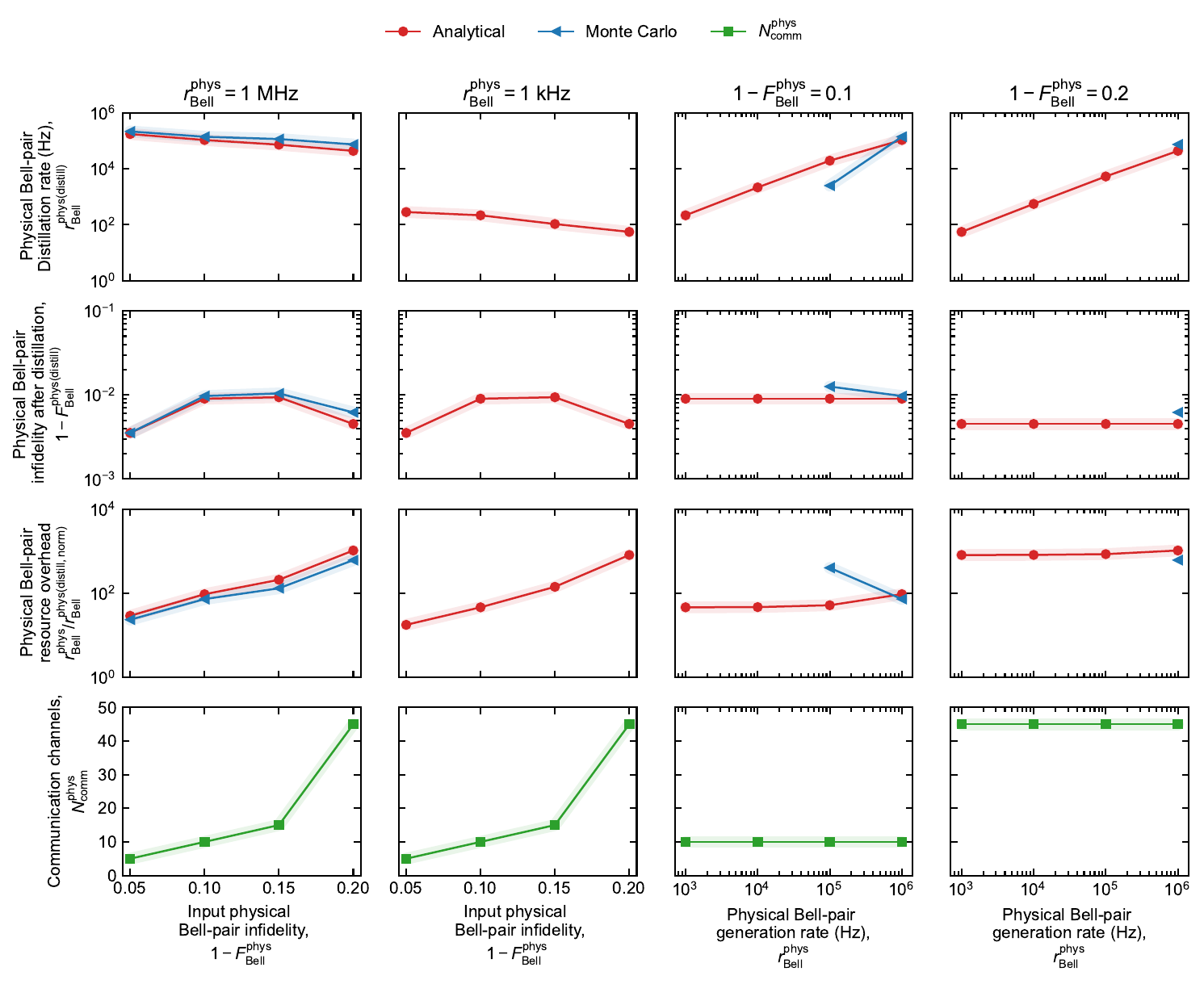}
	\caption
	{Performance analysis of physical Bell-pair distillation for different input physical Bell-pair generation rates $r_\mathrm{Bell}^\mathrm{phys}$ and fidelities $F_\mathrm{Bell}^\mathrm{phys}$. (First row) Physical Bell-pair distillation rate, corresponding to the rate achievable in an actual device with a sufficient number of communication channels. (Second row) Physical Bell-pair infidelity after distillation, $1-F_\mathrm{Bell}^\mathrm{phys(distill)}$. (Third row) Physical Bell-pair resource overhead per distilled physical Bell pair, $r_\mathrm{Bell}^\mathrm{phys}/r_\mathrm{Bell}^\mathrm{phys(distill,norm)}$. Red circles indicate analytical calculations using Eqs.~\eqref{eq:phys_rate_analytical} and \eqref{eq:phys_fidelity_analytical}, and blue triangles indicate Monte Carlo simulations that account for finite connectivity and qubit decoherence. (Fourth row) The number of parallel communication channels required in this model.}
	\label{fig:phys_rate_fid_after_distillation}
\end{figure*}

Since we assume that the physical Bell pairs required for each distillation
step are supplied in parallel through multiple communication channels to
avoid additional waiting-time decoherence associated with sequential
generation, we also introduce the normalized physical Bell-pair distillation rate, $r_\mathrm{Bell}^\mathrm{phys(distill,norm)}$, as
\begin{equation}\label{eq:phys_distill_rate_norm}
	r_\mathrm{Bell}^\mathrm{phys(distill,norm)}
	=
	r_\mathrm{Bell}^\mathrm{phys(distill)}\times\prod_i \frac{k_i}{n_i},
\end{equation}
where $k_i$ and $n_i$ denote the output and input physical Bell-pair numbers, respectively, of the $i$th distillation protocol. The factor $k_i/n_i$ normalizes the number of physical Bell pairs required for distillation across different protocols, enabling a fair comparison. In our case, because $k_i=1$ for all protocols,
$N_\mathrm{comm}^\mathrm{phys}=\prod_i n_i$
corresponds to the number of parallel communication channels assumed
to supply the raw physical Bell pairs for fully parallelized distillation.  Figure~\ref{fig:explanation_of_rate_difinition} illustrates the difference between $r_\mathrm{Bell}^\mathrm{phys(distill)}$ and $r_\mathrm{Bell}^\mathrm{phys(distill,norm)}$.

Equation~\eqref{eq:phys_rate_analytical} is a latency-based analytical
model that estimates the distillation rate from the expected time
required to prepare the input Bell pairs and execute each distillation
step. For a more realistic estimate of the achievable throughput,
including limited qubit connectivity and waiting-time decoherence,
we perform Monte Carlo simulations (see the Appendix for details).
We assume a two-dimensional array-like arrangement of qubits,
allowing nearest-neighbor connections to up to four neighboring
qubits. Decoherence is included by multiplying the fidelity by
$(1/2)[1+\mathrm{exp}(-2\tau_\mathrm{step}/T_1)]$ at each time step
in the simulation, assuming a $\ket{\Phi^+}$ physical Bell pair,
where $\tau_\mathrm{step}$ is the simulation time step and $T_1$
is the qubit relaxation time. We set $T_1=1~\mathrm{ms}$,
referring to state-of-the-art demonstrations with superconducting
qubits \cite{Somoroff2023,Bland2025}.
In the Monte Carlo simulations, Bell pairs waiting in a buffer
are discarded when their waiting time reaches
$T_\mathrm{threshold}=10~\mu\mathrm{s}$.

In this work, to limit computational cost, we perform Monte Carlo simulations only for selected physical Bell-pair distillation protocols rather than all possible combinations. A rigorous optimization would require evaluating all candidate protocols to identify the globally optimal strategy for each set of input parameters. Such exhaustive optimization is left for future work.

\subsection{Numerical Evaluation}

Using Eqs.~\eqref{eq:phys_rate_analytical},
\eqref{eq:phys_fidelity_analytical}, and
\eqref{eq:phys_distill_rate_norm}, we obtain
$r_\mathrm{Bell}^{\mathrm{phys(distill,norm)}}$ as a function of
$F_\mathrm{Bell}^{\mathrm{phys(distill)}}$ for the entire set of
distillation protocols, as shown in
Fig.~\ref{fig:phys_distill_frontier}.
The protocols forming the performance frontier provide the best
trade-off between the normalized distillation rate and output fidelity.
By setting a threshold fidelity, we select the protocol with the
largest $r_\mathrm{Bell}^{\mathrm{phys(distill,norm)}}$ among those
satisfying the threshold.
Table~\ref{tab:set_of_phys_distillation} shows the set of distillation protocols with the highest normalized physical Bell-pair generation rate that satisfy the threshold infidelity, $1-F_\mathrm{Bell}^\mathrm{phys(distill)}<0.01$, which is determined by considering the threshold for a specific remote logical operation \cite{Ramette2024}. As the input $F_\mathrm{Bell}^\mathrm{phys}$ decreases, the number of distillation steps increases. The lower bound on the total number of communication channels, $N_\mathrm{comm}^\mathrm{phys}$, also increases. We further analyze the performance of the selected protocols in Table~\ref{tab:set_of_phys_distillation} under conditions closer to actual experiments using Monte Carlo simulations.

 Based on the considerations above, we plot the physical Bell-pair distillation rate, physical Bell-pair fidelity after distillation, and distillation cost in Fig.~\ref{fig:phys_rate_fid_after_distillation} while varying the input physical Bell-pair generation rate and fidelity, using both the analytical formulas, Eqs.~\eqref{eq:phys_rate_analytical} and \eqref{eq:phys_fidelity_analytical}, and numerical simulations.
Here, we analyze the protocols underlined in Table~\ref{tab:set_of_phys_distillation}, which require the fewest physical Bell pairs for each input $F_\mathrm{Bell}^\mathrm{phys}$. For the distillation rate, we show
$r_\mathrm{Bell}^\mathrm{phys(distill)}$, which corresponds to the
physical Bell-pair distillation rate when the required number of
parallel communication channels is available
(the first row of Fig.~\ref{fig:phys_rate_fid_after_distillation}). The third row of Fig.~\ref{fig:phys_rate_fid_after_distillation} characterizes the communication overhead of the protocols using the physical Bell-pair resource overhead, $r_\mathrm{Bell}^\mathrm{phys}/r_\mathrm{Bell}^\mathrm{phys(distill,norm)}$, which approximately corresponds to the number of consumed physical Bell pairs. The fourth row shows the number of communication channels for physical Bell-pair distillation $N_\mathrm{comm}^\mathrm{phys}$, ranging from $5$ to $\sim50$.

The first and second columns of Fig.~\ref{fig:phys_rate_fid_after_distillation}
compare the distillation performance for a fast microwave link and a
slow optical link. For an input physical Bell-pair generation rate of
$r_\mathrm{Bell}^\mathrm{phys}=1~\mathrm{MHz}$, assuming a high-quality
microwave link, $r_\mathrm{Bell}^\mathrm{phys(distill)}$ decreases
from 170 kHz to approximately 40 kHz as the input
infidelity increases from $0.05$ to $0.2$, while the output infidelity
remains below approximately $10^{-2}$. The physical Bell-pair consumption
factor increases from approximately $10^2$ at an input infidelity of
$0.1$ to approximately $10^3$ at $0.2$.
In contrast, for
$r_\mathrm{Bell}^\mathrm{phys}=1~\mathrm{kHz}$, assuming an optical link,
the analytical model predicts a distillation rate of approximately
$50$--$300~\mathrm{Hz}$, depending on the input infidelity.
However, Monte Carlo simulations at this low generation rate
could not be completed within a practical computation time
because successful output events became extremely rare under
the finite buffer-lifetime constraint.

This trend is also evident in the third and fourth columns of
Fig.~\ref{fig:phys_rate_fid_after_distillation}. The Monte Carlo
simulations indicate that the minimum practical generation rate strongly
depends on the input fidelity: approximately $100~\mathrm{kHz}$ is
required for an input infidelity of $0.1$, whereas a rate close to
$1~\mathrm{MHz}$ is required for an input infidelity of $0.2$.
These are approximate rate scales rather than exact threshold
values; for example, at 100 kHz and an input infidelity of
0.1, the Monte Carlo output infidelity is approximately 0.013.
These results show that physical Bell-pair distillation using kHz-scale
links is strongly limited by waiting-time decoherence for
$T_1=1~\mathrm{ms}$.

Overall, when waiting-time decoherence is taken into account, practical physical Bell-pair distillation requires relatively fast and sufficiently high-fidelity physical Bell-pair generation. For $T_1=1~\mathrm{ms}$, our results indicate that generation rates above approximately $100~\mathrm{kHz}$ are required for $F_\mathrm{Bell}^\mathrm{phys}\gtrsim0.9$, whereas rates approaching $1~\mathrm{MHz}$ are required when $F_\mathrm{Bell}^\mathrm{phys}\sim0.8$, together with on the order of ten parallel communication channels. As a rough criterion for achieving an output fidelity of approximately $99\%$, the waiting-time decoherence should satisfy $1/(r_\mathrm{Bell}^\mathrm{phys}T_1)\lesssim10^{-2}$, although the precise requirement depends on the input fidelity and distillation protocol. According to the link-performance estimates in the previous section, this regime is accessible primarily with microwave links.

\section{Step III: Logical Bell-pair generation}

\subsection{Logical Bell-pair Generation Protocols}

\begin{figure}[t]
	\centering
	\includegraphics[width=85 mm]{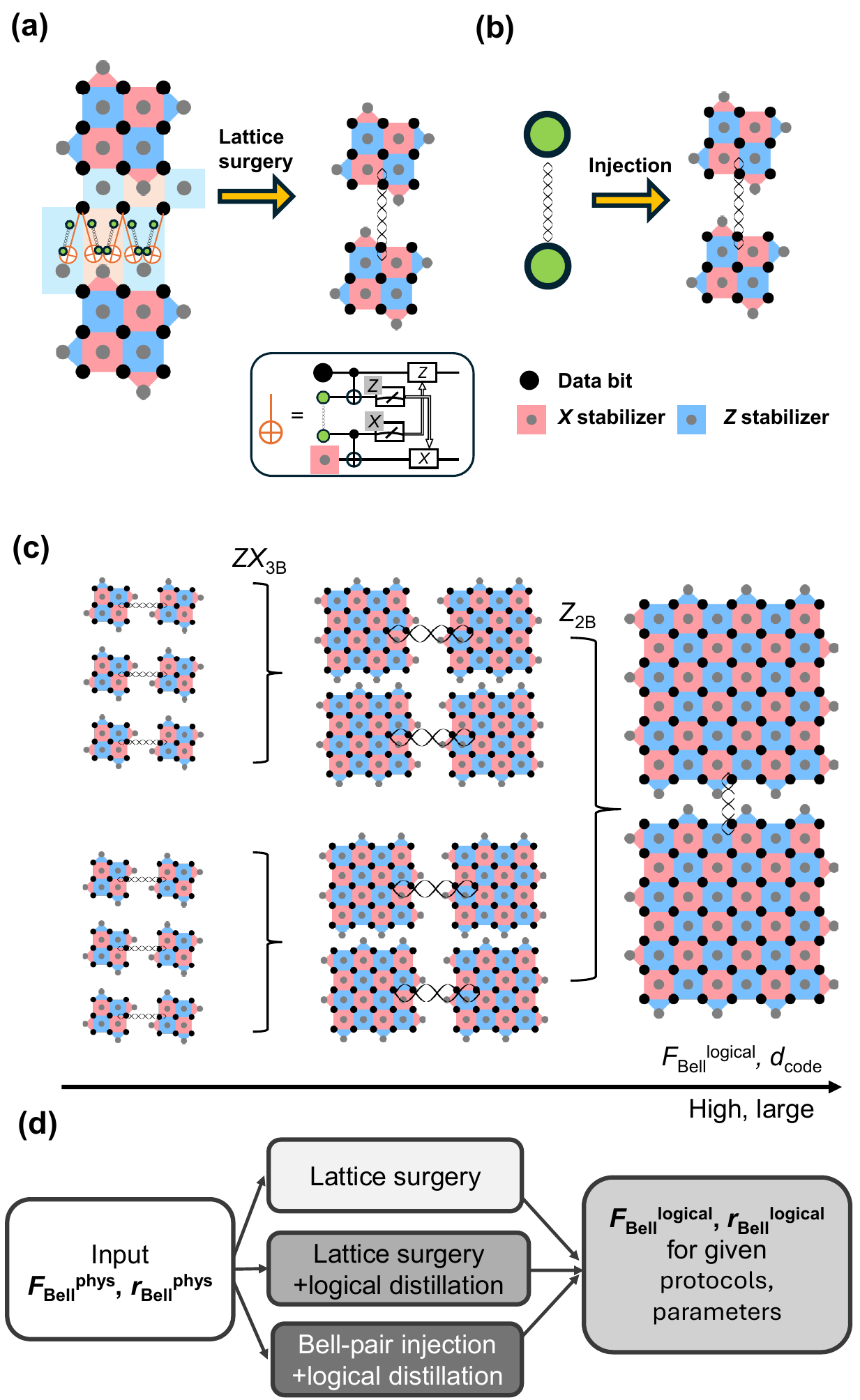}
	\caption
	{Logical Bell-pair generation and distillation for preparing sufficiently low-error logical Bell pairs. (a) Lattice-surgery-based logical Bell-pair generation. Nonlocal CNOT operations between surface-code patches are performed using pre-shared physical Bell pairs and local operations and classical communication (LOCC). (b) Injection of a physical Bell pair to generate a logical Bell pair. (c) Logical Bell-pair distillation cascade. As the target error rate decreases through distillation, the code distance must be increased to suppress the logical error rate. (d) Strategy for logical Bell-pair generation and performance estimation. For a given protocol, we calculate the logical Bell-pair generation rate $r_\mathrm{Bell}^\mathrm{logical}$ and fidelity $F_\mathrm{Bell}^\mathrm{logical}$ from the input physical Bell-pair generation rate $r_\mathrm{Bell}^\mathrm{phys}$ and fidelity $F_\mathrm{Bell}^\mathrm{phys}$.}
	\label{fig:logical_gen_distill}
\end{figure}

Logical Bell pairs can be generated by consuming the physical Bell pairs prepared through Steps~I and II. Although several schemes for generating logical Bell pairs have been proposed \cite{Maeda2025}, we focus on two methods, lattice surgery and state injection, because of their compatibility with the two-dimensional array configuration of superconducting qubits. We estimate the logical Bell-pair generation rate, $r_\mathrm{Bell}^\mathrm{logical}$, and fidelity, $F_\mathrm{Bell}^\mathrm{logical}$, for these two schemes. We also assume the rotated surface code for encoding logical states.

\subsubsection{Lattice Surgery}

We first provide an overview of lattice-surgery-based logical Bell-pair generation (Fig.~\ref{fig:logical_gen_distill}(a)). Lattice surgery is a framework for performing fault-tolerant $n$-qubit logical operations by combining initialization and measurement of data qubits with the activation and deactivation of $X$/$Z$ syndrome measurements \cite{Litinski2019}. Logical Bell-pair generation consists of three steps: (i) initialization of a logical state in each node, i.e., $\ket{+}^\mathrm{logical}$; (ii) a lattice-merge operation; and (iii) a lattice-split operation. This process can be understood as a measurement of $Z^\mathrm{logical}\otimes Z^\mathrm{logical}$ on $\ket{+}^\mathrm{logical}\otimes \ket{+}^\mathrm{logical}=(1/2)(\ket{00}^\mathrm{logical}+\ket{01}^\mathrm{logical}+\ket{10}^\mathrm{logical}+\ket{11}^\mathrm{logical})$, which projects the state onto $\ket{00}^\mathrm{logical}+\ket{11}^\mathrm{logical}$ for the $+1$ measurement outcome and onto $\ket{01}^\mathrm{logical}+\ket{10}^\mathrm{logical}$ for the $-1$ measurement outcome. In these steps, (i) initialization and (iii) lattice splitting can be performed locally and therefore do not require remote operations. In contrast, (ii) lattice merging requires $d_\mathrm{code}$ rounds of syndrome measurements between patches after the intermediate data qubits are initialized, where $d_\mathrm{code}$ is the code distance of the surface code. When two surface-code patches are spatially separated, the edge syndrome measurements must be performed via physical gate teleportation, i.e., remote CNOT operations, consuming pre-shared physical Bell pairs (Fig.~\ref{fig:logical_gen_distill}(a)). Lattice merging consumes $(2d_\mathrm{code}-1)$ physical Bell pairs in each round, repeated $d_\mathrm{code}$ times, with the surface-code cycle time $\tau_\mathrm{code}^\mathrm{cycle}$ characterizing the duration of a single syndrome-measurement round. Thus, it requires $\propto d_\mathrm{code}^2$ physical Bell pairs per logical Bell-pair generation.

For lattice-surgery-based logical Bell-pair generation, each syndrome-measurement round requires \((2d_\mathrm{code}-1)\) remote CNOT. We assume that these checks are supported by \((2d_\mathrm{code}-1)\) parallel communication links, so that the required physical Bell pairs can be generated in parallel. Each remote CNOT consumes one heralded physical Bell pair, which must be available before the corresponding seam check is performed. We set  \(p_\mathrm{success}^\mathrm{Bell}\) to be the success probability of a physical Bell-pair generation attempt on each communication link. Since the attempts are repeated at a rate \(R_\mathrm{rep}=1/\tau_\mathrm{code}^\mathrm{cycle} \), the physical Bell-pair generation rate per link is
$
r_\mathrm{Bell}^\mathrm{phys}
=
p_\mathrm{success}^\mathrm{Bell}R_\mathrm{rep}.
$
The average number of attempts required until all \((2d_\mathrm{code}-1)\) physical Bell pairs are available is
$
\mathbb{E}[\mathcal{N}(p_\mathrm{success}^\mathrm{Bell},2d_\mathrm{code}-1)] .
$
Thus, the average time required for one remote syndrome-measurement round is
$
\tau_\mathrm{code}^\mathrm{cycle} 
\mathbb{E}[\mathcal{N}(p_\mathrm{success}^\mathrm{Bell},2d_\mathrm{code}-1)] .
$
Because lattice merging requires \(d_\mathrm{code}\) such syndrome-measurement rounds, and ignoring the time cost for initialization and lattice splitting \cite{Litinski2019}, the logical Bell-pair generation rate is approximated by
\begin{equation}
	r_\mathrm{Bell}^\mathrm{logical}
	\simeq
	\frac{1}{
		\qty[
		\tau_\mathrm{code}^\mathrm{cycle}
		\mathbb{E}[\mathcal{N}(p_\mathrm{success}^\mathrm{Bell},2d_\mathrm{code}-1)]
		]
		d_\mathrm{code}
	}.
\end{equation}

The logical Bell-pair infidelity, $1-F_\mathrm{Bell}^\mathrm{logical}$, is given by \cite{Ramette2024}
\begin{equation}
	1-F_\mathrm{Bell}^\mathrm{logical}
	\simeq
	\qty(\frac{p_\mathrm{bulk}}{p_\mathrm{bulk}^\mathrm{th}})^{(d_\mathrm{code}+1)/2}+
	\qty(\frac{p_\mathrm{seam}}{p_\mathrm{seam}^\mathrm{th}})^{(d_\mathrm{code}+1)/2},
\end{equation}
where $p_\mathrm{bulk}$ is the operation error probability for local gate operations, $p_\mathrm{seam}$ is the operation error probability for remote gate operations at the edges of a surface-code patch, $p_\mathrm{bulk}^\mathrm{th}$ is the threshold error probability for local gates, and $p_\mathrm{seam}^\mathrm{th}$ is the threshold error probability for remote gate operations. This expression for infidelity is a rough approximation that depends on the details of the decoding algorithm. According to Ref.~\cite{Ramette2024}, $p_\mathrm{bulk}^\mathrm{th}\simeq 10^{-2}\ (1\%)$ and $p_\mathrm{seam}^\mathrm{th}\simeq 0.1\ (10\%)$, indicating that the surface code is more robust to errors on communication channels than to those in the bulk of the surface code. Following Ref.~\cite{Sinclair2025}, the effective bulk and seam error probabilities can be estimated as
\(p_\mathrm{bulk}\simeq 2p_\mathrm{2q}+0.5p_\mathrm{meas}\)
and
\(p_\mathrm{seam}\simeq 0.5 p_\mathrm{Bell}^\mathrm{phys}+2.5p_\mathrm{2q}+1.5p_\mathrm{meas}\)
by averaging the error rates of data and syndrome qubits.
Here, \(p_\mathrm{2q}=1-F_\mathrm{2q}\) is the two-qubit gate error probability, \(p_\mathrm{meas}=1-F_\mathrm{meas}\) is the measurement error probability, and \(p_\mathrm{Bell}^\mathrm{phys}=1-F_\mathrm{Bell}^\mathrm{phys}\) is the physical Bell-pair error probability. We neglect single-qubit gate errors in this estimate.
In the numerical analysis below, however, we use \(p_\mathrm{bulk}\) as a representative effective bulk error probability and sweep \(p_\mathrm{seam}\) independently to characterize the quality of the remote seam.
The above expression for \(p_\mathrm{seam}\) is used only to relate the chosen seam error probabilities to approximate physical Bell-pair error probabilities.

We also include the memory error accumulated during the extended syndrome-measurement cycle caused by the stochastic preparation of physical Bell pairs. In each syndrome-measurement round, \((2d_\mathrm{code}-1)\) physical Bell pairs must be available for the remote CNOT. The additional waiting time is estimated as
\begin{equation}
	t_\mathrm{wait}
	=
	\tau_\mathrm{code}^\mathrm{cycle}
	\qty{
		\mathbb{E}[\mathcal{N}(p_\mathrm{success}^\mathrm{Bell},2d_\mathrm{code}-1)]
		-1
	}.
\end{equation}
Assuming amplitude relaxation with \(T_1=1~\mathrm{ms}\), we model the effective memory-error probability accumulated during the additional waiting time as
\begin{equation}
	p_\mathrm{wait}
	=
	\frac{1}{2}
	\qty[
	1-\exp\qty(
	-\frac{t_\mathrm{wait}}{T_1}
	)
	],
\end{equation}
where the factor \(1/2\) accounts for the average excited-state population.
Because this waiting time effectively extends the syndrome-measurement cycle, we add \(p_\mathrm{wait}\) to both the bulk and seam error probabilities as
\begin{equation}
	p_\mathrm{bulk}\rightarrow p_\mathrm{bulk}+p_\mathrm{wait},
\end{equation}
\begin{equation}
	p_\mathrm{seam}\rightarrow p_\mathrm{seam}+p_\mathrm{wait}.
\end{equation}

\begin{figure*}[t]
	\centering
	\includegraphics[width=170 mm]{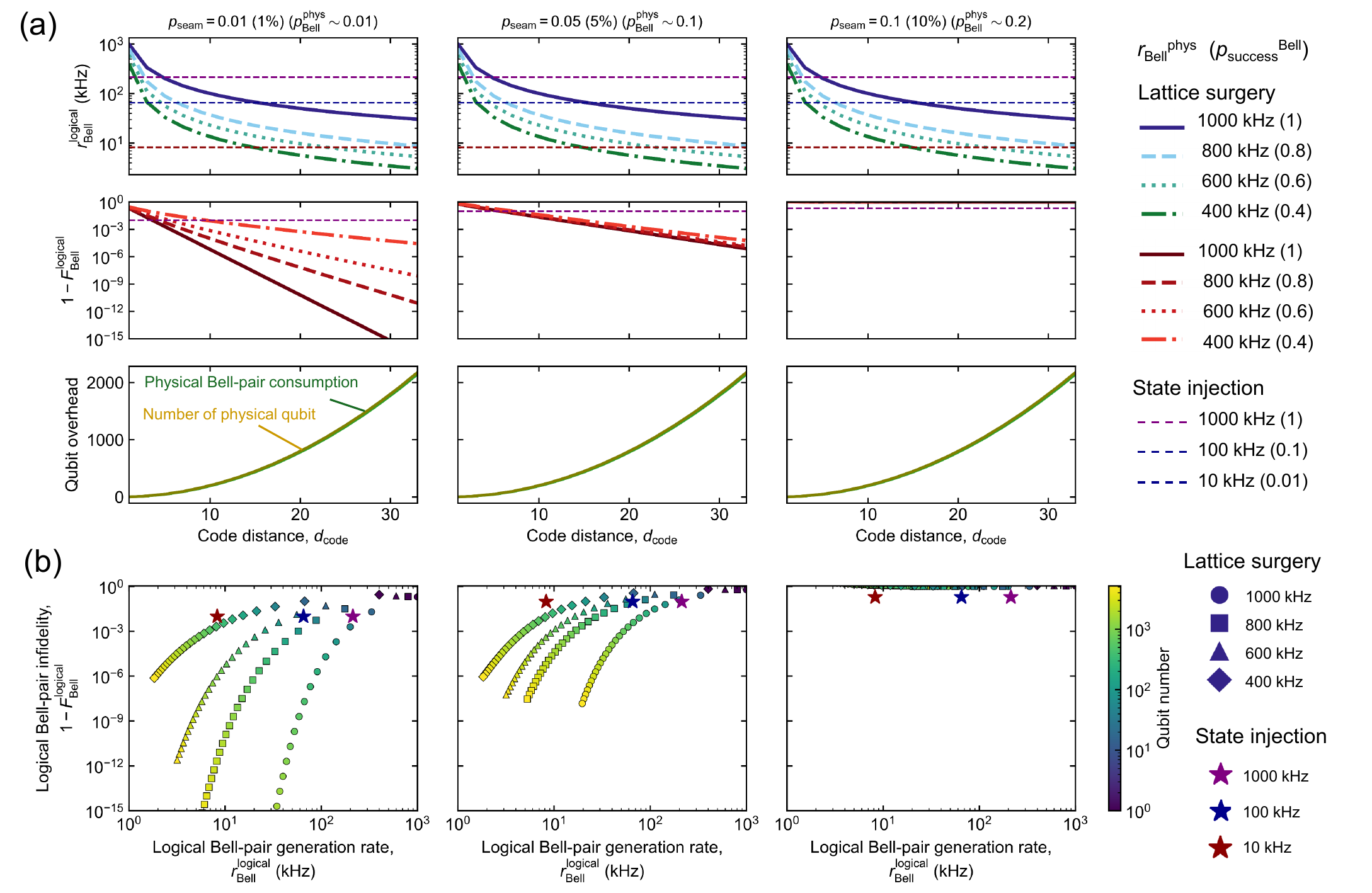}
	\caption
	{Logical Bell-pair generation rate $r_\mathrm{Bell}^\mathrm{logical}$ and fidelity $F_\mathrm{Bell}^\mathrm{logical}$ using lattice surgery and state injection for different seam error rates. (a) (Top) Logical Bell-pair generation rate as a function of the code distance
		$d_\mathrm{code}$ for different physical Bell-pair generation rates
		$r_\mathrm{Bell}^\mathrm{phys}$ and seam error rates $p_\mathrm{seam}$.
		The value of $r_\mathrm{Bell}^\mathrm{phys}$ is varied by changing
		$p_\mathrm{success}^\mathrm{Bell}$ while keeping the physical Bell-pair
		generation trial rate constant. The horizontal dashed lines show the results for state injection
		for different $r_\mathrm{Bell}^\mathrm{phys}$. (Middle) Logical Bell-pair infidelity $1-F_\mathrm{Bell}^\mathrm{logical}$ as a function of code distance for the same parameters. The result for state injection is shown by the purple dotted line. (Bottom) Physical Bell-pair consumption per generated logical Bell pair
		and the number of physical qubits required per logical qubit (per node). (b) Logical Bell-pair generation rate versus logical error rate for $d_\mathrm{code}\le 51$. Each marker corresponds to a different physical Bell-pair generation rate, and each point corresponds to a different code distance. The required number of physical qubits is indicated by the color scale of each data point. The star symbols correspond to state injection for different $r_\mathrm{Bell}^\mathrm{phys}$. The seam error rate, $p_\mathrm{seam}$, is the same within each column.}
	\label{fig:logical_gen_rate_fid}
\end{figure*}

\subsubsection{State Injection}

Next, we consider logical Bell-pair generation via state injection, a process that directly encodes the state of a physical qubit into a logical state. This process is commonly used to generate a logical magic state from a physical magic state \cite{Li2015b,Lao2022}. The injection of a physical Bell pair into a logical Bell pair at both nodes succeeds with probability $p_\mathrm{success}^\mathrm{injection}$, given by
\begin{equation}
	p_\mathrm{success}^\mathrm{injection}
	=(1-p_\mathrm{rejection}^\mathrm{injection})^2,
\end{equation}
where $p_\mathrm{rejection}^\mathrm{injection}$ is the injection failure probability at either node. Thus, the logical Bell-pair generation rate via state injection is given by
\begin{equation}
	\frac{1}{r_\mathrm{Bell}^\mathrm{logical}}
	=\frac{1}{p_\mathrm{success}^\mathrm{injection}}
	\qty(
	\frac{1}{r_\mathrm{Bell}^\mathrm{phys}}+\tau_\mathrm{Bell}^\mathrm{injection}
	).
\end{equation}
According to Ref.~\cite{Pattison2025,Lao2022}, $1-p_\mathrm{success}^\mathrm{injection}\simeq 0.15\ (15\%)$ for a local gate error of $10^{-3}\ (0.1\%)$ and for the rotated surface code. We use this value in the calculation below. Following Refs.~\cite{Lao2022,Pattison2025}, the infidelity of injected logical Bell pair can be written as
\begin{equation}
	1-F_\mathrm{Bell}^\mathrm{logical,injection}
	\simeq
	\frac{4}{3}p_\mathrm{1q}+
	\frac{6}{5}p_\mathrm{2q}+p_\mathrm{Bell}^\mathrm{phys},
\end{equation}
where $p_\mathrm{1q}=1-F_\mathrm{1q}$ is the single-qubit gate error probability. In the numerical analysis, we set \(p_\mathrm{1q}=p_\mathrm{2q}=10^{-3}\). Since these local gate errors are smaller than the physical Bell-pair error probabilities considered below, \(p_\mathrm{Bell}^\mathrm{phys}=10^{-2}\), \(0.1\), and \(0.2\), the injected logical Bell-pair infidelity is approximated as
\begin{equation}
	1-F_\mathrm{Bell}^\mathrm{logical,injection}
	\simeq p_\mathrm{Bell}^\mathrm{phys}.
\end{equation}
The injection time is estimated as \(\tau_\mathrm{Bell}^\mathrm{injection}=3\tau_\mathrm{code}^\mathrm{cycle} \) \cite{Li2015b}, with \(\tau_\mathrm{code}^\mathrm{cycle} =1~\mu\mathrm{s}\).

\subsection{Numerical Analysis}

Using the above definitions, we calculate \(r_\mathrm{Bell}^\mathrm{logical}\) and \(1-F_\mathrm{Bell}^\mathrm{logical}\) for lattice-surgery-based logical Bell-pair generation and state injection for different input \(r_\mathrm{Bell}^\mathrm{phys}\) and seam error probabilities, \(p_\mathrm{seam}\) (Fig.~\ref{fig:logical_gen_rate_fid}(a)). For lattice-surgery-based logical Bell-pair generation, we set \(p_\mathrm{bulk}=10^{-3}\ (0.1\%)\) as a representative effective bulk error probability, rather than deriving it from the above averaging estimate. We then sweep the seam error probability as \(p_\mathrm{seam}=10^{-2}\ (1\%)\), \(5\times10^{-2}\ (5\%)\), and \(0.1\ (10\%)\). Using \(p_\mathrm{2q}=p_\mathrm{meas}=10^{-3}\) in the estimate for \(p_\mathrm{seam}\), these values approximately correspond to \(p_\mathrm{Bell}^\mathrm{phys}\simeq10^{-2}\ (1\%)\), \(0.1\ (10\%)\), and \(0.2\ (20\%)\), respectively.

From Fig.~\ref{fig:logical_gen_rate_fid}(a), we find that, as $d_\mathrm{code}$ increases, the logical Bell-pair error rate, $1-F_\mathrm{Bell}^\mathrm{logical}$, decreases (middle row of Fig.~\ref{fig:logical_gen_rate_fid}(a)) at the expense of a decrease in the logical Bell-pair generation rate, $r_\mathrm{Bell}^\mathrm{logical}$ (top row of Fig.~\ref{fig:logical_gen_rate_fid}(a)), and an increase in the required numbers of physical qubits and physical Bell pairs (bottom row of Fig.~\ref{fig:logical_gen_rate_fid}(a)). Here, the number of physical qubits per node is $2d_\mathrm{code}^2-1$ for the rotated surface code, and the number of consumed physical Bell pairs is $(2d_\mathrm{code}-1)d_\mathrm{code}$. The factor \(2d_\mathrm{code}-1\) corresponds to the number of remote CNOT that must be performed along the merged boundary in one syndrome-measurement round. In addition, as $p_\mathrm{seam}$ increases from the left to the right column, error suppression becomes less efficient. As $p_\mathrm{seam}$ approaches
$p_\mathrm{seam}^\mathrm{th}=0.1$, increasing $d_\mathrm{code}$
no longer effectively suppresses the seam-error contribution within
this threshold-scaling model.

To interpret the results from a different perspective, we plot $r_\mathrm{Bell}^\mathrm{logical}$ versus $1-F_\mathrm{Bell}^\mathrm{logical}$ while varying $r_\mathrm{Bell}^\mathrm{phys}$ (Fig.~\ref{fig:logical_gen_rate_fid}(b)). We find that, to achieve a target error rate of $10^{-12}$ within the range $d_\mathrm{code}\le 51$, $r_\mathrm{Bell}^\mathrm{phys}\ge800~\mathrm{kHz}$ and $p_\mathrm{seam}\sim 10^{-2}\ (1\%)$ are required. The resulting logical Bell-pair generation rate is on the order of 10 kHz. In contrast, for state injection, the error rate cannot be reduced below $\sim 10^{-2}\ (1\%)$, and the logical Bell-pair generation rate is $10$--$200~\mathrm{kHz}$ depending on the input physical Bell-pair generation rate.

In summary, lattice-surgery-based logical Bell-pair generation can generate an extremely low-error, $<10^{-12}$, logical Bell-pair at the cost of a low error, $p_\mathrm{Bell}^\mathrm{phys}\sim 0.01\ (1\%)$, and fast, $\ge800~\mathrm{kHz}$, physical Bell-pair supply rate per communication channel, as well as a large number of consumed physical Bell pairs, on the order of 1000. In other words, to perform lattice surgery, we need $2d_\mathrm{code}-1$ quantum links, each supplying physical Bell pairs at a rate of $\ge800~\mathrm{kHz}$. In contrast, state injection can be performed over a wide range of physical Bell-pair generation rates and fidelities. Although the logical Bell-pair fidelity is nearly the same as the physical Bell-pair fidelity, state injection  provides a viable way to generate noisy logical Bell pairs in situations where lattice-surgery-based logical Bell-pair generation cannot be performed because of a low physical Bell-pair supply rate, a high physical Bell-pair error rate, or an insufficient number of communication channels. Although the logical Bell-pair fidelity is insufficient for meaningful quantum computation in that case, the fidelity of logical Bell pairs can be increased using logical Bell-pair distillation.

\section{Step IV: Logical Bell-pair distillation}
	
	\subsection{Logical Bell-pair distillation calculation}
	
	Similar to physical Bell-pair distillation, logical Bell-pair distillation
	generates $k$ logical Bell pairs from $n$ input logical Bell pairs.
	The difference from physical Bell-pair distillation is that extremely  high-fidelity
	logical Bell pairs can be generated through repeated distillation because the
	logical qubits are protected by quantum error correction.
	Quantum operations for logical Bell-pair distillation consist of Pauli or
	parity measurements on multiple logical qubits and readout of the measurement
	outcomes, and can therefore be described within the same framework as that
	used for physical Bell-pair distillation.
	As discussed in the previous section, distillation is required when logical
	Bell pairs are generated using state injection because of their relatively
	low initial fidelity (Fig. \ref{fig:logical_gen_distill}(b)).
	For the distillation after lattice-surgery-based logical Bell-pair generation (Fig. \ref{fig:logical_gen_distill}(c)), we evaluate its performance in the Appendix, since it typically exhibits intermediate performance between direct lattice surgery and state injection and would otherwise complicate the main discussion.

	We enumerate candidate logical Bell-pair distillation protocols and evaluate
	their output fidelity, generation latency, and resource overhead.
	We consider ordered combinations of the
	$Z_{\mathrm{2B}}$, $ZX_{\mathrm{3B}}$, $[[4,1,2]]$, and $[[5,1,3]]$
	protocols with up to six distillation stages. For an $n_i$-to-1 distillation protocol at the $i$th stage, we denote the
	error-suppression function and success probability by
	$f_i^\mathrm{logical}$ and
	$p_\mathrm{success}^{\mathrm{logical(distill},i)}$, respectively.
	The distillation time is modeled as
	$\tau^{\mathrm{logical(distill},i}
	=c_i d_{\mathrm{code},i}\tau_\mathrm{code}^\mathrm{cycle}$,
	where $c_i$ is a protocol-dependent coefficient (for details of each distillation protocol, see Appendix).
	
	Throughout this section, the analytical distillation rate is defined as
	the inverse of the expected latency required to obtain one final logical
	Bell pair from an initially empty distillation system, rather than the
	steady-state throughput of an indefinitely pipelined process. This
	latency-based definition is consistent with the finite space available
	for logical Bell-pair distillation and buffering in the assumed
	architecture.
	
The analysis of logical Bell-pair distillation differs from that of physical Bell-pair distillation in several respects.
First, all local operations are assumed to be performed fault tolerantly using lattice surgery.
Second, additional logical-operation errors during the distillation circuits are neglected, whereas memory errors accumulated by logical Bell pairs during waiting are explicitly considered.
Third, the surface-code distance $d_{\mathrm{code}}$ is adjusted at each distillation stage according to the required logical error rate and the waiting-induced memory error.
Finally, we neglect additional movement and routing costs of logical code blocks and assume that they can be rearranged into the configuration required for each distillation process without additional overhead.

Because logical Bell pairs are protected by error correction, we consider both single- and multi-channel operation, unlike physical Bell-pair distillation, for which single-channel operation is strongly limited by waiting-time decoherence. We therefore consider two analytical models: a sequential model using a single communication channel and a parallel model using multiple communication channels. Both models include waiting-induced logical memory errors. The analytical models are further validated for selected cases using discrete-event Monte Carlo simulations.

	\subsubsection{Sequential model with memory errors}
	
	In the sequential model, input Bell pairs are generated one by one using a
	single communication channel.
	For an $n_i$-to-1 distillation protocol, $n_i$ successful Bell pairs from the
	preceding stage must be prepared sequentially.
	The output rate is recursively calculated as
	
	\begin{equation}
		\begin{aligned}
			&
			\frac{1}{
				r_{\mathrm{Bell}}^{\mathrm{logical(distill},i=1)}
			}=
			\frac{1}{
			p_{\mathrm{success}}^{\mathrm{logical(distill},i=1)}
			}
			\\
			&\times
			\qty(
			\frac{n_1}{
				r_{\mathrm{Bell}}^{\mathrm{logical(injection)}}
			}
			+
			\tau^{\mathrm{logical(distill},i=1)}
			),
			\\[3pt]
			&
			\frac{1}{
				r_{\mathrm{Bell}}^{\mathrm{logical(distill},i)}
			}=
			\frac{1}{
			p_{\mathrm{success}}^{\mathrm{logical(distill},i)}
			}
			\\
			&\times
			\qty(
			\frac{n_i}{
				r_{\mathrm{Bell}}^{\mathrm{logical(distill},i-1)}
			}
			+
			\tau^{\mathrm{logical(distill},i)}
			).
		\end{aligned}
		\label{eq:logical_distill_sequential_rate}
	\end{equation}
	
	We model the logical error probability per code block per code cycle as
	$
	p_{\mathrm{logical}}(d_{\mathrm{code}})
	=
	(p_{\mathrm{bulk}}/p_{\mathrm{bulk}}^{\mathrm{th}})
	^{(d_{\mathrm{code}}+1)/2}.
	$
	Since a logical Bell pair consists of two logical code blocks, the
	waiting-induced Bell-pair infidelity accumulated over a waiting time
	$t_{\mathrm{wait}}$ is
	$
	\epsilon_{\mathrm{wait}}
	=
	2p_{\mathrm{logical}}(d_{\mathrm{code}})
	\frac{t_{\mathrm{wait}}}{\tau_{\mathrm{code}}^{\mathrm{cycle}}}.
	$
	
	Because the $n_i$ input Bell pairs are prepared sequentially, Bell pairs
	generated earlier remain stored while the remaining inputs are prepared.
	Their average waiting time is
	$t_{\mathrm{wait}}^{\mathrm{seq},i}
	=
	(n_i-1)/
	[2r_{\mathrm{Bell}}^{\mathrm{logical(distill},i-1)}]$.
	For the first stage, the denominator is replaced by
	$r_{\mathrm{Bell}}^{\mathrm{logical(injection)}}$.
	
	The output fidelity is recursively calculated by adding the
	waiting-induced error to the input infidelity before applying the
	distillation map:
	\begin{equation}
		\begin{aligned}
			&
			1-F_{\mathrm{Bell}}^{\mathrm{logical(distill},i=1)}
			\\
			&=
			f_1^{\mathrm{logical}}
			\qty[
			1-F_{\mathrm{Bell}}^{\mathrm{logical(injection)}}
			+
			\frac{2p_{\mathrm{logical}}(d_{\mathrm{code},1})
			}
			{\tau_{\mathrm{code}}^{\mathrm{cycle}}/	t_{\mathrm{wait}}^{\mathrm{seq},i=1}
			}
			],
			\\[3pt]
			&
			1-F_{\mathrm{Bell}}^{\mathrm{logical(distill},i)}
			\\
			&=
			f_i^{\mathrm{logical}}
			\qty[
			1-F_{\mathrm{Bell}}^{\mathrm{logical(distill},i-1)}
			+
			\frac{2p_{\mathrm{logical}}(d_{\mathrm{code},i})
			}{\tau_{\mathrm{code}}^{\mathrm{cycle}}
			/	t_{\mathrm{wait}}^{\mathrm{seq},i}
			}
			].
		\end{aligned}
		\label{eq:logical_distill_sequential_fidelity}
	\end{equation}
	Equations~\eqref{eq:logical_distill_sequential_rate}
	and~\eqref{eq:logical_distill_sequential_fidelity} recursively determine the
	logical Bell-pair generation rate and fidelity after each distillation stage.

	\subsubsection{Parallel model with memory errors}
	
		\begin{figure}[b]
		\centering
		\includegraphics[width=80 mm]{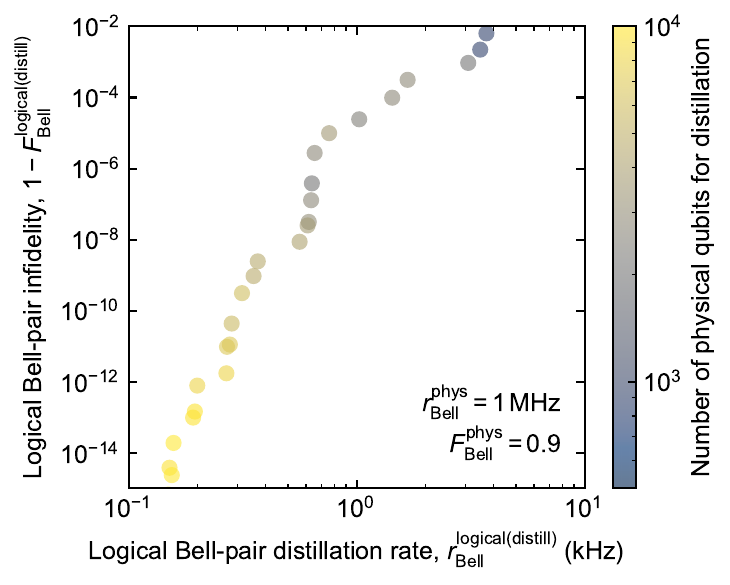}
		\caption{
			Logical Bell-pair infidelity
			$1-F_\mathrm{Bell}^\mathrm{logical(distill)}$
			versus logical Bell-pair distillation rate
			$r_\mathrm{Bell}^\mathrm{logical(distill)}$
			for an input physical Bell-pair generation rate of 1 MHz and fidelity of
			0.9, calculated using the sequential  model.
			For each target infidelity, the fastest feasible protocol
is selected from all ordered combinations of the considered
distillation protocols with up to six stages.
			The color scale represents the peak number of physical qubits required for
			logical Bell-pair distillation.
		}
		\label{fig:logical_distill_tradeoff}
	\end{figure}
	
		\begin{figure*}[bt]
		\centering
		\includegraphics[width=140 mm]{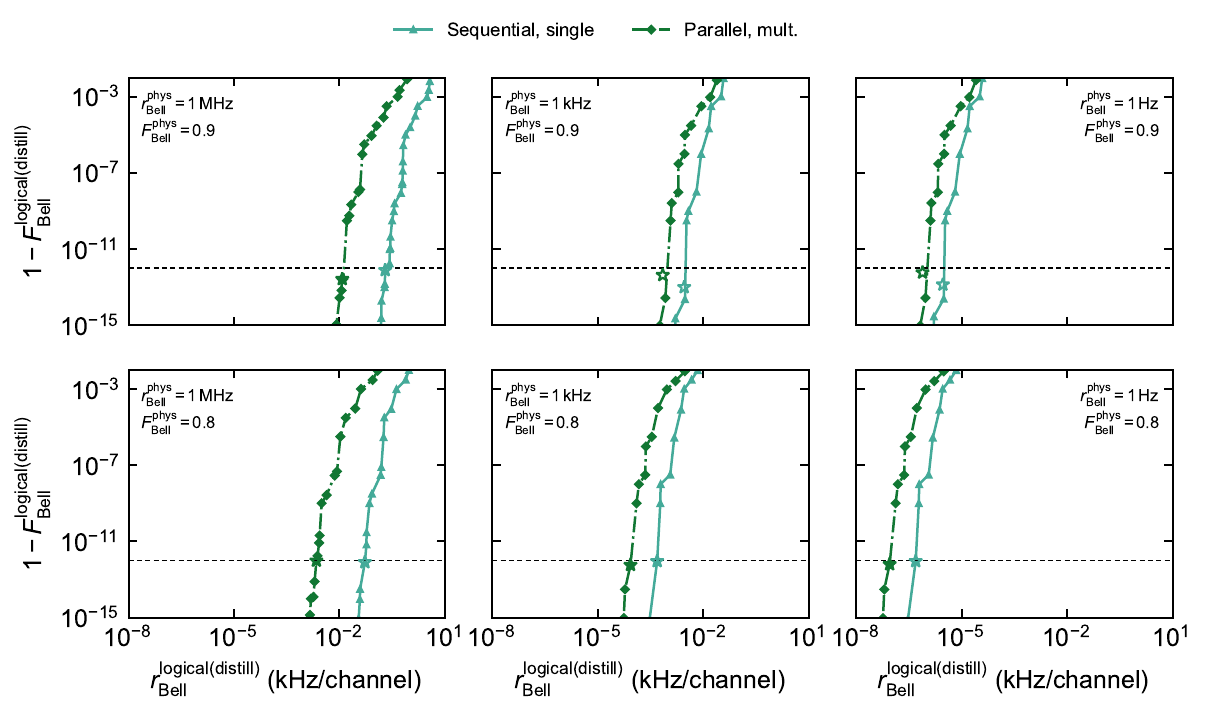}
		\caption
		{Combination of the achievable  logical Bell-pair generation rate $r_\mathrm{Bell}^\mathrm{logical(distill)}$ and its infidelity $1-F_\mathrm{Bell}^\mathrm{logical(distill)}$ with varying input physical Bell-pair parameters. The results for the sequential and parallel communication channel model are shown.  The points correspond to the best distillation protocols among the six-step distillation and different target errors. For the parallel model, $r_\mathrm{Bell}^\mathrm{logical(distill)}$ is normalized by the number of parallel communication channel so that we can compare their performance in a fair manner.  }
		\label{fig:logical_distill_frontier}
	\end{figure*}
	
	\begin{figure}[t]
		\centering
		\includegraphics[width=75 mm]{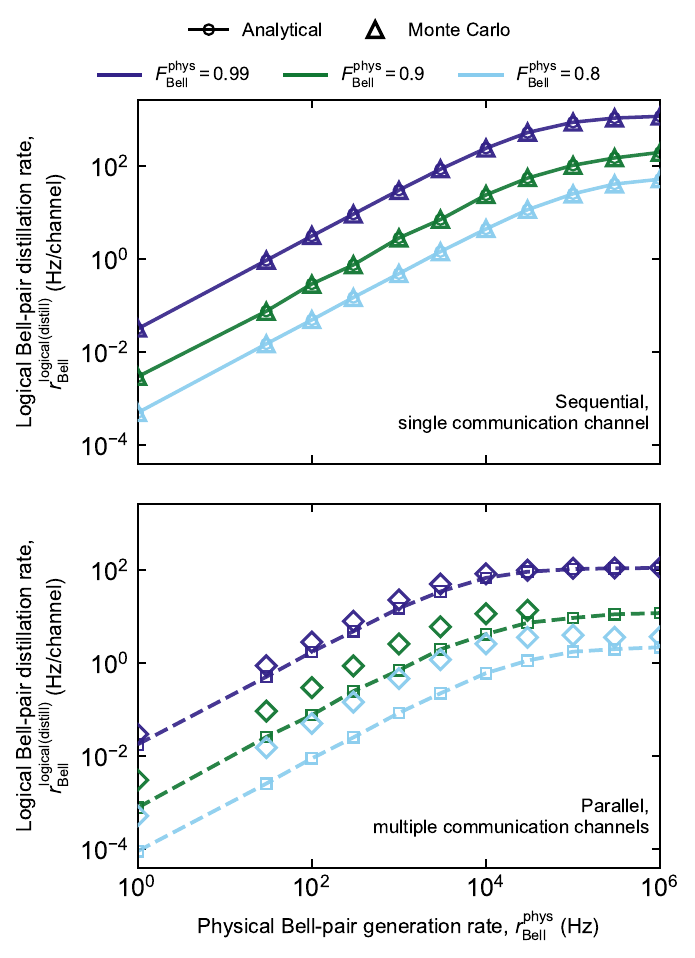}
		\caption{
			Logical Bell-pair distillation rate
			$r_{\mathrm{Bell}}^{\mathrm{logical(distill)}}$
			satisfying
			$1-F_{\mathrm{Bell}}^{\mathrm{logical(distill)}}\le10^{-12}$
			as a function of the input physical Bell-pair generation rate
			$r_{\mathrm{Bell}}^{\mathrm{phys}}$
			for
			$F_{\mathrm{Bell}}^{\mathrm{phys}}=0.99$, $0.9$, and $0.8$.
			The upper and lower panels show the sequential single-channel and parallel
			multiple-channel configurations, respectively.
			For the parallel configuration, the rate is normalized by
			$N_{\mathrm{comm}}^{\mathrm{logical}}$.
			Analytical and discrete-event Monte Carlo results are shown.
			Physical Bell-pair distillation is not applied in this figure.
		}
		\label{fig:logical_distill_rate}
	\end{figure}

	In the parallel model, the branches required for the distillation cascade are
	generated simultaneously using independent communication channels.
	For a distillation sequence with input sizes $n_i$, the total number of
	logical communication channels is
	$
	N_{\mathrm{comm}}^{\mathrm{logical}}
	=
	\prod_i n_i.
	$
	
	As in the physical Bell-pair distillation calculation, we use
	$\mathbb{E}[\mathcal{N}(p,n)]$ to describe the expected number of trials
	required for $n$ independently generated branches to complete, when each
	branch succeeds with probability $p$ per attempt.
	The logical Bell-pair generation rate is recursively calculated as
	
	\begin{equation}
		\begin{aligned}
			&
			\frac{1}{
				r_{\mathrm{Bell}}^{\mathrm{logical(distill},i=1)}
			}
			=
			\frac{1}{
				p_{\mathrm{success}}^{\mathrm{logical(distill},i=1)}
			}
			\\
			&\times
			\Bigg[
			\frac{
					\mathbb{E}
				\qty[
				\mathcal{N}
				\qty(
				p_{\mathrm{success}}^{\mathrm{injection}},
				n_1
				)
				]
			}{\frac{	r_{\mathrm{Bell}}^{\mathrm{logical(injection)}}}{	p_{\mathrm{success}}^{\mathrm{injection}}}
			}
			+
			\tau^{\mathrm{logical(distill},i=1)}
			\Bigg],
			\\[3pt]
			&
			\frac{1}{
				r_{\mathrm{Bell}}^{\mathrm{logical(distill},i)}
			}
			=
			\frac{1}{
				p_{\mathrm{success}}^{\mathrm{logical(distill},i)}
			}
			\\
			&\times
			\Bigg[
			\frac{	\mathbb{E}
				\qty[
				\mathcal{N}
				\qty(
				p_{\mathrm{success}}^{\mathrm{logical(distill},i-1)},
				n_i
				)
				]
			}{\frac{				r_{\mathrm{Bell}}^{\mathrm{logical(distill},i-1)}}{	p_{\mathrm{success}}^{\mathrm{logical(distill},i-1)}}
			}
			+
			\tau^{\mathrm{logical(distill},i)}
			\Bigg].
		\end{aligned}
		\label{eq:logical_distill_parallel_rate}
	\end{equation}

	Because the parallel branches generally complete at different times, Bell
	pairs generated earlier remain stored until the slowest branch is completed.
	Their average waiting time is calculated as
	$t_{\mathrm{wait}}^{\mathrm{para},i}
	=
	\{\mathbb{E}[\mathcal{N}
	(p_{\mathrm{success}}^{\mathrm{logical(distill},i-1)},n_i)]
	-
	1/p_{\mathrm{success}}^{\mathrm{logical(distill},i-1)}\}
	p_{\mathrm{success}}^{\mathrm{logical(distill},i-1)}
	/
	r_{\mathrm{Bell}}^{\mathrm{logical(distill},i-1)}$.
	For the first stage,
	$p_{\mathrm{success}}^{\mathrm{logical(distill},i-1)}$ and
	$r_{\mathrm{Bell}}^{\mathrm{logical(distill},i-1)}$ are replaced by
	$p_{\mathrm{success}}^{\mathrm{injection}}$ and
	$r_{\mathrm{Bell}}^{\mathrm{logical(injection)}}$, respectively.
	
	The waiting-induced infidelity is calculated using the same logical
	memory-error model as in the sequential case.
	The output fidelity is therefore
	\begin{equation}
		\begin{aligned}
			&
			1-F_{\mathrm{Bell}}^{\mathrm{logical(distill},i=1)}
			\\
			&=
			f_1^{\mathrm{logical}}
			\qty[
			1-F_{\mathrm{Bell}}^{\mathrm{logical(injection)}}
			+
			\frac{	2p_{\mathrm{logical}}(d_{\mathrm{code},1})
			}{\tau_{\mathrm{code}}^{\mathrm{cycle}}/
			t_{\mathrm{wait}}^{\mathrm{para},i=1}
			}
			],
			\\[3pt]
			&
			1-F_{\mathrm{Bell}}^{\mathrm{logical(distill},i)}
			\\
			&=
			f_i^{\mathrm{logical}}
			\qty[
			1-F_{\mathrm{Bell}}^{\mathrm{logical(distill},i-1)}
			+
			\frac{	2p_{\mathrm{logical}}(d_{\mathrm{code},i})
						}{\tau_{\mathrm{code}}^{\mathrm{cycle}}
			/	t_{\mathrm{wait}}^{\mathrm{para},i}
			}
			].
		\end{aligned}
		\label{eq:logical_distill_parallel_fidelity}
	\end{equation}
	Thus,
	$r_{\mathrm{Bell}}^{\mathrm{logical(distill},i)}$ and
	$F_{\mathrm{Bell}}^{\mathrm{logical(distill},i)}$ are recursively obtained
	at every stage, with the parallel model additionally accounting for the
	asynchronous completion and waiting of independently generated branches.

\subsubsection{Code-distance and protocol selection}

For each candidate distillation sequence, the surface-code distance is
optimized at each distillation stage so that the requested final Bell-pair
fidelity can be achieved while accounting for waiting-induced logical memory
errors.
The target final infidelity is propagated backward through the distillation
sequence, and the smallest feasible odd code distance is selected at each
stage.

For the rate--infidelity trade-off calculation, the target infidelity is swept
from $10^{-2}$ to $10^{-15}$, and all ordered protocol sequences containing up
to six distillation stages are evaluated.
For the representative resource comparison, the physical-rate sweep, and the
Monte Carlo validation, we instead use
$1-F_{\mathrm{Bell}}^{\mathrm{logical(distill)}}\le10^{-12}$
as a feasibility constraint. For the parallel model, protocol selection is based on the
communication-normalized rate
$
	r_{\mathrm{Bell}}^{\mathrm{logical(distill,norm)}}
	=
			r_{\mathrm{Bell}}^{\mathrm{logical(distill)}}
	/
		N_{\mathrm{comm}}^{\mathrm{logical}}
	,
	\label{eq:normalized_logical_distill_rate}
$
whereas $N_{\mathrm{comm}}^{\mathrm{logical}}=1$ for the sequential model.
Details of the code-distance and protocol-selection procedure are provided in
the Appendix.

\subsubsection{Resource estimation}

We estimate the physical-qubit requirement of the logical Bell-pair
distillation process using
$2(d_{\mathrm{code}}+1)^2$ physical qubits per logical code block.
For the sequential configuration, the peak requirement includes both stored
logical Bell pairs and code blocks used by active distillation operations.
For the parallel configuration, the hardware required for the full parallel
distillation tree is provisioned, including different distillation stages
that can operate simultaneously.
Additional movement and routing resources are neglected.
The detailed resource-counting procedure and the calculation of Bell-pair
consumption are given in the Appendix.

\subsection{Numerical Evaluation}

Based on the procedure described above, the logical Bell-pair distillation
protocol with the highest rate can be selected for a specified physical
Bell-pair generation rate and fidelity.
Figure~\ref{fig:logical_distill_tradeoff} shows the resulting rate--infidelity
trade-off for
$r_{\mathrm{Bell}}^{\mathrm{phys}}=1~\mathrm{MHz}$ and
$F_{\mathrm{Bell}}^{\mathrm{phys}}=0.9$
in the sequential configuration.
To achieve a logical Bell-pair infidelity of approximately $10^{-10}$ or
lower, the logical Bell-pair distillation rate is on the order of
$10^{2}~\mathrm{Hz}$, approximately four orders of magnitude lower than
$r_{\mathrm{Bell}}^{\mathrm{phys}}$.
The color indicates the peak number of physical qubits required for logical
Bell-pair distillation, which ranges from approximately $10^{3}$ to $10^{4}$
over the target-infidelity range considered.

Figure~\ref{fig:logical_distill_frontier} compares the achievable combinations
of
$r_{\mathrm{Bell}}^{\mathrm{logical(distill)}}$
and
$1-F_{\mathrm{Bell}}^{\mathrm{logical(distill)}}$
for different input physical Bell-pair parameters and the two communication
configurations.
Because the parallel configuration uses
$N_{\mathrm{comm}}^{\mathrm{logical}}>1$ communication channels, its rate is
normalized by $N_{\mathrm{comm}}^{\mathrm{logical}}$ for a per-channel
comparison.
The sequential model generally provides a higher communication-normalized rate
than the analytical parallel model.
This rate--infidelity trade-off can therefore be used to select the fastest protocol satisfying
a specified target infidelity, such as
$1-F_{\mathrm{Bell}}^{\mathrm{logical(distill)}}\le10^{-12}$.

Fixing the target infidelity at $10^{-12}$,
Fig.~\ref{fig:logical_distill_rate} shows
$r_{\mathrm{Bell}}^{\mathrm{logical(distill)}}$
as a function of
$r_{\mathrm{Bell}}^{\mathrm{phys}}$
for
$F_{\mathrm{Bell}}^{\mathrm{phys}}=0.99$, $0.9$, and $0.8$.
The sequential analytical model agrees closely with the Monte Carlo results
over the entire range.
For the parallel configuration, the Monte Carlo simulations give higher rates
than the analytical model, particularly at lower
$r_{\mathrm{Bell}}^{\mathrm{phys}}$ for
$F_{\mathrm{Bell}}^{\mathrm{phys}}=0.8$--$0.9$.
This indicates that the recursive analytical treatment of the parallel model
is conservative, because it approximates the completion of later-stage
branches using mean latencies, whereas the Monte Carlo simulation explicitly
resolves asynchronous scheduling.
At low
$r_{\mathrm{Bell}}^{\mathrm{phys}}$,
the output rate decreases approximately in proportion to the input rate,
whereas at high input rates it saturates because the logical-distillation
latency becomes dominant.

To evaluate representative operating points, we perform the same optimization
with
$1-F_{\mathrm{Bell}}^{\mathrm{logical(distill)}}\le10^{-12}$
for the input parameters listed in Table~\ref{tab:table2}.
The table summarizes the selected logical-distillation protocols and the
numbers of physical and logical communication channels.
For lower physical Bell-pair fidelities, particularly
$F_{\mathrm{Bell}}^{\mathrm{phys}}=0.8$, the
$ZX_{3\mathrm{B}}$ protocol is often selected at the first stage to reduce the
initial error.
The $[[5,1,3]]$ protocol is used for most subsequent stages because of its
strong error-suppression capability.
In the full parallel configuration, the selected protocols require
$25$--$375$ logical communication channels for the cases without physical
Bell-pair predistillation.

\begin{table*}[t]
	\centering
	\scriptsize
	\caption{Input parameters and selected logical Bell-pair distillation protocols. The single-communication-channel section uses the sequential model with memory errors, whereas the multiple-communication-channel section uses the parallel model with memory errors. The selected protocols satisfy $1-F_{\rm Bell}^{\rm logical}\le10^{-12}$; no explicit upper bound on $d_{\rm code}$ is imposed.}
	\label{tab:table2}
	\setlength{\tabcolsep}{4pt}
	\begin{tabular}{c c c c c c}
		\hline\hline
		\multicolumn{3}{c|}{Input parameters} & \multicolumn{3}{c}{Derived protocol / results} \\
		\cline{1-3}\cline{4-6}
		$r_{\mathrm{Bell}}^{\mathrm{phys}}$ & $F_{\mathrm{Bell}}^{\mathrm{phys}}$ &
		\begin{tabular}{c}Physical Bell-pair distillation\\$\left(r_{\mathrm{Bell}}^{\mathrm{distill}},\,1-F_{\mathrm{Bell}}^{\mathrm{distill}},\,N_{\mathrm{comm}}^{\mathrm{phys}}\right)$\end{tabular} &
		\begin{tabular}{c}Logical Bell-pair\\distillation protocol\end{tabular} &
		\begin{tabular}{c}Num. of physical\\distillation channels\\$N_{\mathrm{comm}}^{\mathrm{phys}}$\end{tabular} &
		\begin{tabular}{c}Num. of logical\\distillation channels\\$N_{\mathrm{comm}}^{\mathrm{logical}}$\end{tabular} \\
		\hline
		\multicolumn{6}{c}{Lattice surgery-based logical Bell-pair generation} \\
		\hline
		$1~\mathrm{MHz}$ & $0.99$ & No distill. & No distill. & $1$ & $49\ (=2d_{\mathrm{code}}-1)$ \\
		\hline
		\multicolumn{6}{c}{Injection-distillation protocol (single communication channel, sequential model with memory errors)} \\
		\hline
		$1\,\mathrm{MHz}$ & $0.99$ & No distill. & [[5,1,3]] $\rightarrow$ [[5,1,3]] & $1$ & $1$ \\
		$1\,\mathrm{MHz}$ & $0.9$ & No distill. & $ZX_{3\mathrm{B}}$ $\rightarrow$ [[4,1,2]] $\rightarrow$ [[5,1,3]] $\rightarrow$ [[5,1,3]] & $1$ & $1$ \\
		$1\,\mathrm{MHz}$ & $0.8$ & No distill. & $ZX_{3\mathrm{B}}$ $\rightarrow$ [[5,1,3]] $\rightarrow$ [[5,1,3]] $\rightarrow$ [[5,1,3]] & $1$ & $1$ \\
		$1\,\mathrm{MHz}$ & $0.9$ & $(106\,\mathrm{kHz},\,9\times10^{-3},\,10)$ & [[5,1,3]] $\rightarrow$ [[5,1,3]] & $10$ & $1$ \\
		$1\,\mathrm{MHz}$ & $0.8$ & $(43.3\,\mathrm{kHz},\,4.5\times10^{-3},\,45)$ & [[5,1,3]] $\rightarrow$ [[5,1,3]] & $45$ & $1$ \\
		$1\,\mathrm{kHz}$ & $0.9$ & No distill. & [[5,1,3]] $\rightarrow$ [[5,1,3]] $\rightarrow$ [[5,1,3]] & $1$ & $1$ \\
		$1\,\mathrm{kHz}$ & $0.8$ & No distill. & $ZX_{3\mathrm{B}}$ $\rightarrow$ [[5,1,3]] $\rightarrow$ [[5,1,3]] $\rightarrow$ [[5,1,3]] & $1$ & $1$ \\
		$1\,\mathrm{Hz}$ & $0.9$ & No distill. & [[5,1,3]] $\rightarrow$ [[5,1,3]] $\rightarrow$ [[5,1,3]] & $1$ & $1$ \\
		$1\,\mathrm{Hz}$ & $0.8$ & No distill. & $ZX_{3\mathrm{B}}$ $\rightarrow$ [[5,1,3]] $\rightarrow$ [[5,1,3]] $\rightarrow$ [[5,1,3]] & $1$ & $1$ \\
		\hline
		\multicolumn{6}{c}{Injection-distillation protocol (multiple communication channels, parallel model with memory errors)} \\
		\hline
		$1\,\mathrm{MHz}$ & $0.99$ & No distill. & [[5,1,3]] $\rightarrow$ [[5,1,3]] & $1$ & $25$ \\
		$1\,\mathrm{MHz}$ & $0.9$ & No distill. & [[5,1,3]] $\rightarrow$ [[5,1,3]] $\rightarrow$ [[5,1,3]] & $1$ & $125$ \\
		$1\,\mathrm{MHz}$ & $0.8$ & No distill. & $ZX_{3\mathrm{B}}$ $\rightarrow$ [[5,1,3]] $\rightarrow$ [[5,1,3]] $\rightarrow$ [[5,1,3]] & $1$ & $375$ \\
		$1\,\mathrm{MHz}$ & $0.9$ & $(106\,\mathrm{kHz},\,9\times10^{-3},\,10)$ & [[5,1,3]] $\rightarrow$ [[5,1,3]] & $10$ & $25$ \\
		$1\,\mathrm{MHz}$ & $0.8$ & $(43.3\,\mathrm{kHz},\,4.5\times10^{-3},\,45)$ & [[5,1,3]] $\rightarrow$ [[5,1,3]] & $45$ & $25$ \\
		$1\,\mathrm{kHz}$ & $0.9$ & No distill. & [[5,1,3]] $\rightarrow$ [[5,1,3]] $\rightarrow$ [[5,1,3]] & $1$ & $125$ \\
		$1\,\mathrm{kHz}$ & $0.8$ & No distill. & $ZX_{3\mathrm{B}}$ $\rightarrow$ [[5,1,3]] $\rightarrow$ [[5,1,3]] $\rightarrow$ [[5,1,3]] & $1$ & $375$ \\
		$1\,\mathrm{Hz}$ & $0.9$ & No distill. & [[5,1,3]] $\rightarrow$ [[5,1,3]] $\rightarrow$ [[5,1,3]] & $1$ & $125$ \\
		$1\,\mathrm{Hz}$ & $0.8$ & No distill. & $ZX_{3\mathrm{B}}$ $\rightarrow$ [[5,1,3]] $\rightarrow$ [[5,1,3]] $\rightarrow$ [[5,1,3]] & $1$ & $375$ \\
		\hline\hline
	\end{tabular}
\end{table*}

\begin{table*}[t]
	\centering
	\scriptsize
	\caption{
    Comparison of logical Bell-pair generation rates obtained from the
    analytical models including waiting-induced memory errors and
    discrete-event Monte Carlo simulations.
    For the analytical calculation, the protocol sequence and stage code distances
    are optimized under
    $1-F_{\rm Bell}^{\rm logical}\le10^{-12}$ without imposing an upper bound
    on $d_{\rm code}$.
    $d_{\rm code}^{\rm final}$ denotes the code distance used in the final
    distillation stage of the analytically selected protocol.
    For the multiple-channel configuration, the multiplexing gain
    $G_{\rm mult}
    =
    r_{\rm multiple}/
    (r_{\rm single}N_{\rm comm}^{\rm logical})$
    is shown in parentheses after the corresponding rate.
    Monte Carlo rates are reported when the mean simulated logical Bell-pair
    infidelity is no larger than $2\times10^{-12}$.
    Monte Carlo validation is restricted to the protocol sequences
    selected by the analytical model.
    Therefore, failure to meet the fidelity target in Monte Carlo
    does not rule out other feasible protocols.
}
	\label{tab:table3_rate}
	\setlength{\tabcolsep}{4pt}
	
	\begin{tabular}{c c c c c c}
		\hline\hline
		
		\multicolumn{3}{c|}{Input parameters}
		&
		\multicolumn{3}{c}{Derived results}
		\\
		
		\cline{1-3}\cline{4-6}
		
		$r_{\mathrm{Bell}}^{\mathrm{phys}}$
		&
		$F_{\mathrm{Bell}}^{\mathrm{phys}}$
		&
		\begin{tabular}{c}
			Physical Bell-pair distillation\\
			$\left(
			r_{\mathrm{Bell}}^{\mathrm{distill}},
			\,1-F_{\mathrm{Bell}}^{\mathrm{distill}},
			\,N_{\mathrm{comm}}^{\mathrm{phys}}
			\right)$
		\end{tabular}
		&
		$d_{\mathrm{code}}^{\mathrm{final}}$
		&
		\begin{tabular}{c}
			Analytical\\rate
		\end{tabular}
		&
		\begin{tabular}{c}
			Monte Carlo\\rate
		\end{tabular}
		\\
		
		\hline
		
		\multicolumn{6}{c}{
			Lattice surgery-based logical Bell-pair generation
		}
		\\
		
		\hline
		$1~\mathrm{MHz}$ & $0.99$ & No distill. & $25$ & $40~\mathrm{kHz}$ & -- \\
		
		\hline
		
		\multicolumn{6}{c}{
			Injection-distillation protocol
			(single communication channel: sequential / MC single)
		}
		\\
		
		\hline
		$1\,\mathrm{MHz}$ & $0.99$ & No distill. & $23$ & $1.19\,\mathrm{kHz}$ & $1.2\,\mathrm{kHz}$ \\
		$1\,\mathrm{MHz}$ & $0.9$ & No distill. & $23$ & $200\,\mathrm{Hz}$ & $200\,\mathrm{Hz}$ \\
		$1\,\mathrm{MHz}$ & $0.8$ & No distill. & $23$ & $52.9\,\mathrm{Hz}$ & $53.5\,\mathrm{Hz}$ \\
		$1\,\mathrm{MHz}$ & $0.9$ & $(106\,\mathrm{kHz},\,9\times10^{-3},\,10)$ & $23$ & $908\,\mathrm{Hz}$ & -- \\
		$1\,\mathrm{MHz}$ & $0.8$ & $(43.3\,\mathrm{kHz},\,4.5\times10^{-3},\,45)$ & $23$ & $625\,\mathrm{Hz}$ & -- \\
		$1\,\mathrm{kHz}$ & $0.9$ & No distill. & $23$ & $2.87\,\mathrm{Hz}$ & $2.85\,\mathrm{Hz}$ \\
		$1\,\mathrm{kHz}$ & $0.8$ & No distill. & $23$ & $0.488\,\mathrm{Hz}$ & $0.489\,\mathrm{Hz}$ \\
		$1\,\mathrm{Hz}$ & $0.9$ & No distill. & $27$ & $0.00294\,\mathrm{Hz}$ & $0.00297\,\mathrm{Hz}$ \\
		$1\,\mathrm{Hz}$ & $0.8$ & No distill. & $29$ & $0.000494\,\mathrm{Hz}$ & $0.000493\,\mathrm{Hz}$ \\
		
		\hline
		
		\multicolumn{6}{c}{
			Injection-distillation protocol
			(multiple communication channels: parallel / MC multiple)
		}
		\\
		
		\hline
		$1\,\mathrm{MHz}$ & $0.99$ & No distill. & $23$ & $2.81\,\mathrm{kHz}\;(G_{\rm mult}=0.094)$ & $2.84\,\mathrm{kHz}\;(G_{\rm mult}=0.095)$ \\
		$1\,\mathrm{MHz}$ & $0.9$ & No distill. & $23$ & $1.5\,\mathrm{kHz}\;(G_{\rm mult}=0.06)$ & -- \\
		$1\,\mathrm{MHz}$ & $0.8$ & No distill. & $23$ & $828\,\mathrm{Hz}\;(G_{\rm mult}=0.042)$ & $1.39\,\mathrm{kHz}\;(G_{\rm mult}=0.069)$ \\
		$1\,\mathrm{MHz}$ & $0.9$ & $(106\,\mathrm{kHz},\,9\times10^{-3},\,10)$ & $23$ & $2.69\,\mathrm{kHz}\;(G_{\rm mult}=0.12)$ & -- \\
		$1\,\mathrm{MHz}$ & $0.8$ & $(43.3\,\mathrm{kHz},\,4.5\times10^{-3},\,45)$ & $23$ & $2.43\,\mathrm{kHz}\;(G_{\rm mult}=0.16)$ & -- \\
		$1\,\mathrm{kHz}$ & $0.9$ & No distill. & $23$ & $88\,\mathrm{Hz}\;(G_{\rm mult}=0.25)$ & $325\,\mathrm{Hz}\;(G_{\rm mult}=0.91)$ \\
		$1\,\mathrm{kHz}$ & $0.8$ & No distill. & $23$ & $31.9\,\mathrm{Hz}\;(G_{\rm mult}=0.17)$ & $175\,\mathrm{Hz}\;(G_{\rm mult}=0.96)$ \\
		$1\,\mathrm{Hz}$ & $0.9$ & No distill. & $23$ & $0.0965\,\mathrm{Hz}\;(G_{\rm mult}=0.26)$ & $0.381\,\mathrm{Hz}\;(G_{\rm mult}=1.02)$ \\
		$1\,\mathrm{Hz}$ & $0.8$ & No distill. & $25$ & $0.0336\,\mathrm{Hz}\;(G_{\rm mult}=0.18)$ & $0.194\,\mathrm{Hz}\;(G_{\rm mult}=1.05)$ \\
		
		\hline\hline
	\end{tabular}
\end{table*}

\begin{table*}[tb]
	\centering
	\scriptsize
	\caption{Resource overheads of logical Bell-pair generation and distillation including waiting-induced memory errors. The sequential model is used for the single-communication-channel case, whereas the parallel model is used for the multiple-communication-channel case. The distillation protocols are selected under $1-F_{\rm Bell}^{\rm logical}\le10^{-12}$ without an explicit upper bound on $d_{\rm code}$. For cases involving physical Bell-pair distillation, the reported physical Bell-pair consumption is an effective count that includes Bell pairs that could have been generated during waiting and distillation, rather than only those actually consumed.
}
	\label{tab:table4_overhead}
	\setlength{\tabcolsep}{4pt}
	\begin{tabular}{c c c c c c c}
		\hline\hline
		\multicolumn{3}{c|}{Input parameters} & \multicolumn{4}{c}{Resource overheads} \\
		\cline{1-3}\cline{4-7}
		$r_{\mathrm{Bell}}^{\mathrm{phys}}$ &
		$F_{\mathrm{Bell}}^{\mathrm{phys}}$ &
		\begin{tabular}{c}Physical Bell-pair distillation\\$\left(r_{\mathrm{Bell}}^{\mathrm{distill}},\,1-F_{\mathrm{Bell}}^{\mathrm{distill}},\,N_{\mathrm{comm}}^{\mathrm{phys}}\right)$\end{tabular} &
		\begin{tabular}{c}Total num. of\\communication channels\\$N_{\mathrm{comm}}^{\mathrm{total}}$\end{tabular} &
		\begin{tabular}{c}Consumed num. of\\physical Bell pairs\\$N_{\mathrm{consumed}}^{\mathrm{phys}}$\end{tabular} &
		\begin{tabular}{c}Peak physical-qubit count per node\\for logical Bell-pair distillation\end{tabular} &
		\begin{tabular}{c}Total overhead, normalized\\by lattice surgery\end{tabular} \\
		\hline
		\multicolumn{7}{c}{Lattice surgery-based logical Bell-pair generation} \\
		\hline
		$1~\mathrm{MHz}$ & $0.99$ & No distill. & $49$ & $1.2\times10^{3}$ & $1.2\times10^{3}$ & $1.00$ \\
		\hline
		\multicolumn{7}{c}{Injection-distillation protocol (single communication channel, sequential model with memory errors)} \\
		\hline
		$1\,\mathrm{MHz}$ & $0.99$ & No distill. & $1$ & $32$ & $8.1\times10^{3}$ & $0.121$ \\
		$1\,\mathrm{MHz}$ & $0.9$ & No distill. & $1$ & $6.6\times10^{2}$ & $8.1\times10^{3}$ & $15$ \\
		$1\,\mathrm{MHz}$ & $0.8$ & No distill. & $1$ & $2\times10^{3}$ & $8.1\times10^{3}$ & $174$ \\
		$1\,\mathrm{MHz}$ & $0.9$ & $(106\,\mathrm{kHz},\,9\times10^{-3},\,10)$ & $10$ & $3\times10^{3}$ & $8.1\times10^{3}$ & $150$ \\
		$1\,\mathrm{MHz}$ & $0.8$ & $(43.3\,\mathrm{kHz},\,4.5\times10^{-3},\,45)$ & $45$ & $3.2\times10^{4}$ & $8.1\times10^{3}$ & $1.05\times10^{4}$ \\
		$1\,\mathrm{kHz}$ & $0.9$ & No distill. & $1$ & $3.4\times10^{2}$ & $8.7\times10^{3}$ & $584$ \\
		$1\,\mathrm{kHz}$ & $0.8$ & No distill. & $1$ & $2\times10^{3}$ & $1.1\times10^{4}$ & $2.54\times10^{4}$ \\
		$1\,\mathrm{Hz}$ & $0.9$ & No distill. & $1$ & $3.4\times10^{2}$ & $1.5\times10^{4}$ & $9.65\times10^{5}$ \\
		$1\,\mathrm{Hz}$ & $0.8$ & No distill. & $1$ & $2\times10^{3}$ & $1.9\times10^{4}$ & $4.4\times10^{7}$ \\
		\hline
		\multicolumn{7}{c}{Injection-distillation protocol (multiple communication channels, parallel model with memory errors)} \\
		\hline
		$1\,\mathrm{MHz}$ & $0.99$ & No distill. & $25$ & $32$ & $1.5\times10^{4}$ & $2.39$ \\
		$1\,\mathrm{MHz}$ & $0.9$ & No distill. & $1.2\times10^{2}$ & $3.4\times10^{2}$ & $2.8\times10^{4}$ & $445$ \\
		$1\,\mathrm{MHz}$ & $0.8$ & No distill. & $3.8\times10^{2}$ & $2\times10^{3}$ & $9.8\times10^{4}$ & $5.12\times10^{4}$ \\
		$1\,\mathrm{MHz}$ & $0.9$ & $(106\,\mathrm{kHz},\,9\times10^{-3},\,10)$ & $2.5\times10^{2}$ & $3\times10^{3}$ & $1.5\times10^{4}$ & $2.36\times10^{3}$ \\
		$1\,\mathrm{MHz}$ & $0.8$ & $(43.3\,\mathrm{kHz},\,4.5\times10^{-3},\,45)$ & $1.1\times10^{3}$ & $3.2\times10^{4}$ & $1.8\times10^{4}$ & $1.52\times10^{5}$ \\
		$1\,\mathrm{kHz}$ & $0.9$ & No distill. & $1.2\times10^{2}$ & $3.7\times10^{2}$ & $5.7\times10^{4}$ & $1.68\times10^{4}$ \\
		$1\,\mathrm{kHz}$ & $0.8$ & No distill. & $3.8\times10^{2}$ & $2\times10^{3}$ & $2.7\times10^{5}$ & $3.62\times10^{6}$ \\
		$1\,\mathrm{Hz}$ & $0.9$ & No distill. & $1.2\times10^{2}$ & $3.7\times10^{2}$ & $1.3\times10^{5}$ & $3.39\times10^{7}$ \\
		$1\,\mathrm{Hz}$ & $0.8$ & No distill. & $3.8\times10^{2}$ & $2\times10^{3}$ & $5.9\times10^{5}$ & $7.36\times10^{9}$ \\
		\hline\hline
	\end{tabular}
\end{table*}

The resulting logical Bell-pair distillation rates are listed in
Table~\ref{tab:table3_rate}.
The sequential single-channel model represents the minimum-communication
configuration, whereas the parallel model represents the full-multiplexing
limit.
For comparison, lattice-surgery-based logical Bell-pair generation gives a
rate of $40~\mathrm{kHz}$ for $d_{\mathrm{code}}=25$.
For the injection-distillation protocol, the final-stage code distance
required to reach the $10^{-12}$ target ranges from 23 to 29.
At
$r_{\mathrm{Bell}}^{\mathrm{phys}}=1~\mathrm{Hz}$,
the sequential model requires
$d_{\mathrm{code}}^{\mathrm{final}}=27$--$29$ because of the longer
waiting-induced memory error.
For the sequential configuration,
$(r_{\mathrm{Bell}}^{\mathrm{phys}},F_{\mathrm{Bell}}^{\mathrm{phys}})
=(1~\mathrm{MHz},0.99)$
gives a logical Bell-pair rate of approximately $1~\mathrm{kHz}$, whereas
$F_{\mathrm{Bell}}^{\mathrm{phys}}=0.8$--$0.9$ gives rates on the order of
$10^{2}~\mathrm{Hz}$.
The rate then decreases approximately linearly with
$r_{\mathrm{Bell}}^{\mathrm{phys}}$ in the low-rate regime.
For the parallel configuration, Table~\ref{tab:table3_rate} reports the
total multiple-channel rate, whereas
Fig.~\ref{fig:logical_distill_rate} shows the communication-normalized rate.
The Monte Carlo total rates are on the order of $10^{3}~\mathrm{Hz}$ at
$r_{\mathrm{Bell}}^{\mathrm{phys}}=1~\mathrm{MHz}$,
$10^{2}~\mathrm{Hz}$ at $1~\mathrm{kHz}$, and
$10^{-1}~\mathrm{Hz}$ at $1~\mathrm{Hz}$ for the representative
$F_{\mathrm{Bell}}^{\mathrm{phys}}=0.8$--$0.99$ cases.

The benefit of communication multiplexing depends strongly on the input
physical Bell-pair generation rate.
We define the multiplexing gain as
$G_{\mathrm{mult}}
=
r_{\mathrm{multiple}}/
(r_{\mathrm{single}}N_{\mathrm{comm}}^{\mathrm{logical}})$.
Thus, $G_{\mathrm{mult}}=1$ corresponds to an ideal proportional increase in
the total logical Bell-pair generation rate with the number of logical
communication channels.
At
$r_{\mathrm{Bell}}^{\mathrm{phys}}\sim1~\mathrm{MHz}$,
where the logical-distillation rate is close to saturation,
$G_{\mathrm{mult}}$ is only on the order of $0.05$--$0.1$, indicating that
additional channels provide a sublinear rate improvement.
In contrast, the Monte Carlo results give
$G_{\mathrm{mult}}\simeq0.9$--$1.0$ at
$r_{\mathrm{Bell}}^{\mathrm{phys}}\le1~\mathrm{kHz}$,
showing nearly ideal utilization of the multiplexed channels in the
generation-limited regime.
The analytical parallel model gives a smaller gain in this regime, consistent
with its conservative rate estimate discussed above.
An intermediate degree of multiplexing may therefore provide a more favorable
rate--resource trade-off than either limiting configuration; optimizing this
degree is left for future work.

The communication and space overheads are summarized in
Table~\ref{tab:table4_overhead}.
We define the total number of communication channels as
$N_{\mathrm{comm}}^{\mathrm{total}}
=
N_{\mathrm{comm}}^{\mathrm{phys}}
N_{\mathrm{comm}}^{\mathrm{logical}}$.
For the sequential configuration,
$N_{\mathrm{comm}}^{\mathrm{total}}=1$ without physical Bell-pair
distillation and
$N_{\mathrm{comm}}^{\mathrm{total}}
=
N_{\mathrm{comm}}^{\mathrm{phys}}$
when physical Bell-pair distillation is used.
The full parallel configuration requires between 25 and approximately
$10^{3}$ communication channels for the cases considered here.
For comparison, lattice-surgery-based logical Bell-pair generation requires
$2d_{\mathrm{code}}-1=49$ channels for $d_{\mathrm{code}}=25$.
Thus, the ability to generate logical Bell pairs using only a single
communication channel is a distinct advantage of the sequential
injection-distillation configuration.

The consumed number of physical Bell pairs,
$N_{\mathrm{consumed}}^{\mathrm{phys}}$, is typically on the order of
$10^{2}$--$10^{3}$ for the cases without physical Bell-pair predistillation,
comparable to the $1.2\times10^{3}$ Bell pairs required by the
lattice-surgery-based scheme.
For
$F_{\mathrm{Bell}}^{\mathrm{phys}}=0.99$,
$N_{\mathrm{consumed}}^{\mathrm{phys}}=32$, more than an order of magnitude
smaller than the lattice-surgery value.
Physical Bell-pair predistillation can substantially increase the logical
Bell-pair generation rate, but it also increases the communication-channel
and Bell-pair-consumption overheads.

We also evaluate the physical-qubit requirement for logical Bell-pair
distillation as the space overhead.
The sequential configuration requires approximately
$8\times10^{3}$--$2\times10^{4}$ physical qubits per node, whereas the full
parallel configuration requires approximately
$1.5\times10^{4}$--$6\times10^{5}$.
The larger requirement in the parallel configuration arises because the
distillation hardware for the full parallel tree, including simultaneously
active stages, is provisioned concurrently.
By contrast, the sequential configuration limits the number of simultaneously
active distillation resources through backpressure.
Compared with the approximately $1.2\times10^{3}$ physical qubits required for
the lattice-surgery-based reference case, the sequential scheme therefore
requires roughly one order of magnitude more space, while full multiplexing
can require substantially larger space overhead.

To compare the combined communication, consumption, space, and time costs
with lattice-surgery-based logical Bell-pair generation, we define
$
	\mathcal{O}_{\mathrm{total}}
	=
		N_{\mathrm{comm}}^{\mathrm{total}}
		N_{\mathrm{consumed}}^{\mathrm{phys}}
		N_{\mathrm{buffer}}^{\mathrm{phys}}
	/
		r_{\mathrm{Bell}}^{\mathrm{logical(distill)}}
	,
$
and report this quantity normalized by the corresponding value for lattice
surgery,
$\mathcal{O}_{\mathrm{total}}^{\mathrm{norm}}$.
For the sequential model at
$r_{\mathrm{Bell}}^{\mathrm{phys}}=1~\mathrm{MHz}$,
$\mathcal{O}_{\mathrm{total}}^{\mathrm{norm}}$ is $0.121$, $15$, and $174$
for
$F_{\mathrm{Bell}}^{\mathrm{phys}}=0.99$, $0.9$, and $0.8$, respectively.
Thus, the injection-distillation scheme can have a lower total overhead than
the lattice-surgery reference in the high-fidelity, high-rate regime, but the
overhead increases rapidly as the physical Bell-pair fidelity or generation
rate is reduced.
At lower
$r_{\mathrm{Bell}}^{\mathrm{phys}}$,
the increase in
$\mathcal{O}_{\mathrm{total}}^{\mathrm{norm}}$
is dominated by the reduced logical Bell-pair generation rate, while the
communication-channel count and Bell-pair consumption change comparatively
little.

In summary, the injection-distillation protocol can generate logical Bell
pairs with infidelities below $10^{-12}$ over a broad range of physical-link
parameters.
The single-channel configuration minimizes communication hardware and requires
roughly one order of magnitude more physical qubits than the lattice-surgery
reference, but its logical Bell-pair generation rate is substantially lower.
Full multiplexing increases the total logical Bell-pair generation rate, particularly in
the low-input-rate regime, but requires many more communication channels and
can increase the physical-qubit requirement by more than two orders of
magnitude relative to the lattice-surgery reference.
The principal advantages of injection and distillation are that
(1) physical Bell-pair fidelities of approximately $0.8$--$0.9$ and physical
Bell-pair generation rates well below $100~\mathrm{kHz}$ can be tolerated,
(2) logical Bell-pair generation remains possible with a single communication
channel, and
(3) the physical Bell-pair consumption can be substantially reduced in the
high-fidelity regime.
If a much larger physical-qubit space is available, as shown in
Ref.~\cite{Lao2022}, the number of physical Bell pairs required per logical
Bell pair can be further reduced.
Because preparing high-fidelity physical Bell pairs at high rates over remote
communication links is technically challenging, these trade-offs can relax
the requirements on the quantum communication hardware.

\section{Step V: Execution of Quantum Algorithms and Estimation of Their Resource Overheads}
The resource overheads required for a distributed fault-tolerant quantum computer depend primarily on the target applications.
To determine whether the overheads discussed in the previous section are acceptable, we performed extensive benchmarking using practical problem instances.

\subsection{Fault-tolerant Quantum Computer Node and Network}
\label{step5:benchmark}
\begin{figure}[bp]
    \centering
    \includegraphics[width=\linewidth]{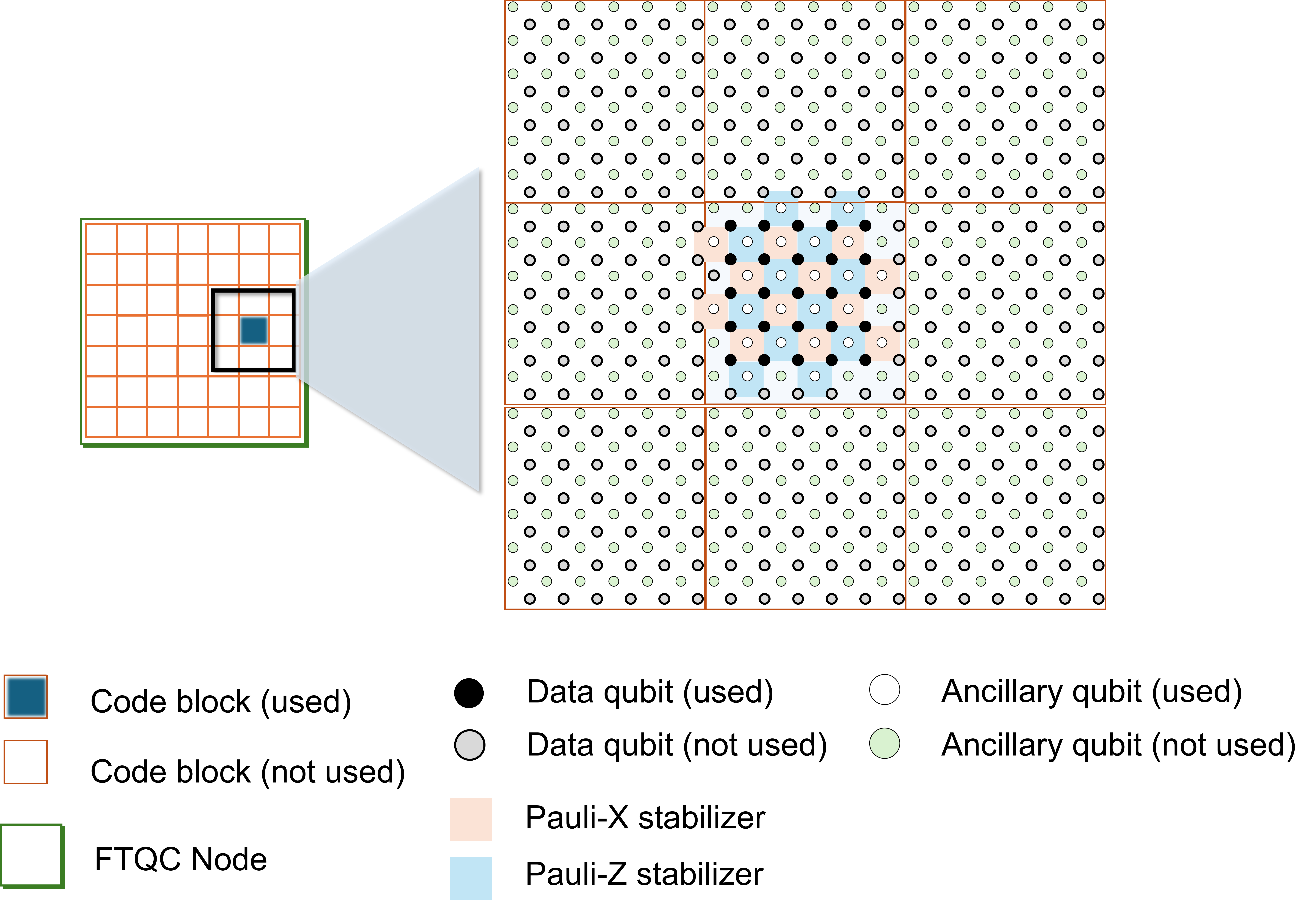}
    \caption{Structure of a fault-tolerant quantum computer (FTQC) node. The left schematic shows a fault-tolerant quantum computer node partitioned into code blocks (orange-outlined squares), each of which can encode a logical qubit. A blue-filled code block denotes a used block participating in logical-state encoding, whereas an unfilled (white) code block denotes an inactive block. The enlarged view shows the underlying two-dimensional array of data qubits (black and gray circles) and ancillary qubits (white and green circles) used for syndrome measurements. Red and blue plaquettes denote Pauli-X and Pauli-Z stabilizers, respectively. Black and white circles indicate data and ancillary qubits, respectively, within the used code block, whereas gray and green circles indicate those in inactive code blocks. The green outline marks the boundary of the fault-tolerant quantum computer node. }
    \label{fig:dftqc_node}
\end{figure}

\begin{figure}[bp]
    \centering
    \includegraphics[width=\linewidth]{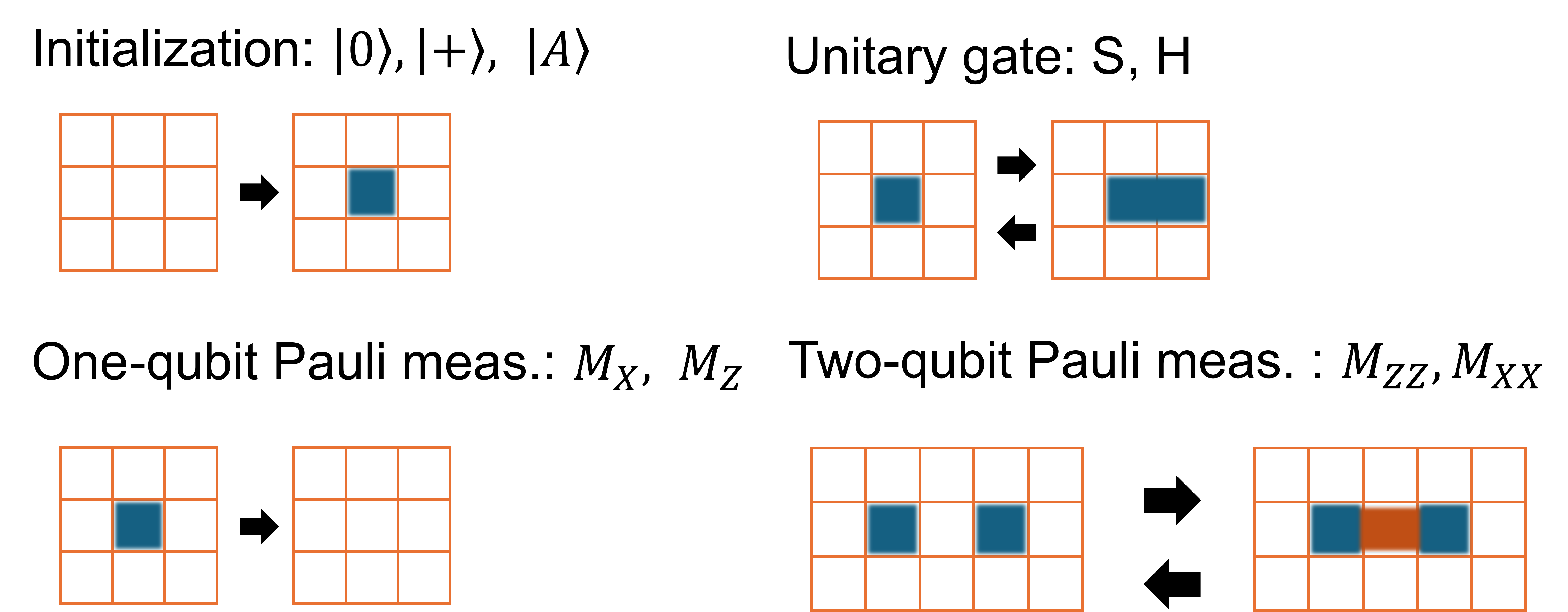}
    \caption{Schematic illustrations of logical operations considered in this work: initialization into ($\ket{0}$), ($\ket{+}$), and ($\ket{A}$); single-qubit unitary gates (S) and (H); single-qubit Pauli measurements ($M_X$) and ($M_Z$); and two-qubit Pauli measurements ($M_{ZZ}$) and ($M_{XX}$).
    }
    \label{fig:dftqc_operation}
\end{figure}

This section describes the numerical settings used to evaluate a distributed fault-tolerant quantum computer.
We focus on a surface-code-based fault-tolerant quantum computer architecture in which distance-$d$ code blocks are arranged on two-dimensional grids as shown in Fig.\,\ref{fig:dftqc_node}.
Here, the code distance $d_\mathrm{code}$ is treated as a variable and is optimized such that the total logical error rate remains below the allowable error threshold $\epsilon$.
We assume that logical operations are performed using lattice surgery~\cite{horsman2012surface,fowler2018low}, and logical S and H gates are implemented using code deformation~\cite{fowler2018low,Litinski2019,bombin2023logical,beverland2022assessing}.
During computation, stabilizer measurements are performed periodically. Each such period is called a code cycle and has a duration of $\tau_{\mathrm{code}}^\mathrm{cycle}$.
To perform typical logical Clifford operations using lattice surgery and code deformation, stabilizer measurements must be repeated $d_\mathrm{code}$ times to ensure reliable error estimation~\cite{horsman2012surface}.
We refer to the latency of lattice surgery, i.e., $d_\mathrm{code}$ cycles of stabilizer measurements, as a code beat~\cite{yoshioka2024hunting}. The latency in units of code beats and the required space for each logical operation are listed in Table~\ref{tab:dftqc_operation} and illustrated in Fig.\,\ref{fig:dftqc_operation}.
\begin{table}
    \begin{tabular}{|c|c|c|}
        \hline
        Operation & Latency & Space \\
        \hline \hline
        \makecell{$\ket{0}, \ket{+}$\\Initialization} & 0 & in place \\
        \hline
        \makecell{Magic state $\ket{A}$\\Initialization} & Random (1) & in place \cite{gidney2024magic} \\
        \hline
        S gate & 2 & horizontally adjacent cell \cite{bombin2006topological} \\
        \hline
        H gate & 3 & adjacent cell \cite{Litinski2019} \\
        \hline
        CNOT gate & 2 & path between the target nodes~\cite{fowler2018low} \\
        \hline
        \makecell{Pauli-X/Z\\Measurement} & 0 & in place \\
        \hline
        \makecell{Pauli-XX/ZZ\\Measurement} & 1 & path between target nodes \cite{horsman2012surface} \\
        \hline
    \end{tabular}
    \caption{Logical operations and their spacetime costs. Latency is expressed in units of code beats. Although magic-state initialization is probabilistic and its latency is therefore stochastic, we assume a latency of one code beat in the calculations presented in this paper. }
    \label{tab:dftqc_operation}
\end{table}

\begin{figure}[thbp]
    \centering
    \includegraphics[width=\linewidth]{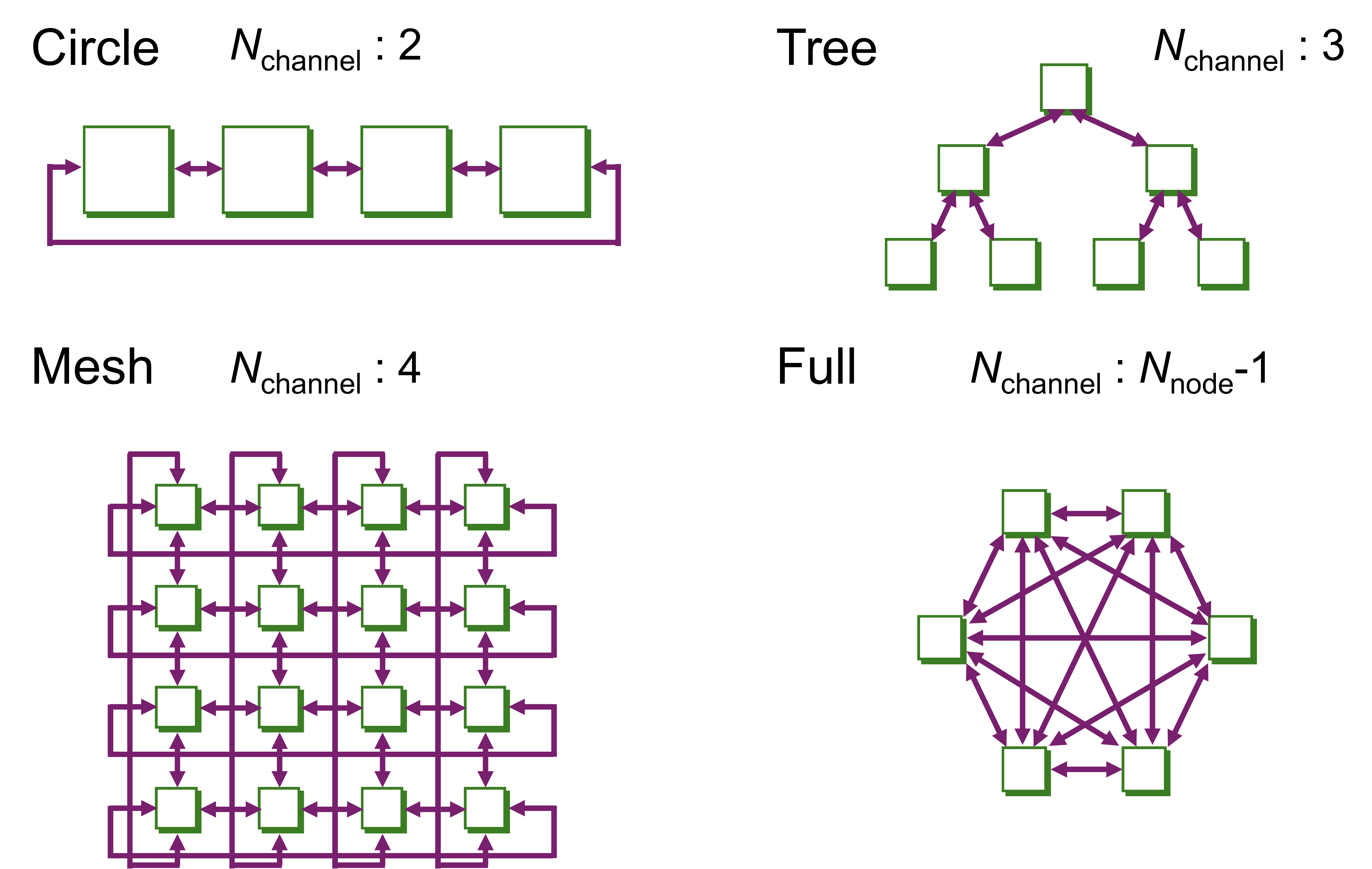}
    \caption{Distributed fault-tolerant quantum computer network structures. In the circle network, each fault-tolerant quantum computer node is connected to its two neighboring nodes, forming a ring. In the tree network, each parent node is connected to two child nodes, and this branching is repeated until the required number of nodes is reached. In the mesh network, fault-tolerant quantum computer nodes are arranged in a two-dimensional grid with connections across the boundaries. In the full network, every node is directly connected to every other node. }
    \label{fig:dftqc_network}
\end{figure}

We assume that several logical qubits with sufficiently long lifetimes can be constructed within each node and consider a distributed fault-tolerant quantum computer consisting of several nodes connected by logical Bell-pair generators. 
We consider four types of fixed fault-tolerant quantum computer networks: a circle graph, a 2D mesh grid, a tree graph, and a fully connected graph. We refer to these topologies as circle, mesh, tree, and full, respectively. Figure~\ref{fig:dftqc_network} shows examples of these network structures. 

Although the network structure could be reconfigured during computation, we focus on fixed network topologies and leave dynamic reconfiguration for future work because it would substantially complicate the compilation process.

We denote the number of nodes in the network by $N_{\rm node}$. 
We assume a maximum of $2$, $4$, $3$, and $N_{\rm node}-1$ communication channels per node for the circle, mesh, tree, and full networks, respectively, for generating logical Bell pairs with other nodes. We denote the number of channels per node by $N_{\rm channel}$. Although $N_{\rm channel}$ for the tree network ranges from 1 to 3 depending on the node, we assume three channels per node for simplicity in the calculations. The number of Bell-pair generators is fixed at one per node, irrespective of $N_{\rm channel}$.

We assume that each node has a finite number of physical quantum channels and can distill one logical Bell pair every $T_{\rm Bell}^{\mathrm{period}}$ code beats. Although logical Bell-pair generation is inherently probabilistic, we approximate it as a periodic process for simplicity. When a node is connected to $N_{\rm channel}$ other nodes, we assume that logical Bell pairs for the individual connections are generated sequentially. We phenomenologically account for this sharing of the Bell-pair generation capability by increasing the generation period for each connection from $T_{\rm Bell}^{\mathrm{period}}$ to $N_{\rm channel}T_{\rm Bell}^{\mathrm{period}}$. Accordingly, the generation period is $2T_{\rm Bell}^{\mathrm{period}}$ for the circle network considered here and $(N_{\rm node}-1)T_{\rm Bell}^{\mathrm{period}}$ for the full network. In the latter case, the generation period therefore increases linearly with the number of nodes. This model captures the trade-off associated with network connectivity: a denser network reduces the number of logical Bell pairs required to connect a pair of nodes, while decreasing the Bell-pair generation rate available to each connection as the network size increases.

We reserve $N_{\rm channel}$ code blocks at each node for storing distilled Bell pairs, corresponding to its $N_{\rm channel}$ connections. Once an entanglement distiller has generated a logical Bell pair, it cannot begin generating the next pair until the stored pair has been consumed. 

We further assume that each node has one magic-state generator.
The generator occupies a single surface-code block and produces one
distilled magic state per code beat. We do not explicitly include the
physical-qubit footprint of the logical Bell-pair generator in the
following algorithm-level resource estimates, because it depends on
the choice of Bell-pair generation and distillation protocols and on
the required output error rate, as discussed in the previous sections.
The physical-qubit counts presented below therefore include the
logical Bell-pair storage blocks but exclude the physical qubits
required for logical Bell-pair generation and distillation. For the
representative protocols considered above, this additional overhead is
typically $O(10^4)$ physical qubits per node.

\begin{figure*}[t]
    \centering
    \includegraphics[width=\linewidth]{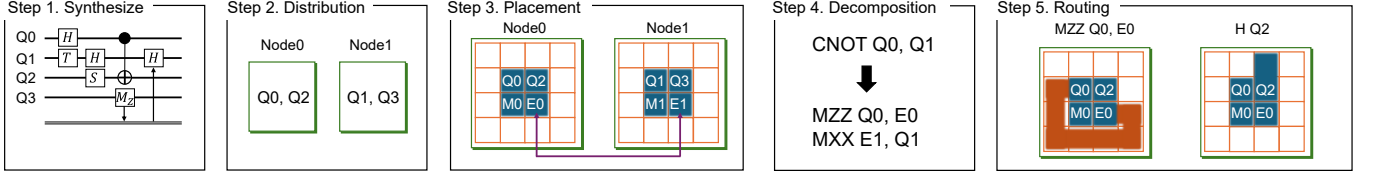}
    \caption{Overview of the compilation flow. In Step 1, a quantum circuit consisting of Clifford + T gates is synthesized for a given quantum algorithm without considering remote operations. In Step 2, logical qubits are distributed among the fault-tolerant quantum computer nodes. In Step 3, the logical qubits are placed within each node together with code blocks for magic-state generation and logical Bell-pair generation/distillation. In Step 4, the operations synthesized in Step 1 are decomposed into logical operations executable on the distributed architecture. Finally, in Step 5, these operations are routed onto the code-block layout determined in Step 3. }
    \label{fig:dftqc_compile}
\end{figure*}

\subsection{Benchmark and Compilation}
We evaluated the execution time for each combination of network topology and node configuration using the following procedure.
We selected three application subroutines, SELECT, PREPARE, and ADDER, as benchmark instance types that represent a typical structure of integer factoring and ground-energy estimation.
SELECT and PREPARE are subroutines used in qubitization-based quantum phase estimation~\cite{low2019hamiltonian,babbush2018encoding} and quantum singular value transformation~\cite{martyn2021grand}. Given a Hamiltonian $\hat{H} = \sum_i \alpha_i P_i$ with positive coefficients $\alpha_i$ and $n$-qubit Pauli operators $P_i \in \pm \{I,\ X,\ Y,\ Z\}^{\otimes n}$, SELECT acts as $U_{\rm SEL} = \sum_i \ket{i}\bra{i} \otimes P_i$, whereas PREPARE is defined by $U_{\rm PRE}$ such that $U_{\rm PRE}\ket{0} = \sum_i \sqrt{\frac{\alpha_i}{\sum_j \alpha_j}} \ket{i}$. See Figs.~5 and 11 of Ref.~\cite{babbush2018encoding} for efficient implementations of these operations. ADDER is a unitary operator $U_{\rm ADD}$ with action $U_{\rm ADD} \ket{x}\ket{y} = \ket{x} \ket{x+y}$. We chose an ADDER circuit as a representative arithmetic computation and used the circuit shown in Fig.~1 of Ref.~\cite{gidney2018halving}. In the following discussion, we denote the number of spins in the Hamiltonian of PREPARE and SELECT, and the bitwidth of ADDER as instance sizes. We used one-dimensional transverse-field Ising models for generating instances of PREPARE and SELECT. 

These benchmark tasks are compiled into executable operations of a distributed fault-tolerant quantum computer listed in Table~\ref{tab:dftqc_operation} using the network type, the number of nodes $N_{\rm node}$, the instance type, and the instance size $N_{\rm instance}$ as input parameters. This process consists of five phases: synthesis, distribution, placement, decomposition, and routing. An overview of the compilation flow is shown in Fig.~\ref{fig:dftqc_compile}.

In the synthesis phase, we generate algorithms expressed in the Clifford+T formalism, i.e., using CNOT, H, S, and T gates and Pauli measurements in the Z and X bases. We refer to the qubits in this representation, i.e., the individual wires in a quantum circuit, as algorithmic qubits. We define the number of qubits in the quantum circuit as $n_\mathrm{circuit}$.

\begin{figure}[b]
    \centering
    \includegraphics[width=\linewidth]{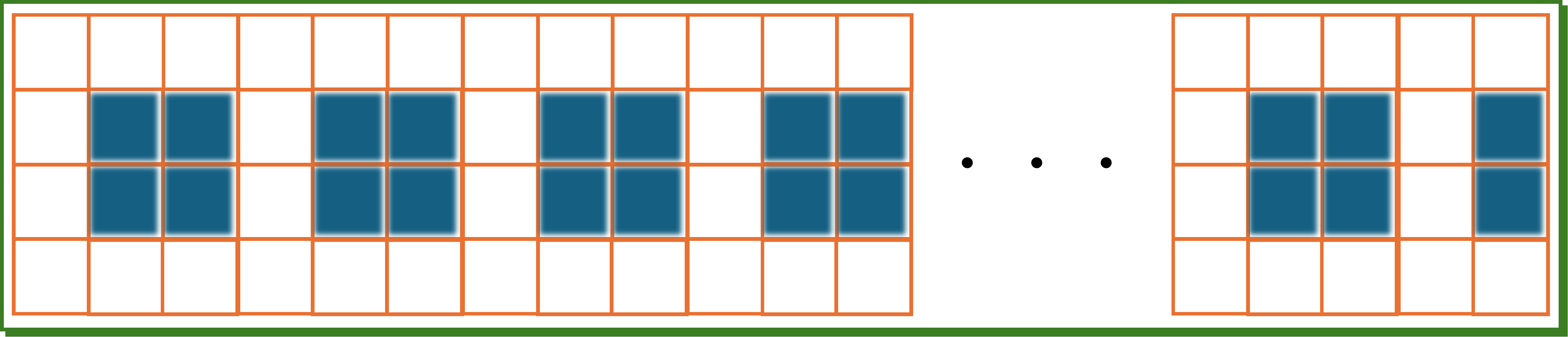}
    \caption{Layout of a 4-in-12 fault-tolerant quantum computer node. Four logical qubits are placed in each group of 12 code blocks, with the remaining code blocks reserved for routing and logical operations.}
    \label{fig:dftqc_layout}
\end{figure}

In the distribution phase, each algorithmic qubit is mapped to a node in the given network. Suppose the target network
consists of $N_{\rm node}$ nodes. We assign the $i$-th algorithmic qubit to the $(i \bmod N_{\rm node})$-th node.
Consequently, each node contains at most $\lceil n_\mathrm{circuit}/N_{\rm node} \rceil$ algorithmic qubits, $N_{\rm channel}$ code blocks for storing logical Bell pairs, and one code block for magic-state generation. We assume one logical Bell-pair distiller per
node, but do not include its physical-qubit cost at this stage, as it depends on the choice of distillation protocol.
We refer to the Bell-pair storage and magic-state blocks as reserved code blocks.

In the placement phase, we assign each reserved code block to a specific location within its node. In our evaluation, we use the 4-in-12 layout shown in Fig.\,\ref{fig:dftqc_layout}, which determines the number of code blocks in each fault-tolerant quantum computer node. Note that this rectangular layout may cause routing conflicts in some cases, but we leave the optimization of code-block placement for future work. This procedure determines the number of code blocks, $N_{\rm block}$, in each node.

\begin{figure}[b]
    \centering
    \includegraphics[width=\linewidth]{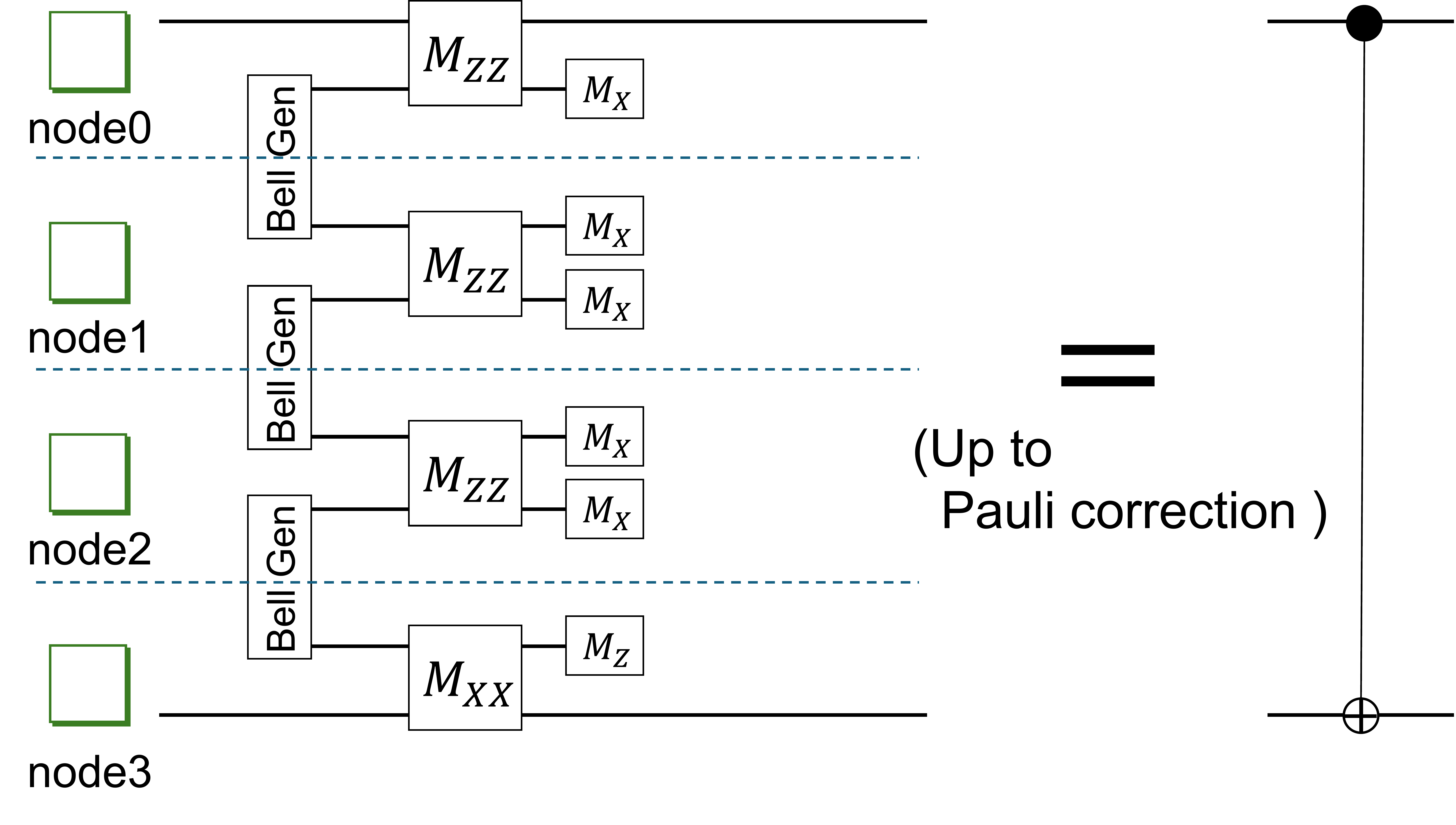}
    \caption{Decomposition of remote CNOT. A logical CNOT operation between distant nodes can be implemented by consuming logical Bell pairs in the connection path and performing local logical operations in each node, i.e. logical gate teleportation.}
    \label{fig:dftqc_decompose}
\end{figure}

In the decomposition phase, we decompose logical operations into the executable instructions shown in Table~\ref{tab:dftqc_operation}. Each T gate is decomposed into a combination of magic-state generation, a two-qubit Pauli-$ZZ$ measurement, a single-qubit Pauli-$X$ measurement, and a conditional S gate, as shown in Fig.~12 of Ref.~\cite{fowler2018low}. Remote CNOT gates, i.e., CNOT gates acting on algorithmic qubits assigned to different nodes, are also decomposed into local operations as follows. Each remote CNOT is decomposed into local lattice-surgery operations using the equivalence shown in Fig.\,\ref{fig:dftqc_decompose}~\cite{beverland2022surface,hamada2026efficient}. 
To determine which logical Bell pairs to use, we simulate the execution of the quantum programs while considering only the latency and generation of logical Bell pairs. Let the control and target qubits of a remote CNOT be located at nodes $n_c$ and $n_t$, respectively. Let $G$ be a graph whose vertices correspond to fault-tolerant quantum computer nodes, with an edge between two vertices whenever a distilled Bell pair is available between the corresponding nodes. We search for a shortest path between $n_c$ and $n_t$ in $G$. If no path exists, we postpone the execution of the remote CNOT and wait for the generation of Bell pairs. If a path is found, we implement the remote CNOT as a chain of two-body Pauli-product measurements.

\begin{figure*}[thbp]
	\centering
	\includegraphics[width=0.75\linewidth]{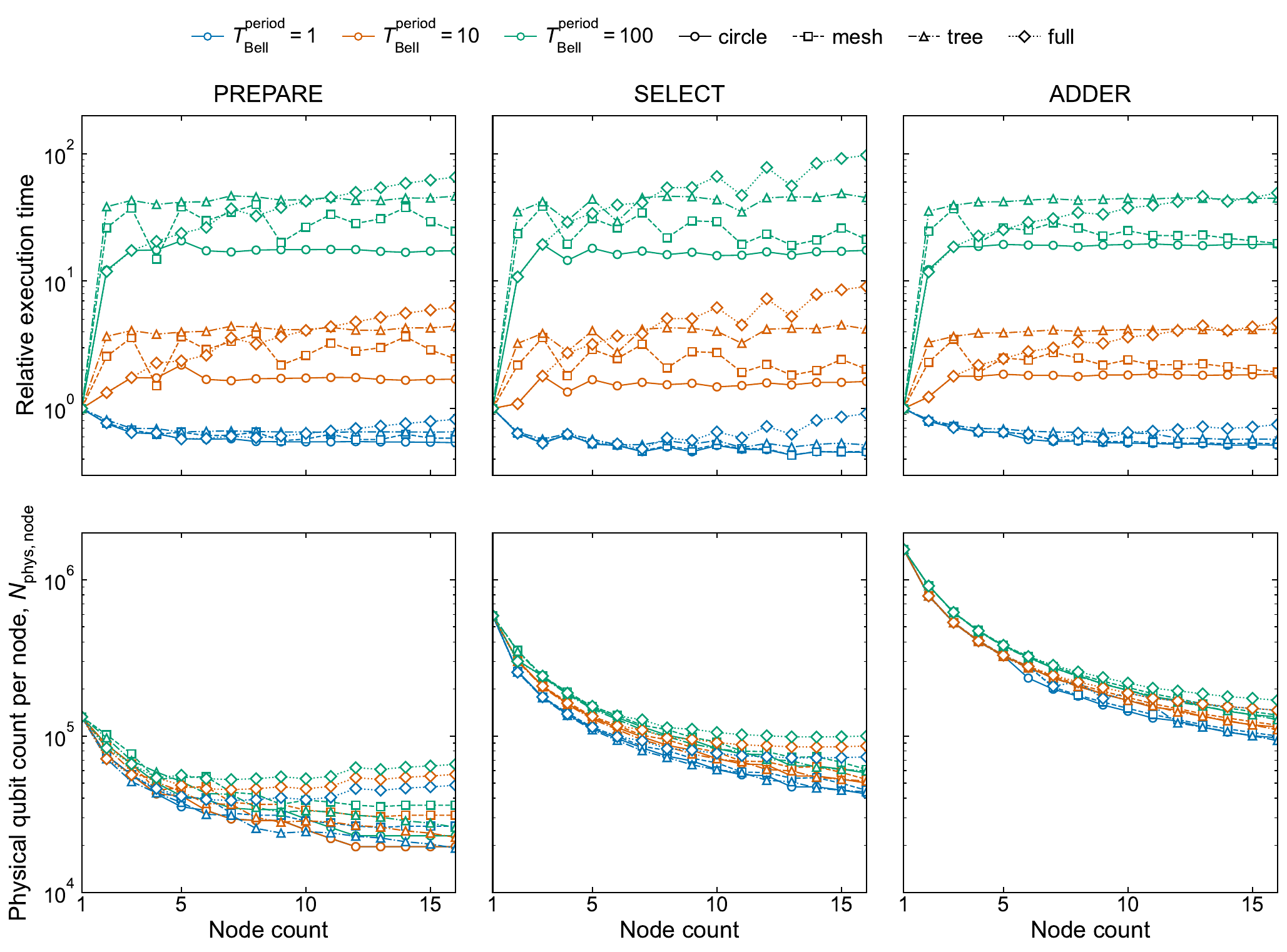}
	\caption{Execution time and physical qubit count per node $N_\mathrm{phys,node}$ as functions of the node count. The upper and lower rows show the execution-time overhead and $N_\mathrm{phys,node}$, respectively. The columns correspond to the instance types. Four network configurations are considered for three different periods ($T_\mathrm{Bell}^\mathrm{period}$= 1, 10, and 100). The instance size is $N_\mathrm{instance}=128$.}
	\label{fig:result_execution_time}
\end{figure*}

Finally, in the routing phase, we determine the ancillary space used by each local logical operation. We greedily assign neighboring ancillary code blocks to H and S gates. For lattice surgery and CNOT gates, we assign one of the shortest paths between the code blocks corresponding to the operands.

After generating an executable program for a distributed fault-tolerant quantum computer, we perform code-beat-accurate simulation of it. From this evaluation, we obtain the execution time $T_\mathrm{exec}^\mathrm{norm}$ in code beats and the normalized qubit volume $V_\mathrm{qubit}^\mathrm{norm}$. The normalized qubit volume is defined as the number of syndrome values obtained during the computation divided by $d_\mathrm{code}^3$, i.e., the time-space cost for $d$-round idling, and can be calculated without determining the code distance $d_\mathrm{code}$. We assume that the total logical failure rate of the circuit is given by $p_{\rm circuit}(d_\mathrm{code}) = V_\mathrm{qubit}^\mathrm{norm} d_\mathrm{code} p_{\rm logical}(d_\mathrm{code})$, where $p_{\rm logical}(d_\mathrm{code})$ is the logical error rate per code block per code cycle and is modeled as $p_{\rm logical} = (p_{\rm bulk}/p_{\rm bulk}^{\rm th})^{(d_\mathrm{code}+1)/2}$. 
Strictly speaking, a prefactor in this scaling relation depends on the physical error rate and the decoding algorithm~\cite{Litinski2019,yoshioka2024hunting}. We determine the code distance $d_\mathrm{code}$ as the minimum integer satisfying $p_{\rm circuit}(d_\mathrm{code}) \leq \epsilon$. Note that this evaluation assumes the distilled logical Bell pairs have sufficiently low error rates. The analysis integrating the trade-off between distilled fidelity, size of distillers, and rate of distillation will be analyzed in the next subsection.
The execution time in seconds is then given by $T_\mathrm{exec}^\mathrm{norm} d_\mathrm{code} \tau_{\rm code}^\mathrm{cycle}$, and the number of physical qubits per node is given by $N_{\rm block} 2(d_\mathrm{code}+1)^2$ as each code block contains $2(d_\mathrm{code}+1)^2$ qubits (see Fig.\,\ref{fig:dftqc_node}).

In the following numerical evaluation, we fix several parameters as follows:
\begin{itemize}
    \item allowed failure probability $\epsilon = 10^{-6}$
    \item logical error rate per code block per round $p_{\rm bulk}/p_{\rm bulk}^{\rm th} = 10^{-1}$
    \item code cycle $\tau_\mathrm{code}^{\rm cycle} = 1\,\mu{\rm s}$
\end{itemize}
The instance type, instance size $N_\mathrm{instance}$, Bell-pair generation period $T_{\rm Bell}^\mathrm{period}$, network type, and node count $N_{\rm node}$, are varied in the numerical experiments presented in the following section. The other parameters are determined from them.

\subsection{Numerical Analysis: Application demands}
First, we evaluated the execution time and the number of physical qubits per node by varying the number of nodes $N_{\rm node}$. Because we are interested in the overhead introduced by distributed fault-tolerant quantum computation, we plot the relative execution time, defined as the execution time divided by that of the single-node ($N_{\rm node} = 1$) case. The results are shown in Fig.\,\ref{fig:result_execution_time}. We fix the instance size at $N_{\rm instance} = 128$ and show results for $T_{\rm Bell}^\mathrm{period} \in \{1, 10, 100\}$, measured in code beats. In this figure, circle, square, triangle, and diamond correspond to the circle, mesh, tree, and full network types, respectively. The blue, orange, and green markers correspond to logical Bell-pair generation periods of 1, 10, and 100 code beats, respectively.

\begin{figure*}[thb]
	\centering
	\includegraphics[width=0.75\linewidth]{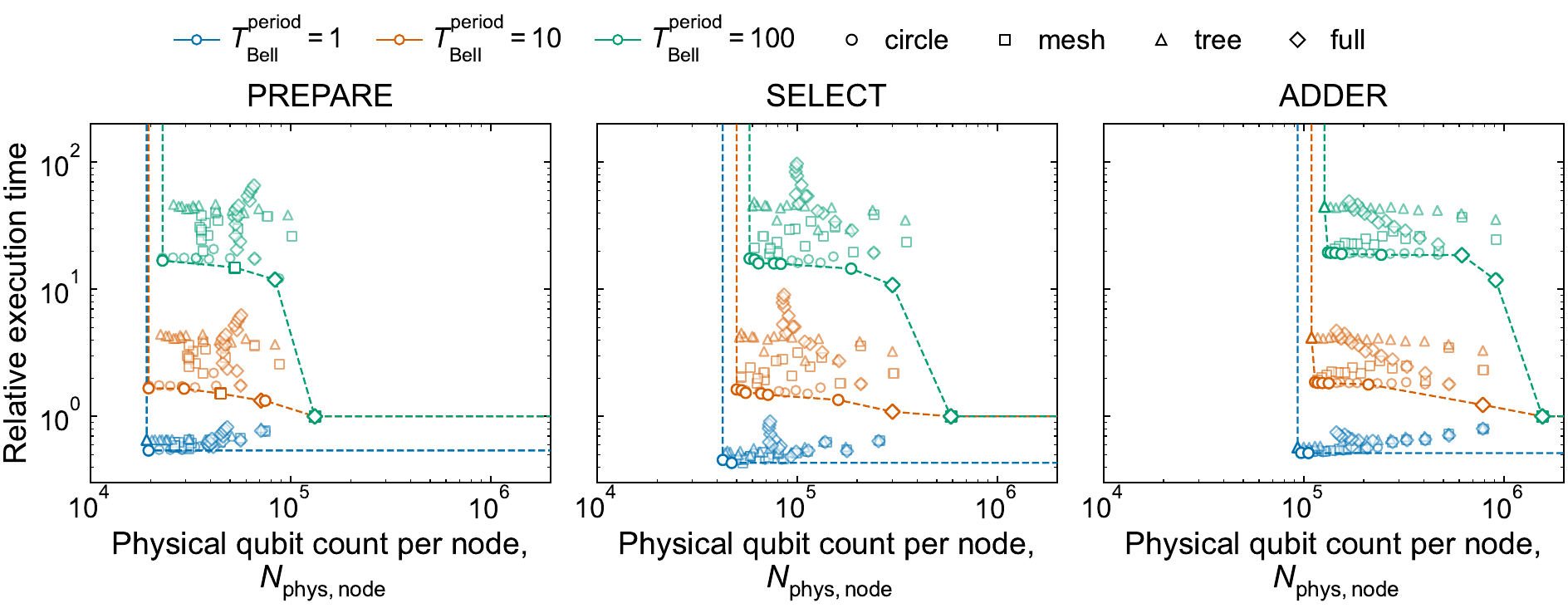}
	\caption{Achievable configurations and frontier curves in terms of relative execution time and physical qubit count per node $N_\mathrm{phys,node}$ for different periods ($T_\mathrm{Bell}^\mathrm{period}$= 1, 10, and 100) and network configurations. The columns correspond to the instance types. The instance size is $N_\mathrm{instance}=128$.}
	\label{fig:result_frontier_line}
\end{figure*}

\begin{figure*}[thbp]
	\centering
	\includegraphics[width=0.75\linewidth]{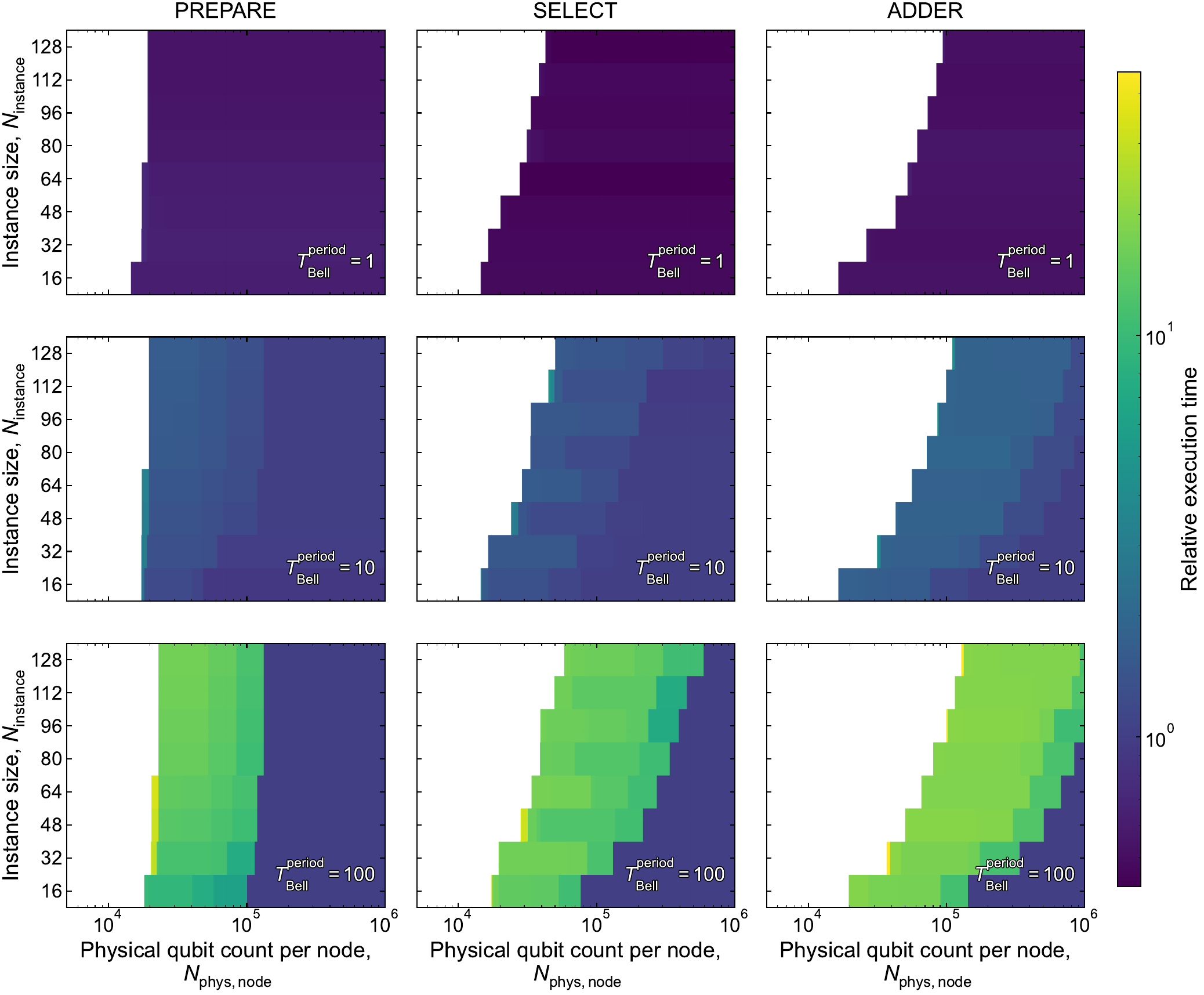}
\caption{Heatmap of the execution-time overhead as a function
of the number of physical qubits per node
$N_\mathrm{phys,node}$ and the instance size
$N_\mathrm{instance}$ for different periods
($T_\mathrm{Bell}^\mathrm{period}=$ 1, 10, 100).
The columns correspond to the instance types.
The results are obtained by varying the number of nodes
up to $N_\mathrm{node}=16$.
Thus, the blank regions do not necessarily indicate
fundamental limits on the minimum physical-qubit
requirements per node.}
	\label{fig:result_time_heatmap}
\end{figure*}

\begin{figure*}[thb]
	\centering
	\includegraphics[width=0.75\linewidth]{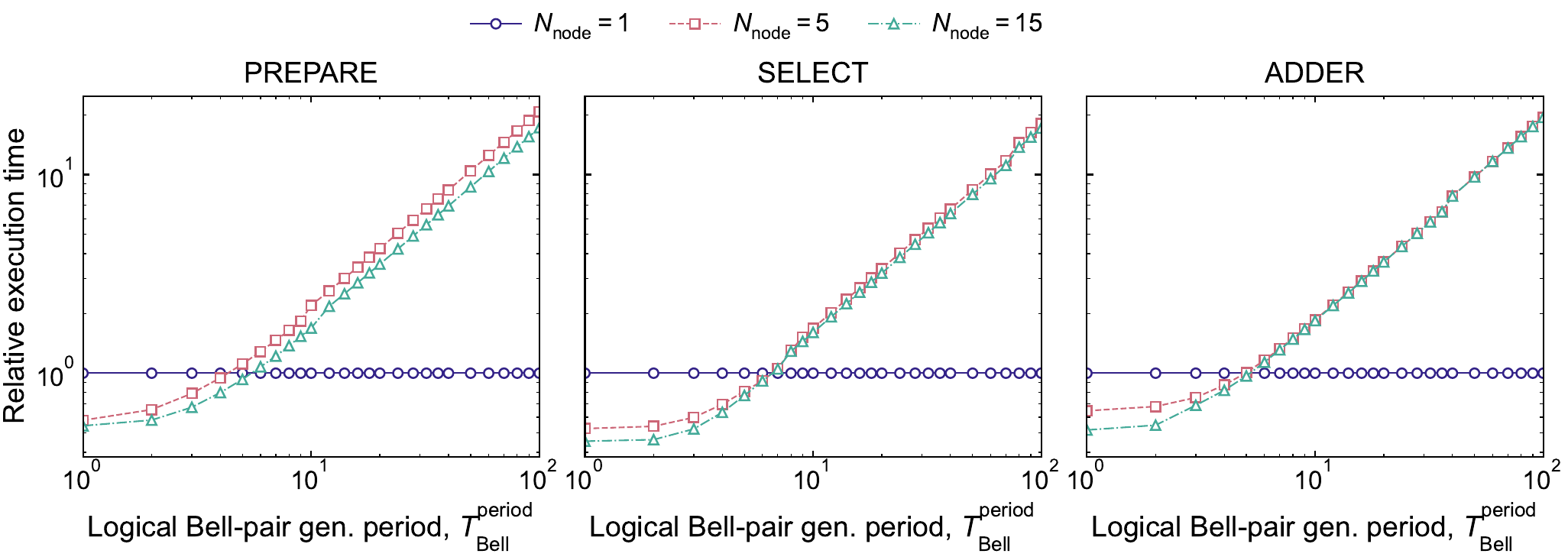}
	\caption{The relative execution time compared to the logical Bell-pair generation period, $T_\mathrm{Bell}^\mathrm{period}$, for different numbers of nodes ($N_\mathrm{node}=$ 1, 5, 15). The columns correspond to the instance types. The instance size is $N_\mathrm{instance}=$ 128.}
	\label{fig:result_period_vs_time}
\end{figure*}

\begin{figure*}[thb]
	\centering
	\includegraphics[width=0.75\linewidth]{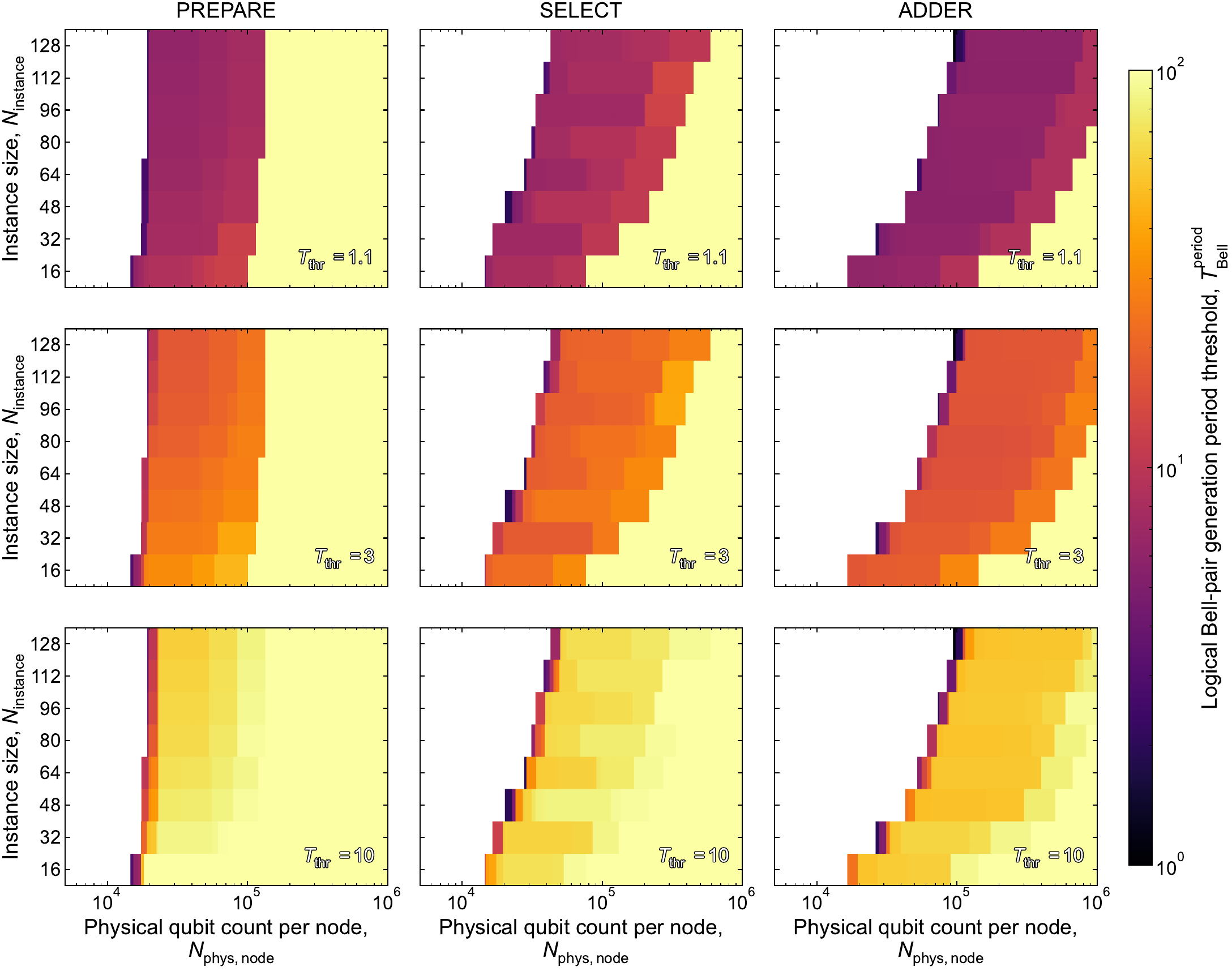}
	\caption{Logical Bell-pair generation period threshold required for achieving the target relative execution time, $T_\mathrm{thr}$ (=1.1, 3, 10) with varying physical qubit count per node $N_\mathrm{phys,node}$ and instance size $N_\mathrm{instance}$. The results are obtained by varying the number of nodes
up to $N_\mathrm{node}=16$.
Thus, the blank regions do not necessarily indicate
fundamental limits on the minimum physical-qubit
requirements per node.}
	\label{fig:result_period_heatmap}
\end{figure*}

As expected, increasing $N_{\rm node}$ reduces the required number of physical qubits per node. This reduction gradually saturates because each node requires reserved code blocks independent of the number of algorithmic qubits, including one code block for magic-state generation and $N_{\rm channel}$ code blocks for storing logical Bell pairs. 
For a fully connected network, the number of physical qubits per node increases at large $N_{\rm node}$ because $N_{\rm channel}=N_{\rm node}-1$, and hence the number of logical Bell-pair buffer blocks increases with the node count. 
For $T_{\rm Bell}^\mathrm{period} \in \{10, 100\}$, the relative execution time increases with $N_{\rm node}$ because of communication overhead. The execution time also depends on the choice of network topology. In this experiment, the circle network typically performs best. We attribute this result to the fact that these benchmark programs do not require parallel instruction execution and therefore cannot exploit densely connected channels. When $T_{\rm Bell}^\mathrm{period}=1$, distribution reduces, rather than increases, the execution time. We attribute this behavior to communication not becoming a bottleneck in this regime, while distribution provides a speedup by reducing routing conflicts and magic-state shortages.
We expect instances with high parallelism, such as random circuits or a Trotter step of an all-to-all Hamiltonian, might show different behavior, but we leave such extensive analysis as future work.

\begin{table*}[tb]
    \centering
    \scriptsize
    \caption{Numerical results for different instances, node counts $N_\mathrm{node}$, and logical Bell-pair generation periods $T_\mathrm{Bell}^\mathrm{period}$. The instance size and the failure probability are fixed as $N_\mathrm{instance}=128$ and $\epsilon=10^{-6}.$}
    \label{tab:numerical_result_table}
    \setlength{\tabcolsep}{4pt}
    \begin{tabular}{c c c c c c c c}
        \hline\hline
        \multicolumn{2}{c|}{Input parameters} &
        \multicolumn{6}{c}{Output parameters} \\
        \cline{1-2}\cline{3-8}
        $N_{\rm node}$ &
        $T_{\rm Bell}^{\rm period}$ &
        $d_{\rm code}$ &
        \begin{tabular}{c}
            Logical Bell-pair\\
            generation rate, $r_\mathrm{Bell}^\mathrm{logical}$,\\
            $(T_{\rm Bell}^{\rm period} d_{\rm code}
            \tau_{\rm code}^{\rm cycle})^{-1}$\\
            $[\mathrm{kHz}]$
        \end{tabular} &
        \begin{tabular}{c}
            Physical qubits\\
            per node\\
            $N_{\rm phys,node}$
        \end{tabular} &
        \begin{tabular}{c}
            Total physical\\
            qubits\\
            $N_{\rm phys,total}$
        \end{tabular} &
        \begin{tabular}{c}
            Execution\\
            time [s]
        \end{tabular} &
        \begin{tabular}{c}
            Relative\\
            execution time
        \end{tabular}
        \\
        \hline

        \multicolumn{8}{c}{Instance: PREPARE} \\
        \hline
        1  & 1   & 23 & 43.48 & $1.3\times10^{5}$ & $1.3\times10^{5}$ & 0.48  & 1.00 \\
        1  & 10  & 23 & 4.35  & $1.3\times10^{5}$ & $1.3\times10^{5}$ & 0.48  & 1.00 \\
        1  & 100 & 23 & 0.43  & $1.3\times10^{5}$ & $1.3\times10^{5}$ & 0.48  & 1.00 \\
        5  & 1   & 23 & 43.48 & $3.5\times10^{4}$ & $1.8\times10^{5}$ & 0.28  & 0.58 \\
        5  & 10  & 25 & 4.00  & $4.1\times10^{4}$ & $2.1\times10^{5}$ & 1.06  & 2.19 \\
        5  & 100 & 25 & 0.40  & $4.1\times10^{4}$ & $2.1\times10^{5}$ & 10.00 & 20.73 \\
        15 & 1   & 23 & 43.48 & $2.0\times10^{4}$ & $2.9\times10^{5}$ & 0.26  & 0.54 \\
        15 & 10  & 23 & 4.35  & $2.0\times10^{4}$ & $2.9\times10^{5}$ & 0.81  & 1.69 \\
        15 & 100 & 25 & 0.40  & $2.3\times10^{4}$ & $3.4\times10^{5}$ & 8.30  & 17.21 \\
        \hline

        \multicolumn{8}{c}{Instance: SELECT} \\
        \hline
        1  & 1   & 25 & 40.00 & $5.9\times10^{5}$ & $5.9\times10^{5}$ & 0.34 & 1.00 \\
        1  & 10  & 25 & 4.00  & $5.9\times10^{5}$ & $5.9\times10^{5}$ & 0.34 & 1.00 \\
        1  & 100 & 25 & 0.40  & $5.9\times10^{5}$ & $5.9\times10^{5}$ & 0.34 & 1.00 \\
        5  & 1   & 23 & 43.48 & $1.1\times10^{5}$ & $5.6\times10^{5}$ & 0.18 & 0.53 \\
        5  & 10  & 25 & 4.00  & $1.3\times10^{5}$ & $6.6\times10^{5}$ & 0.57 & 1.68 \\
        5  & 100 & 27 & 0.37  & $1.5\times10^{5}$ & $7.6\times10^{5}$ & 6.14 & 18.09 \\
        15 & 1   & 23 & 43.48 & $4.5\times10^{4}$ & $6.8\times10^{5}$ & 0.15 & 0.45 \\
        15 & 10  & 25 & 4.00  & $5.3\times10^{4}$ & $7.9\times10^{5}$ & 0.54 & 1.60 \\
        15 & 100 & 27 & 0.37  & $6.1\times10^{4}$ & $9.2\times10^{5}$ & 5.82 & 17.15 \\
        \hline

        \multicolumn{8}{c}{Instance: ADDER} \\
        \hline
        1  & 1   & 25 & 40.00 & $1.6\times10^{6}$ & $1.6\times10^{6}$ & 0.19 & 1.00 \\
        1  & 10  & 25 & 4.00  & $1.6\times10^{6}$ & $1.6\times10^{6}$ & 0.19 & 1.00 \\
        1  & 100 & 25 & 0.40  & $1.6\times10^{6}$ & $1.6\times10^{6}$ & 0.19 & 1.00 \\
        5  & 1   & 25 & 40.00 & $3.3\times10^{5}$ & $1.6\times10^{6}$ & 0.12 & 0.65 \\
        5  & 10  & 25 & 4.00  & $3.3\times10^{5}$ & $1.6\times10^{6}$ & 0.34 & 1.86 \\
        5  & 100 & 27 & 0.37  & $3.8\times10^{5}$ & $1.9\times10^{6}$ & 3.60 & 19.41 \\
        15 & 1   & 23 & 43.48 & $1.0\times10^{5}$ & $1.5\times10^{6}$ & 0.10 & 0.52 \\
        15 & 10  & 25 & 4.00  & $1.2\times10^{5}$ & $1.8\times10^{6}$ & 0.34 & 1.84 \\
        15 & 100 & 27 & 0.37  & $1.4\times10^{5}$ & $2.0\times10^{6}$ & 3.60 & 19.39 \\
        \hline\hline
    \end{tabular}
\end{table*}

Using these results, we plot the frontier lines for solving a target problem. Figure~\ref{fig:result_frontier_line} shows the relationship between the number of physical qubits per node and the relative execution time. The plotted points correspond to designs achievable by varying the network type and the number of nodes, and the regions above and to the right of the frontier curves are feasible. The figure clearly shows that a minimum number of physical qubits per node is required and that the execution-time overhead can be reduced by increasing the number of physical qubits available per node.

Next, we perform the same analysis for instance sizes $N_{\rm instance} \in \{16, 32, 48, 64, 80, 96, 112, 128\}$ and create a heatmap whose X and Y axes represent the number of physical qubits per node and the instance size, respectively. Each value represents the relative execution time minimized by optimizing the network type and node count. The results are shown in Fig.\,\ref{fig:result_time_heatmap}. Here, the white regions on the left indicate that the program cannot be executed with the corresponding number of physical qubits per node. 

For logical Bell-pair generation periods $T_{\rm Bell}^\mathrm{period} \in \{10, 100\}$, the relative execution time approaches one when a sufficient number of physical qubits for a given instance size is provided per node. 
The case $T_{\rm Bell}^\mathrm{period} = 1$ is exceptional because the distribution reduces the execution time. In this case, the speed-up compared to the single-node case increases as the instance size grows.

Finally, we estimate the necessary logical Bell-pair generation period to suppress the relative execution time below a given threshold $T_{\rm thr}$. The relative execution time as a function of logical Bell-pair generation period $T_{\rm Bell}^\mathrm{period}$ is shown in Fig.\,\ref{fig:result_period_vs_time}. Here, we fixed the instance size to 128 and used a circle network. This figure indicates that the relative execution time is reduced by distribution if the logical Bell-pair generation period is sufficiently short. On the other hand, if the logical Bell-pair generation period exceeds a few code beats, the relative execution time grows linearly with the logical Bell-pair generation period. If we limit the allowed execution time overhead compared to the single-node execution time, this provides the allowed logical Bell-pair generation period to satisfy the condition.

We perform the same analysis for instance sizes $N_{\rm instance} \in \{16, 32, 48, 64, 80, 96, 112, 128\}$ and generation periods $T_{\rm Bell}^\mathrm{period}$ between 1 and 100. We then create a heatmap whose X and Y axes represent the number of physical qubits per node and the instance size, respectively. Each value represents the logical Bell-pair generation period, expressed in code beats, required to keep the relative execution time below a given threshold $T_{\rm thr} \in \{1.1, 3, 10\}$. Thus, this graph indicates the logical Bell-pair generation period required to limit the execution-time overhead to an allowed value. The results are shown in Fig.\,\ref{fig:result_period_heatmap}. Similarly to the last figure, the white regions on the left indicate that the program cannot be executed with the corresponding number of physical qubits per node. The bottom right regions allow any long logical Bell-pair generation period as a single node can store all the reserved code blocks.

This figure indicates that shorter logical Bell-pair generation periods are required as the instance size and the number of physical qubits per node decrease and as the threshold of relative execution time $T_{\rm thr}$ decreases. According to this figure, we need approximately one logical Bell pair per 10 code beats to suppress the relative execution time to a negligible level ($\simeq 1.1$). As one code beat corresponds to $d_\mathrm{code} \tau_\mathrm{code}^\mathrm{cycle}$, which is typically a few tens of microseconds in typical superconducting qubits, this corresponds to logical Bell-pair generation rates of a few kHz. If the generation frequency is slower than this value, the relative execution time to the single-node case increases proportionally to the logical Bell-pair generation period.

Table \ref{tab:numerical_result_table} summarizes representative numerical results. The input parameters are the instance type, node count \(N_{\rm node}\), and logical Bell-pair generation period \(T_{\rm Bell}^{\rm period}\). The table also lists the resulting code distance \(d_{\rm code}\), logical Bell-pair generation rate, physical-qubit counts, execution time, and relative execution time. Note that we assume a circle network topology, $N_\mathrm{instance}=128$, and set the failure probability of programs to $\epsilon = 10^{-6}$.

Based on the above analysis, distributed fault-tolerant quantum computation can be performed with realistic space, time, and communication overheads for specific parameter regimes. In the parameter ranges and problem sizes considered here, corresponding to a total of 0.1--2 million physical qubits, the number of physical qubits per node can be reduced by a factor of 5--10. The minimum achievable number is limited by the physical qubits reserved for magic-state preparation and logical Bell-pair buffering. A further reduction in the number of physical qubits per node is expected for larger problem sizes.

The execution time can remain close to that of the single-node case when $T_\mathrm{Bell}^\mathrm{period}$ is below 10 code beats. Further increasing $T_\mathrm{Bell}^\mathrm{period}$ leads to an approximately linear increase in the execution time. When $T_\mathrm{Bell}^\mathrm{period}=100$, the execution time becomes $\sim20$ times longer than that of the single-node case. Assuming a typical code-beat duration of 20--30 $\mu$s for superconducting quantum computers, $T_\mathrm{Bell}^\mathrm{period}=1,\ 10,\ 100$ correspond to logical Bell-pair generation rates of approximately 40 kHz, 4 kHz, and 0.4 kHz, respectively. Thus, the communication requirement is not necessarily set by the code-beat rate itself; logical Bell-pair generation rates in the sub-kHz to several-kHz range can already enable distributed computation with moderate execution-time overheads.

\begin{figure*}[thb]
	\centering
	\includegraphics[width=0.95\linewidth]{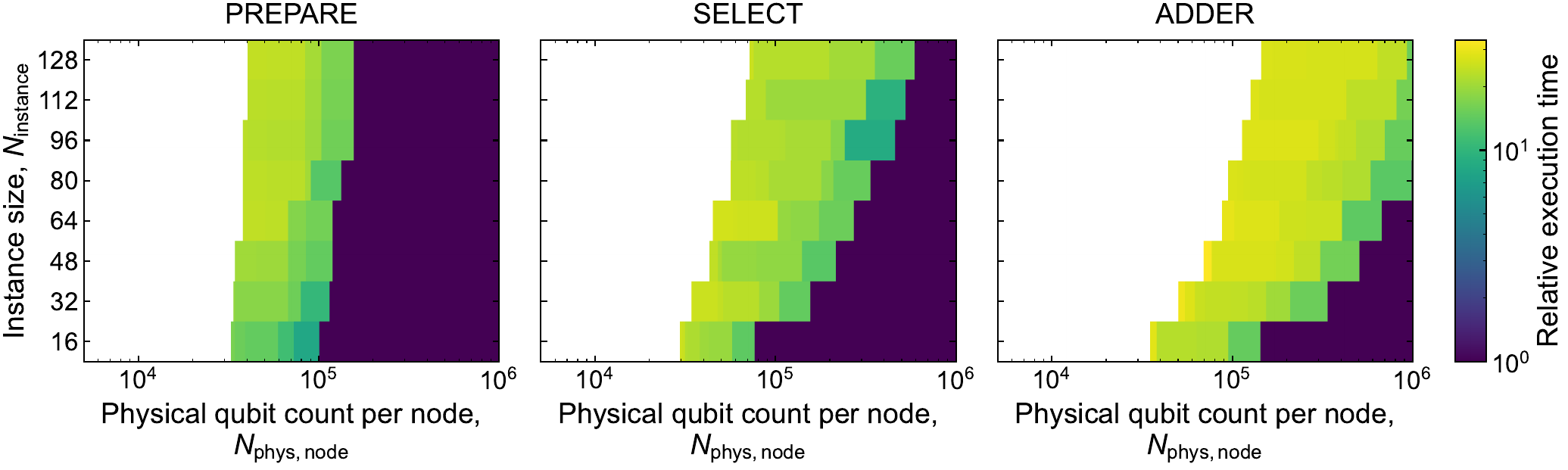}
	\caption{Heatmap of the execution-time overhead as a function of the number of physical qubits per node $N_{\rm phys, node}$ and the instance size $N_{\rm instance}$ for available protocols and networks. The results are obtained by varying the number of nodes
up to $N_\mathrm{node}=16$.
Thus, the blank regions do not necessarily indicate
fundamental limits on the minimum physical-qubit
requirements per node.}
	\label{fig:combined_analysis_execution_time}
\end{figure*}

\begin{figure*}[thb]
	\centering
	\includegraphics[width=0.95\linewidth]{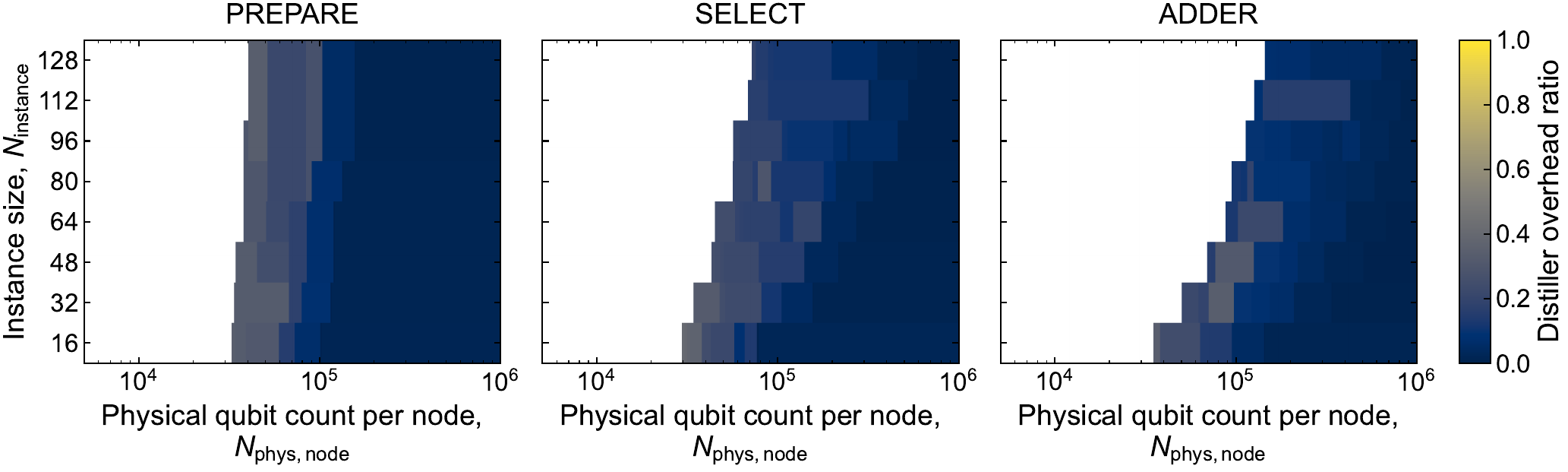}
	\caption{Heatmap of the physical-qubit overhead ratio as a function of the number of physical qubits per node $N_{\rm phys, node}$ and the instance size $N_{\rm instance}$ for available protocols and networks. The results are obtained by varying the number of nodes
up to $N_\mathrm{node}=16$.
Thus, the blank regions do not necessarily indicate
fundamental limits on the minimum physical-qubit
requirements per node.}
	\label{fig:combined_analysis_ratio}
\end{figure*}

\subsection{Numerical Analysis: Combined evaluation}
Finally, we integrate the hardware-side trade-offs obtained in Steps~I-IV into the application-level resource analysis in Step~V. From the hardware side, we collect candidate configurations for logical Bell-pair generation, characterized by the logical Bell-pair generation rate $r^{\rm logical}_{\rm Bell}$, the logical Bell-pair fidelity $F^{\rm logical(distill)}_{\rm Bell}$, and the number of physical qubits per entanglement distiller $N_{\rm phys,distill}$. We retain only the configurations on the Pareto frontier. For simplicity, we focus on the physical Bell-pair generation parameters $(r^{\rm phys}_{\rm Bell}, F^{\rm phys}_{\rm Bell}) = (1~\mathrm{MHz}, 0.9)$ with a single physical channel.

On the software side, we evaluate the execution times of PREPARE, SELECT, and ADDER instances with sizes $N_{\rm instance} \in \{16, 32, 48, 64, 80, 96, 112, 128\}$. We consider four network topologies: ring, mesh, tree, and fully connected networks, with the number of nodes $N_{\rm node} \in \{1, 2, \ldots, 16\}$. For each combination of benchmark instance, network topology, node count, and entanglement distillation protocol, we perform a code-beat-accurate simulation to obtain the execution time normalized to the single-node case, the normalized qubit volume $V^{\rm norm}_{\rm qubit}$ defined in Step~V, the number of physical qubits per node $N_{\rm phys,node}$ including entanglement distillers, and the number of consumed logical Bell pairs $N^{\rm consumed}_{\rm Bell}$.

In this evaluation, we again choose the allowed failure probability as $\epsilon=10^{-6}$. This evaluation considers cases where the provided logical entanglement might be too noisy. So, we choose the surface-code distance $d_{\rm code}$ for computation such that the circuit failure probability satisfies $p_{\rm circuit}(d_{\rm code}) < \epsilon/2$, and we require the failure probability due to imperfect Bell-pair distillation to satisfy $N^{\rm consumed}_{\rm Bell}(1-F^{\rm logical(distill)}_{\rm Bell}) < \epsilon/2$. Configurations that cannot satisfy these conditions are excluded from the plots.

A subtlety in this evaluation is the circular dependency between the simulation parameters. Code-beat-accurate simulation of distributed FTQC requires the logical Bell-pair generation period $T^{\rm period}_{\rm Bell}$ expressed in code beats. However, converting the logical Bell-pair generation rate $r^{\rm logical}_{\rm Bell}$ in hertz into $T^{\rm period}_{\rm Bell}$ requires the code distance $d_{\rm code}$, which must itself be determined from the simulation results. We resolve this dependency by performing code-beat-accurate simulations for multiple candidate code distances and selecting a distance that satisfies the above conditions.

The results of this analysis are shown in Fig.~\ref{fig:combined_analysis_execution_time}. While the qualitative behavior is similar to that of the $T_\mathrm{Bell}^\mathrm{period}=100$ case in Fig.~\ref{fig:result_time_heatmap}, the required physical qubit count per node to achieve the same relative execution time increases due to the additional resources required for entanglement distillation.

To illustrate the distiller overhead, Fig.~\ref{fig:combined_analysis_ratio} shows the distiller overhead ratio, defined as the number of physical qubits required for entanglement distillers divided by the total number of physical qubits per node. The distiller overhead ratio increases for smaller problem sizes and smaller physical qubit counts per node. Since logical Bell-pair distillation requires approximately $10^4$ physical qubits, the distiller overhead ratio reaches approximately 0.5 when $N_\mathrm{phys,node}$ is on the order of a few times $10^4$. As the number of nodes increases, the resources required for entanglement distillation and magic-state generation eventually limit the minimum achievable physical qubit count per node to approximately $10^4$.

These results quantitatively characterize the trade-offs between hardware resources and application-level performance, providing end-to-end design guidelines for distributed FTQC systems from physical-level analysis to application-level compilation.

\newcommand{\Y}{\ensuremath{\checkmark}}
\newcommand{\X}{Fixed}
\newcommand{\Z}{\ensuremath{\triangle}}

\begin{table*}[t]
\centering
\caption{Coverage of the five design stages in related work.
Here, \Y, \Z, and -- indicate substantial analysis,
limited analysis, and no independent analysis, respectively.
For Step III, direct encoding includes injection-based and
other physical-to-logical encoding protocols.
}
\label{tab:related_work}
\small
\setlength{\tabcolsep}{3pt}
\begin{tabular}{|c||l|c|c|c|c|c|c|c|c|c|}
\hline
Year & Paper & Step I & Step II
& \multicolumn{2}{c|}{Step III} & Step IV
& \multicolumn{4}{c|}{Step V} \\
\hline
& & Device & Phys. dist.
& Remote LS & Direct enc. & Log. dist.
& Application & Network & Compile & Schedule \\
\hline \hline
2023 & Ramette et al. ~\cite{Ramette2024}
& -- & -- & \Y & -- & -- & -- & -- & -- & -- \\
2023 & Wu et al. ~\cite{Wu2023QuComm}
& -- & -- & -- & -- & -- & \Y & \Y & \Y & -- \\
2024 & Saadatmand et al. ~\cite{Saadatmand2026Modularity}
& -- & -- & -- & -- & -- & \Y & \Y & \Y & \Y \\
2024 & Leone et al. ~\cite{Leone2025}
& \Z & \Y & \Y & -- & -- & -- & -- & -- & -- \\
2024 & Shi et al. ~\cite{Shi2025StabilizerDistillation}
& -- & \Y & -- & -- & -- & -- & -- & -- & -- \\
2024 & Sinclair et al. ~\cite{Sinclair2025}
& \Y & -- & \Y & -- & -- & -- & -- & -- & -- \\
2024 & Pattison et al. ~\cite{Pattison2025}
& -- & -- & -- & \Y & \Y & \Z & -- & -- & -- \\
2024 & Sakuma et al. ~\cite{Sakuma2024QFly}
& \Y & -- & -- & \Z & \Z & \Y & \Y & \Y & \Y \\
2025 & Bonilla Ataides et al. ~\cite{BonillaAtaides2025HighRateDistillation}
& -- & -- & -- & \Y & -- & -- & -- & -- & -- \\
2025 & Maeda et al. ~\cite{Maeda2025}
& \Y & -- & -- & \Y & \Y & -- & -- & -- & -- \\
2025 & Jacinto et al. ~\cite{Jacinto2026NetworkRequirements}
& -- & -- & \Y & -- & -- & \Y & -- & -- & -- \\
2025 & Marqversen et al. ~\cite{Marqversen2025}
& \Y & \Y & \Y & \Y & \Y & -- & -- & -- & -- \\
2025 & Sunami et al. ~\cite{Sunami2026EntanglementBoosting}
& -- & -- & \Y & \Y & \Y & -- & -- & -- & -- \\
2025 & Filippov et al. ~\cite{Filippov2025}
& -- & \Y & -- & \Z & \Y & \Y & \Z & \Y & -- \\
2026 & Naito et al. ~\cite{Naito2026NBQC}
& -- & -- & -- & -- & -- & \Y & \Y & \Y & \Y \\
2026 & Liu et al. ~\cite{Liu2026ToDistill}
& -- & \Y & \Y & -- & -- & -- & -- & -- & -- \\
2026 & Kaur et al. ~\cite{Kaur2026NetworkConstraints}
& -- & -- & \Y & -- & -- & -- & \Y & \Y & \Y \\
2026 & Stein et al. ~\cite{Stein2026S2PAR}
& \Y & -- & \Y & -- & -- & \Y & \Y & \Y & \Z \\
2026 & Yu and Liu ~\cite{Yu2026SpacetimeOverhead}
& -- & -- & \Y & -- & -- & \Y & \Z & -- & -- \\
\hline \hline
\textbf{2026} & \textbf{This work}
& \Y & \Y & \Y & \Y & \Y
& \Y & \Y & \Y & \Y \\
\hline
\end{tabular}
\end{table*}

\section{Discussion}
\subsection{Feasibility evaluation}

Based on the above analysis, we evaluate the feasibility of
distributed fault-tolerant quantum computation using
superconducting-qubit systems (Fig.~\ref{fig:calc_flow_summary}).

For the forward resource estimation, we consider two
representative sets of physical communication channel parameters:
$(r_\mathrm{Bell}^\mathrm{phys}, F_\mathrm{Bell}^\mathrm{phys})
= (1~\mathrm{MHz}, 0.9)$ for a microwave quantum link and
$(1~\mathrm{kHz}, 0.8$--$0.9)$ for an optical quantum link
(Fig.~\ref{fig:phys_bell_pair_gen_rate_fid}).

The microwave-link parameters are motivated by experiments
using flying microwave photons
\cite{Kurpiers2018,Kurpiers2019,Qiu2025}.
Ref.~\cite{Kurpiers2018} demonstrated a photon-transfer
protocol lasting only 180 ns, with an overall transfer
efficiency of approximately 0.7, suggesting that MHz-scale
trial rates are possible.
For example, a trial rate of 1.5 MHz with a heralded success
probability of 0.7 would yield approximately 1 MHz of usable
Bell pairs.
More recently, Ref.~\cite{Qiu2025} demonstrated a
readout-corrected Bell-pair fidelity of $0.942\pm0.006$
over 64 m using flying microwave photons.
The demonstrated sub-microsecond operation times suggest
that MHz-scale Bell-pair generation is feasible in principle.
The remaining challenge is to achieve a sustained generation
rate of 1 MHz while maintaining a fidelity of 0.9.

Another approach uses standing-wave modes of superconducting
transmission lines as quantum buses.
A Bell-pair fidelity of 0.989 and state transfer within
66 ns were demonstrated over a 25-cm link \cite{Niu2023}.
A heralded scheme achieved a Bell-pair generation rate of
43 kHz with a fidelity of 0.86, while a fidelity of 0.92
was obtained under different operating conditions \cite{Teoh2025}.
More recently, four 80-cm links enabled state transfer
within 256--312 ns with an average fidelity of 0.990 and
remote CZ gates with a fidelity of 0.9475 \cite{Sundaresan2026}.
However, these schemes primarily rely on coherent interactions
mediated by shared standing-wave modes rather than
flying-photon-based Bell-pair distribution, and are therefore
not explicitly modeled in this work.
Furthermore, increasing the link length reduces the free
spectral range, making selective coupling to individual modes
more challenging.
Thus, standing-wave-based schemes are particularly promising
for intra-cryostat connections over meter-scale distances,
whereas flying-photon-based schemes are better suited
for longer-distance inter-cryostat communication.

For optical links, the assumed 1 kHz generation rate
and fidelity of 0.8--0.9 are optimistic targets.
They require substantial improvements in microwave--optical
transduction efficiency and added-noise performance.
For example, at a total transmission efficiency of
$\eta_\mathrm{total}=0.1$, achieving a physical Bell-pair
fidelity of 0.8 requires
$P_\mathrm{noise}\sim10^{-3}$ for the single-photon
protocol and $\sim5\times10^{-3}$ for the two-photon
protocol (Fig.~\ref{fig:phys_bell_pair_gen_rate_fid}).
Higher target fidelities impose more stringent noise
requirements.
Thus, both parameter sets represent demanding
communication-hardware specifications, particularly
for optical links.

Since lattice-surgery-based logical Bell-pair generation
requires substantially higher physical Bell-pair fidelity
and/or extensive multiplexing under these conditions,
we employ an injection-distillation protocol to supply
low-error logical Bell pairs.
For a single microwave communication channel with
$(r_\mathrm{Bell}^\mathrm{phys},
F_\mathrm{Bell}^\mathrm{phys})
= (1~\mathrm{MHz}, 0.9)$,
the resulting logical Bell-pair generation rate is
200--600 Hz for target logical Bell-pair error rates
of $10^{-12}$--$10^{-8}$
(Fig.~\ref{fig:logical_distill_tradeoff}).
For an optical link with
$(r_\mathrm{Bell}^\mathrm{phys},
F_\mathrm{Bell}^\mathrm{phys})
= (1~\mathrm{kHz}, 0.8$--$0.9)$,
multiplexing over 100--500 channels enables logical
Bell-pair generation rates on the order of 100 Hz.
These rates roughly correspond to the
$T_\mathrm{Bell}^\mathrm{period}=100$ case in
Fig.~\ref{fig:result_time_heatmap},
as also supported by the combined analysis in
Fig.~\ref{fig:combined_analysis_execution_time}.

The heatmaps indicate that distributing the computation
across $\sim10$ nodes can reduce the required physical-qubit
count per node by a factor of $\sim10$, at the cost of an
approximately tenfold increase in execution time.
Although this increase is non-negligible, the result
demonstrates that distributed fault-tolerant computation
can remain feasible with only a single physical
communication channel per node under the considered
microwave-link conditions.
At the same time, it highlights the substantial
execution-time penalty that remains even with a
high-performance physical quantum link.
Furthermore, although the size of the logical Bell-pair
distillation unit ultimately imposes a lower bound of
$\sim10^4$ on the physical-qubit count per node,
the distillation unit accounts for at most $\sim50\%$
of the total physical-qubit count when the computational
resources of each node are also on the order of
$10^4$ physical qubits
(Fig.~\ref{fig:combined_analysis_ratio}).

We also perform backward resource requirement estimation.
To execute quantum algorithms with an execution-time
overhead of only a factor of 1--3, the logical Bell-pair
generation period must be reduced to
$T_\mathrm{Bell}^\mathrm{period}\lesssim10$
(Fig.~\ref{fig:result_period_heatmap}).
This corresponds to a logical Bell-pair generation rate
of $\sim4$ kHz for a representative logical error rate
of $10^{-12}$ and a code distance of
$d_\mathrm{code}\sim25$.

Such a rate can be achieved in principle using
lattice-surgery-based logical Bell-pair generation,
which can supply logical Bell pairs approximately
every code beat, providing substantial headroom
above the required kHz-scale rates
(Fig.~\ref{fig:logical_gen_rate_fid}).
However, this requires physical communication channels
with $(r_\mathrm{Bell}^\mathrm{phys},
F_\mathrm{Bell}^\mathrm{phys})
= (1~\mathrm{MHz}, 0.99)$
and multiplexing over $2d_\mathrm{code}-1$ channels.
Alternatively, using an injection-distillation protocol,
a logical Bell-pair generation rate on the order of kHz
can be achieved even with a single channel if
$(r_\mathrm{Bell}^\mathrm{phys},
F_\mathrm{Bell}^\mathrm{phys})
= (1~\mathrm{MHz}, 0.99)$.
At a lower fidelity of 0.9, achieving a comparable rate
requires multiplexing over $\sim100$ channels.
These results highlight the importance of improving
physical Bell-pair fidelity and/or multiplexing to reduce
the execution-time overhead of distributed quantum
algorithms.

Together, the forward and backward resource estimations
establish the trade-off between communication-hardware
requirements and computational performance.
They show both the feasibility and the limitations of
distributed fault-tolerant quantum computation under
restricted communication resources and provide quantitative
targets for the development of quantum interconnects.


\subsection{Related work}

Distributed fault-tolerant quantum computation has been
extensively studied from both quantum-interconnect and
system-architecture perspectives.
A central question is not simply whether fault-tolerant
computation can be distributed with sufficient communication
resources, but how much the communication requirements can
be reduced and what quantum-link performance is necessary
to execute useful quantum algorithms with acceptable space
and time overheads.
These requirements are particularly important because
generating high-fidelity entanglement between remote QPUs
at high rates, especially with many parallel communication
channels, remains technically challenging.
Table~\ref{tab:related_work} summarizes representative
studies explicitly targeting distributed fault-tolerant
quantum computation.
Studies limited to noisy intermediate-scale quantum
computing or general entanglement-distillation theory
are not included.

One line of research focuses on logical Bell-pair generation
through remote lattice surgery.
Ref.~\cite{Ramette2024} analyzed remote lattice surgery
with noisy physical Bell pairs and demonstrated its
tolerance to seam errors.
Ref.~\cite{Leone2025} evaluated remote lattice surgery
for trapped-ion systems, including physical Bell-pair
purification.
Ref.~\cite{Sinclair2025} analyzed remote lattice surgery
using cavity-enhanced optical interconnects for neutral atoms.
More recently, Ref.~\cite{Liu2026ToDistill} compared
remote lattice surgery with and without physical Bell-pair
distillation, explicitly accounting for stochastic Bell-pair
generation and waiting-induced decoherence.

Another line of research investigates physical Bell-pair
distillation and direct encoding into logical Bell pairs.
Ref.~\cite{Shi2025StabilizerDistillation} developed
stabilizer-based entanglement-distillation protocols
with efficient fault-tolerant encoders.
Ref.~\cite{BonillaAtaides2025HighRateDistillation}
proposed direct generation of logical Bell pairs using
high-rate quantum LDPC codes.
Ref.~\cite{Maeda2025} analyzed surface-code encoding
using parallel physical Bell-pair generation and
postselection based on estimated error counts.
Ref.~\cite{Sunami2026EntanglementBoosting} improved the
resource efficiency of logical Bell-pair preparation
through code expansion and soft-information postselection
based on the complementary gap.
Both Refs.~\cite{Maeda2025,Sunami2026EntanglementBoosting}
also evaluated subsequent logical Bell-pair distillation.

Logical Bell-pair distillation under finite physical-qubit
and communication resources has also been investigated.
Ref.~\cite{Pattison2025} optimized sequences of
logical entanglement-distillation protocols under
finite buffer constraints and estimated the communication
bandwidth required by representative quantum circuits.
Ref.~\cite{Marqversen2025} compared multiple logical
communication methods, including remote lattice surgery,
direct encoding, physical Bell-pair distillation,
and logical Bell-pair distillation, while accounting
for generation rates and memory constraints.
Ref.~\cite{Filippov2025} developed a multilevel
entanglement-distillation resource model and connected
the resulting logical communication costs to the
resource requirements of distributed fault-tolerant
quantum algorithms.

On the system-architecture and software sides,
Ref.~\cite{Wu2023QuComm} developed a distributed quantum
compilation and communication-optimization framework.
Ref.~\cite{Kim2024MillionQubit} investigated error
syndrome measurement and decoding in large-scale
fault-tolerant architectures spanning multiple
dilution refrigerators.
Ref.~\cite{Saadatmand2026Modularity} analyzed
modular superconducting fault-tolerant architectures,
including compilation and execution scheduling.
Ref.~\cite{Sakuma2024QFly} proposed the Q-Fly optical
interconnect architecture and evaluated distributed
quantum Fourier transform execution using a simplified
logical Bell-pair purification model.
Ref.~\cite{Jacinto2026NetworkRequirements} connected
remote lattice surgery and its logical-error model
to resource estimation for Shor's factoring algorithm.
Ref.~\cite{Naito2026NBQC} proposed network-based quantum
computing and optimized network structures, compilation,
and scheduling under given logical Bell-pair generation
times.
Ref.~\cite{Kaur2026NetworkConstraints} analyzed
network-constrained compilation and scheduling based on
remote lattice surgery.
Ref.~\cite{Stein2026S2PAR} proposed the S2PAR architecture,
which uses graph-state generation on superconducting-qubit
routing cards to support remote lattice surgery, and
evaluated network-dependent logical errors for quantum
algorithm benchmarks.
Ref.~\cite{Yu2026SpacetimeOverhead} developed a
latency-decoupled inter-chip syndrome-extraction
scheme and evaluated its impact on the spacetime
resources required for factoring.

These studies have established important connections
between quantum communication, fault-tolerant operations,
network architecture, and application-level resource
requirements.
In particular, several studies have already integrated
multiple hardware and software stages.
Nevertheless, the communication assumptions remain
a critical consideration: application-level analyses
often rely on prescribed entanglement-generation
performance, while detailed communication-protocol
studies do not always quantify their implications for
algorithm execution under restricted communication
resources.
The practical question is therefore how to minimize
the required communication channels, generation rates,
and fidelities without incurring prohibitive computation
overheads.

In this work, we address this question by connecting
physically motivated quantum-link models, physical and
logical Bell-pair generation and distillation, and
code-beat-accurate execution simulations of representative
fault-tolerant quantum-algorithm subroutines.
We quantify the trade-offs among physical Bell-pair
generation rate, fidelity, communication-channel count,
physical-qubit overhead, and execution time.
In particular, we evaluate the extent to which
distributed computation remains feasible with a
single physical communication channel and determine,
conversely, the quantum-link performance and multiplexing
required to meet specified computational targets.
This analysis provides quantitative communication-hardware
requirements for practical distributed fault-tolerant
quantum computation rather than assuming that
sufficient communication resources are available.

\section{Conclusion}

A central challenge in distributed fault-tolerant quantum
computation is determining how much computation can be
performed with limited quantum communication resources
and what communication performance is required to achieve
acceptable computational overheads.
To address this challenge, we have performed a comprehensive
resource analysis connecting communication hardware to
quantum-algorithm execution.
We evaluated physical Bell-pair generation and distillation
(Steps I and II), logical Bell-pair generation and distillation
(Steps III and IV), and the resulting space, time, and
communication overheads of distributed quantum computation
(Step V).

In Steps I and II, we estimate that physical Bell-pair
generation rates of 10--1000 kHz with fidelities above
80\% are potentially achievable for microwave links,
whereas optical links could reach $\sim1$ kHz with
80--90\% fidelity under optimistic assumptions for
microwave--optical transduction.
Achieving these performance levels requires high-efficiency
quantum communication, particularly for optical links.
Considering qubit decoherence, physical Bell-pair distillation
becomes impractical at low physical Bell-pair generation
rates, typically below $\sim100$ kHz in the parameter
ranges considered.
Although physical Bell-pair predistillation can improve
the subsequent logical Bell-pair generation rate, it does
not offer a clear overall resource advantage once the
additional communication channels and Bell-pair consumption
are included.

In Steps III and IV, we find that low-error logical Bell
pairs can be generated at useful rates even with imperfect
physical quantum links.
For a single microwave communication channel with
$(r_\mathrm{Bell}^\mathrm{phys},F_\mathrm{Bell}^\mathrm{phys})
=(1~\mathrm{MHz},0.9)$,
an injection-distillation protocol provides logical Bell-pair
generation rates of 200--600 Hz for target error rates of
$10^{-12}$--$10^{-8}$.
For an optical link with
$(r_\mathrm{Bell}^\mathrm{phys},F_\mathrm{Bell}^\mathrm{phys})
=(1~\mathrm{kHz},0.8$--$0.9)$,
multiplexing over 100--500 channels enables logical
Bell-pair generation rates on the order of 100 Hz.
Regarding the space overhead, logical Bell-pair distillation
requires approximately $10^4$ physical qubits, which
imposes a practical lower bound on the physical-qubit
count per node for architectures relying on this protocol.

In Step V, we find that distributing a computation
across $\sim10$ nodes can reduce the required physical-qubit
count per node by approximately an order of magnitude,
at the cost of a roughly tenfold increase in execution
time under the considered single-channel microwave-link
conditions.
This result demonstrates that distributed fault-tolerant
quantum computation can remain feasible with only a single
physical communication channel per node, while also
revealing the substantial execution-time penalty that
persists even with a high-performance quantum link.
For million-qubit-scale computations, such distribution
can reduce the required physical-qubit count per node
to the order of $10^5$, substantially relaxing practical
constraints associated with wiring, cryogenic cooling,
fabrication yield, and system integration.

Conversely, backward resource requirement estimation
allows us to determine the communication performance
needed to achieve a target execution time.
To limit the execution-time overhead to a factor of 1--3,
the logical Bell-pair generation period must be reduced
to $T_\mathrm{Bell}^\mathrm{period}\lesssim10$,
corresponding to a logical Bell-pair generation rate
of $\sim4$ kHz for a representative logical error rate
of $10^{-12}$ and a code distance of
$d_\mathrm{code}\sim25$.
This rate can be achieved in principle using
lattice-surgery-based logical Bell-pair generation
with $(r_\mathrm{Bell}^\mathrm{phys},
F_\mathrm{Bell}^\mathrm{phys})=(1~\mathrm{MHz},0.99)$
and multiplexing over $2d_\mathrm{code}-1$ channels.
Alternatively, kHz-scale generation using an
injection-distillation protocol requires approximately
100 parallel channels with
$(r_\mathrm{Bell}^\mathrm{phys},
F_\mathrm{Bell}^\mathrm{phys})=(1~\mathrm{MHz},0.9)$.
These requirements highlight the substantial
communication-hardware resources needed to reduce
the execution-time overhead to only a fewfold.

Taken together, our analysis quantifies the trade-offs
among physical Bell-pair generation rate, fidelity,
communication-channel count, physical-qubit overhead,
and execution time.
It demonstrates that distributed fault-tolerant quantum
computation is feasible under restricted communication
resources, while identifying the physical-link performance
required to achieve faster computation.
Although advances in quantum error correction, including
more efficient codes such as qLDPC codes, may further
reduce these requirements, the framework provides a
quantitative basis for assessing communication constraints
and guiding the development of quantum interconnects
for practical distributed fault-tolerant quantum computation.




\section*{acknowledgments}
H. Kosaka acknowledges the funding support from the Japan Science and Technology Agency (JST) Moonshot R$\&$D grant (JPMJMS2062) and a JST CREST grant (JPMJCR1773). H. Kosaka also acknowledges the Ministry of Internal Affairs and Communications (MIC) for funding the research and development to construct a global quantum cryptography network  (JPMI00316), and the Japan Society for the Promotion of Science (JSPS) Grants-in-Aid for Scientific Research (20H05661, 20K20441). 
This work is also supported by Japan Science and Technology Agency (JST) as part of Adopting Sustainable Partnerships for Innovative Research Ecosystem (ASPIRE) (JPMJAP24C1). 
Y. Suzuki is supported by MEXT Q-LEAP Grant No.~JPMXS0120319794, JPMXS0118068682, JST Moonshot R\&D Grant No.~JPMJMS256E, JST CREST Grant No.~JPMJCR23I4, JPMJCR24I4, JPMJCR25I4, MEXT Feasibility Study on the future HPCI.

\section*{AI Use}
The human authors developed all research ideas, technical contributions, scientific analyses, and conclusions presented in this paper. The manuscript was also written by the human authors.
The authors used Grammarly and ChatGPT (GPT-6 Astra, GPT-6, GPT-5.6 sol) for grammatical corrections and language polishing. 
ChatGPT (GPT-5.6 sol) was used to develop and organize analysis codes for Step I--IV. 
GitHub Copilot was used for code suggestions, and Codex (GPT-6 Astra) was used for checking potential bugs in the code for Step V. 
All AI-assisted suggestions and code were reviewed and validated by the authors, who take full responsibility for the scientific content, implementation, and results presented in this paper.

\clearpage	
\appendix

\makeatletter
\@addtoreset{equation}{section}
\@addtoreset{figure}{section}
\@addtoreset{table}{section}
\makeatother

\renewcommand{\theequation}{\Alph{section}\arabic{equation}}
\renewcommand{\thefigure}{\Alph{section}\arabic{figure}}
\renewcommand{\thetable}{\Alph{section}\arabic{table}}

\setcounter{equation}{0}
\setcounter{figure}{0}
\setcounter{table}{0}
	
	\section{Summary of parameters}
	\label{app:data}
	
	Tables \ref{tab:para-list} and \ref{tab:para-list2} summarize the symbols and parameters used in this paper.
	\begin{table*}[]

		\centering
		\caption{Parameter list. The right column denotes typical values assumed in the main text.}
		\label{tab:para-list}
		\begin{tabular}{lll}
			\hline
			$r_\mathrm{Bell}^\mathrm{phys}$                                                         & Physical Bell-pair generation rate                                                                                                                                                                                              &                       \\
			$r_\mathrm{Bell}^\mathrm{phys(distill)}$                                                & Physical Bell-pair distillation rate                                                                                                                                                                                            &                       \\
			$r_\mathrm{Bell}^\mathrm{phys(distill,norm)}$                                           & \begin{tabular}{l} Physical Bell-pair distillation rate\\ normalized by the number of \\minimum communication channel required \end{tabular} &                       \\
			$r_\mathrm{Bell}^\mathrm{logical}$                                                      & Logical Bell-pair generation rate                                                                                                                                                                                               &                       \\
			$r_\mathrm{Bell}^\mathrm{logical(distill)}$                                             & Logical Bell-pair distillation rate                                                                                                                                                                                             &                       \\
			$r_\mathrm{Bell}^\mathrm{logical(distill,norm)}$                                        & \begin{tabular}{l} Logical Bell-pair distillation rate\\ normalized by the number of\\ minimum communication channel required \end{tabular}    &                       \\
			$F_\mathrm{Bell}^\mathrm{phys}$                                                         & Physical Bell-pair fidelity                                                                                                                                                                                                     &                       \\
			$F_\mathrm{Bell}^\mathrm{phys(distill)}$                                                & Distilled physical Bell-pair fidelity                                                                                                                                                                                           &                       \\
			$F_\mathrm{Bell}^\mathrm{logical}$                                                      & Logical Bell-pair fidelity                                                                                                                                                                                                      &                       \\
			$F_\mathrm{Bell}^\mathrm{logical(distill)}$                                             & Distilled logical Bell-pair fidelity                                                                                                                                                                                            &                       \\
			$p_\mathrm{Bell}^\mathrm{phys}=1-F_\mathrm{Bell}^\mathrm{phys}$                         & Error rate of physical Bell-pair                                                                                                                                                                                                &                       \\
			$p_\mathrm{Bell}^\mathrm{phys(distill)}=1-F_\mathrm{Bell}^\mathrm{phys(distill)}$       & Error rate of distilled physical Bell-pair                                                                                                                                                                                      &                       \\
			$p_\mathrm{Bell}^\mathrm{logical}=1-F_\mathrm{Bell}^\mathrm{logical}$                   & Error rate of logical Bell-pair                                                                                                                                                                                                 &                       \\
			$p_\mathrm{Bell}^\mathrm{logical(distill)}=1-F_\mathrm{Bell}^\mathrm{logical(distill)}$ & Error rate of distilled logical Bell-pair                                                                                                                                                                                       &                       \\
			$\tau^{\mathrm{phys(distill)}}$ 
			& Physical Bell-pair distillation time
			& \\
			
			$\tau_\mathrm{Bell}^\mathrm{injection}$ 
			& State-injection time for generating an injected logical Bell-pair
			& \(3\tau_\mathrm{code}^\mathrm{cycle} \) \\
			
			$\tau^{\mathrm{logical}(\mathrm{distill})}$ 
			& Logical Bell-pair distillation time
			& \\
			$p_\mathrm{success}^\mathrm{Bell}$                                                   & Success probability of physical Bell-pair generation                                                                                                                                                                   &                       \\			
			$p_\mathrm{success}^\mathrm{distill}$                                                   & Success probability of physical Bell-pair distillation                                                                                                                                                                          &                       \\
			$p_\mathrm{success}^\mathrm{injection}$                                                 & Success probability of two-node state injection                                                                                                 & 85 $\%$ \cite{Lao2022,Pattison2025}              \\
			$p_\mathrm{rejection}^\mathrm{injection}$       & Rejection probability of state injection at a single node                                                                                                                                            &                       \\
			&                                                                                                                                                                                                                                 &                       \\
			$\eta_\mathrm{total}$                                  & Total transmission efficiency of an entangled photon                                                                                                                                                                                  &                       \\
			$R_\mathrm{rep}$                                                                        & Repetition rate of an entangled photon generation                                                                                                                                                                               & 1 MHz  \cite{Kurpiers2018,Kurpiers2019}               \\
			$P_\mathrm{e}$                                                                          & Photon emission probability from a communication qubit                                                                                                                                                                          &                       \\
			$P_\mathrm{noise}$                                                                      & Noise-click probability within the detection window                                                                                                                                                               &                       \\
			&                                                                                                                                                                                                                                 &                       \\
			$p_\mathrm{1q}=1-F_\mathrm{1q}$                                                         & A single-qubit error probability  (infidelity)                                                                                                                                                                                                 & 10$^{-3}$ (0.1$\%$) \\
			$p_\mathrm{2q}=1-F_\mathrm{2q}$                                                         & A two-qubit error probability  (infidelity)                                                                                                                                                                                                    & 10$^{-3}$ (0.1$\%$)  \\
			$p_\mathrm{meas}=1-F_\mathrm{meas}$                                                     & Measurement error probability  (infidelity)                                                                                                                                                                                                    & 10$^{-3}$ (0.1$\%$)  \\
			$T_1$
			& Qubit energy-relaxation time
			& 1 ms \cite{Bland2025}			
			          \\
			&                                                                                                                                                                                                                                 &                       \\
			$p_\mathrm{bulk}$                                                                       & Bulk area error in the surface code                                                                                                                                                                                             &                       \\
			$p_\mathrm{bulk}^\mathrm{th}$                                                           & Surface code error threshold at the bulk area                                                                                                                                                                                   & 10$^{-2}$ (1$\%$) \cite{Ramette2024}   \\
			$p_\mathrm{seam}$                                                                       & Seam (edge) error in the surface code                                                                                                                                                                                           &                       \\
			$p_\mathrm{seam}^\mathrm{th}$                                                           & Surface code error threshold at the seam                                                                                                                                                                                        & 10$^{-1}$ (10$\%$)  \cite{Ramette2024}  \\
			&                                                                                                                                                                                                                                 &                       \\
			$n_{(i)}$                                                                                     & Number of input Bell pairs for distillation (of $i$th distillation)                                                                                                                                                                                     &                       \\
			$k_{(i)}$                                                                                     & Number of output Bell pairs after distillation (of $i$th distillation)                                                                                                                                                                                 &                       \\
			$d_{(i)}$                                                                                     & Code distance of an error correction code (of $i$th distillation)                                                                                                                                                                                      &                       \\
			$i$& the index for physical (logical) distillation round&
			\\
			&                                                                                                                                                                                                                                 &                       \\
			$d_\mathrm{code}$                                                                       & Code distance of the surface code                                                                                                                                                                                               & 1--51                 \\
			$\tau_\mathrm{code}^\mathrm{cycle}$                                                        & Code cycle of the surface code                                                                                                                                                                                                  & 1 $\mu$s              \\ 
			&                                                                                                                                                                                                                                 &                       \\
			$N_\mathrm{comm}^\mathrm{phys}=\Pi_in_i$                                                    & The number of communication channels for physical Bell-pair distillation                                                                                                                                                                                                &             \\ 	
			$N_\mathrm{comm}^\mathrm{logical}$                                                    & The number of communication channels for logical Bell-pair distillation                                                                                                                                                                                                & 1 or $\Pi_in_i$            \\ 	
			$N_\mathrm{comm}^\mathrm{total}=N_\mathrm{comm}^\mathrm{phys}N_\mathrm{comm}^\mathrm{logical}$                                            & The number of communication channels for logical Bell-pair  generation                                                                                                                                                                                                &            \\ 
			$N_\mathrm{consumed}^\mathrm{phys}$                                            & The consumed number of physical Bell pairs for a logical Bell-pair generation                                                                                                                                                                       &            \\ 					
			$N_\mathrm{buffer}^\mathrm{phys}$                                            & 
			\begin{tabular}{l}
				Peak number of physical qubits required for\\
				logical Bell-pair distillation per node
			\end{tabular}
		    \\ 
		    $G_\mathrm{mult}$
		    & Multiplexing gain of logical Bell-pair distillation
		    & \\
			\hline
		\end{tabular}
	\end{table*}
	
	\begin{table*}[]
		\centering
		\caption{List of symbols and their definitions used in Step V.}
		\label{tab:para-list2}
		\begin{tabular}{lll}
			\hline
			
			$N_\mathrm{node}$ 
			& Number of nodes in a distributed fault-tolerant quantum computer network
			& \\
			
			$N_\mathrm{channel}$ 
			& \begin{tabular}{l}
				Number of communication channels\\
				in each node, determined by the network structure
			\end{tabular}
			& \\
			
			$N_\mathrm{block}$ 
			& Number of code blocks in each node
			& \\
			
			$N_\mathrm{instance}$ 
			& Problem size, depending on the choice of quantum algorithm
			& \\

            $N_\mathrm{phys,node}$ 
			& Number of physical qubits per node
			& \\

            $N_\mathrm{phys,total}$ 
			& Number of total physical    qubits
			& \\

			$n_\mathrm{circuit}$ 
			& Number of qubits required in the synthesized quantum circuit
			& \\
			
			& & \\
			
			$\epsilon$ 
			& Error threshold for the entire quantum algorithm
			& \\
			
			$V_\mathrm{qubit}^\mathrm{norm}$ 
			& \begin{tabular}{l}
				Number of syndrome values obtained during the computation divided by
				$d_\mathrm{code}^3$,\\
				corresponding to the number of logical operations per logical qubit
				during execution
			\end{tabular}
			& \\
			
			$p_\mathrm{circuit}
			=V_\mathrm{qubit}^\mathrm{norm} d_\mathrm{code}
			p_\mathrm{logical}(d_\mathrm{code})$ 
			& Total error probability after execution of the quantum algorithm
			& \\

			$G$ 
			& Graph whose vertices correspond to fault-tolerant quantum computer nodes
			& \\
			
			$n_c$ 
			& Node hosting the control qubit
			& \\
			
			$n_t$ 
			& Node hosting the target qubit
			& \\
			
			& & \\
			
			$T_\mathrm{exec}^\mathrm{norm}$ 
			& Total execution time of a quantum algorithm normalized by the code beat
			& \\
			
			$\tau_\mathrm{code}^\mathrm{cycle}d_\mathrm{code}$ 
			& Code beat
			& \\
			
			$T_\mathrm{Bell}^\mathrm{period}$ 
			& Logical Bell-pair generation period normalized by the code beat
			& \\
			
			$T_\mathrm{thr}$ 
			& Threshold of the relative execution time used to estimate the required logical Bell-pair generation rate
			& \\
			
			\hline
		\end{tabular}
	\end{table*}

	\section{Details of distillation protocols}

    \begin{table*}[t]
	\centering
	\scriptsize
	\caption{
		Distillation protocols and their error-suppression properties considered
		in this paper.
		The leading-order expressions are used for the $[[4,1,2]]$ and
		$[[5,1,3]]$ protocols following Ref.~\cite{Aschauer2004}.
	}
	\label{tab:distillation_protocol_property}
	\setlength{\tabcolsep}{5pt}
	\begin{tabular}{ccccc}
		\hline\hline
		Protocol
		&
		Stabilizer
		&
		\begin{tabular}{c}
			Fidelity improvement,\\
			$1-F'=f(1-F)$
		\end{tabular}
		&
		\begin{tabular}{c}
			Success probability,\\
			$p_\mathrm{success}^\mathrm{distill}$
		\end{tabular}
		&
		Ref.
		\\
		\hline
		
		$Z_\mathrm{2B}$
		&
		$ZZ$
		&
		$\displaystyle
		\frac{
			p_\mathrm{success}^\mathrm{distill}
			-
			\left[
			1-2(1-F)+\frac{10}{9}(1-F)^2
			\right]
		}{
			p_\mathrm{success}^\mathrm{distill}
		}$
		&
		\begin{tabular}{c}
			$1-\frac{4}{3}(1-F)$\\
			$+\frac{8}{9}(1-F)^2$
		\end{tabular}
		&
		\cite{Bennett1996,Siddhu2025}
		\\
		
		$ZX_\mathrm{3B}$
		&
		\begin{tabular}{c}
			$ZZZ$,\\
			$IXX$
		\end{tabular}
		&
		\begin{tabular}{c}
			$\displaystyle
			1-\frac{1}{p_\mathrm{success}^\mathrm{distill}}
			\left[
			1-3(1-F)
			\right.$\\[-1mm]
			$\displaystyle\left.
			+\frac{28}{9}(1-F)^2
			-\frac{28}{27}(1-F)^3
			\right]$
		\end{tabular}
		&
		\begin{tabular}{c}
			$1-\frac{8}{3}(1-F)$\\
			$+\frac{28}{9}(1-F)^2
			-\frac{32}{27}(1-F)^3$
		\end{tabular}
		&
		\cite{Fujii2009,Siddhu2025}
		\\
		
		$[[4,1,2]]$
		&
		\begin{tabular}{c}
			$ZZII$,\\
			$IIZZ$,\\
			$XXXX$
		\end{tabular}
		&
		$6(1-F)^2$
		&
		$1-4(1-F)$
		&
		\cite{Aschauer2004,Pattison2025}
		\\
		
		$[[5,1,3]]$
		&
		\begin{tabular}{c}
			$XZXXI$, $ZIZXZ$,\\
			$ZXXIX$, $IZZZX$
		\end{tabular}
		&
		$10(1-F)^3$
		&
		$1-5(1-F)$
		&
		\cite{Aschauer2004,Pattison2025}
		\\
		
		\hline\hline
	\end{tabular}
\end{table*}

\begin{table*}[t]
	\centering
	\scriptsize
	\caption{
		Physical- and logical-operation costs of the distillation protocols
		considered in this paper.
		The logical distillation time is expressed as
		$\tau^{\mathrm{logical(distill},i)}
		=c_i d_{\mathrm{code},i}\tau_\mathrm{code}^\mathrm{cycle}$.
		The number of code blocks denotes the number of logical code blocks
		required to execute one logical Bell-pair distillation unit.
	}
	\label{tab:distillation_protocol_cost}
	\setlength{\tabcolsep}{5pt}
	\begin{tabular}{cccccc}
		\hline\hline
		Protocol
		&
		\begin{tabular}{c}
			Physical Bell-pair\\
			distillation time
		\end{tabular}
		&
		\begin{tabular}{c}
			Number of\\
			stabilizers
		\end{tabular}
		&
		\begin{tabular}{c}
			Logical Bell-pair\\
			distillation time
		\end{tabular}
		&
		$c_i$
		&
		\begin{tabular}{c}
			Number of\\
			code blocks
		\end{tabular}
		\\
		\hline
		
		$Z_\mathrm{2B}$
		&
		$\tau_\mathrm{meas}+\tau_\mathrm{2q}$
		&
		1
		&
		$\tau_\mathrm{code}^\mathrm{cycle}d_\mathrm{code}$
		&
		1
		&
		2
		\\
		
		$ZX_\mathrm{3B}$
		&
		$\tau_\mathrm{meas}+\tau_\mathrm{1q}+2\tau_\mathrm{2q}$
		&
		2
		&
		$\tau_\mathrm{code}^\mathrm{cycle}d_\mathrm{code}$
		&
		1
		&
		5
		\\
		
		$[[4,1,2]]$
		&
		$\tau_\mathrm{meas}+\tau_\mathrm{1q}+2\tau_\mathrm{2q}$
		&
		3
		&
		$2\tau_\mathrm{code}^\mathrm{cycle}d_\mathrm{code}$
		&
		2
		&
		6
		\\
		
		$[[5,1,3]]$
		&
		$2\tau_\mathrm{meas}
		+6\tau_\mathrm{1q}
		+9\tau_\mathrm{2q}$
		&
		4
		&
		$10\tau_\mathrm{code}^\mathrm{cycle}d_\mathrm{code}$
		&
		10
		&
		7
		\\
		
		\hline\hline
	\end{tabular}
\end{table*}
	
\begin{figure*}[t]
		\centering
		\includegraphics[width=160 mm]{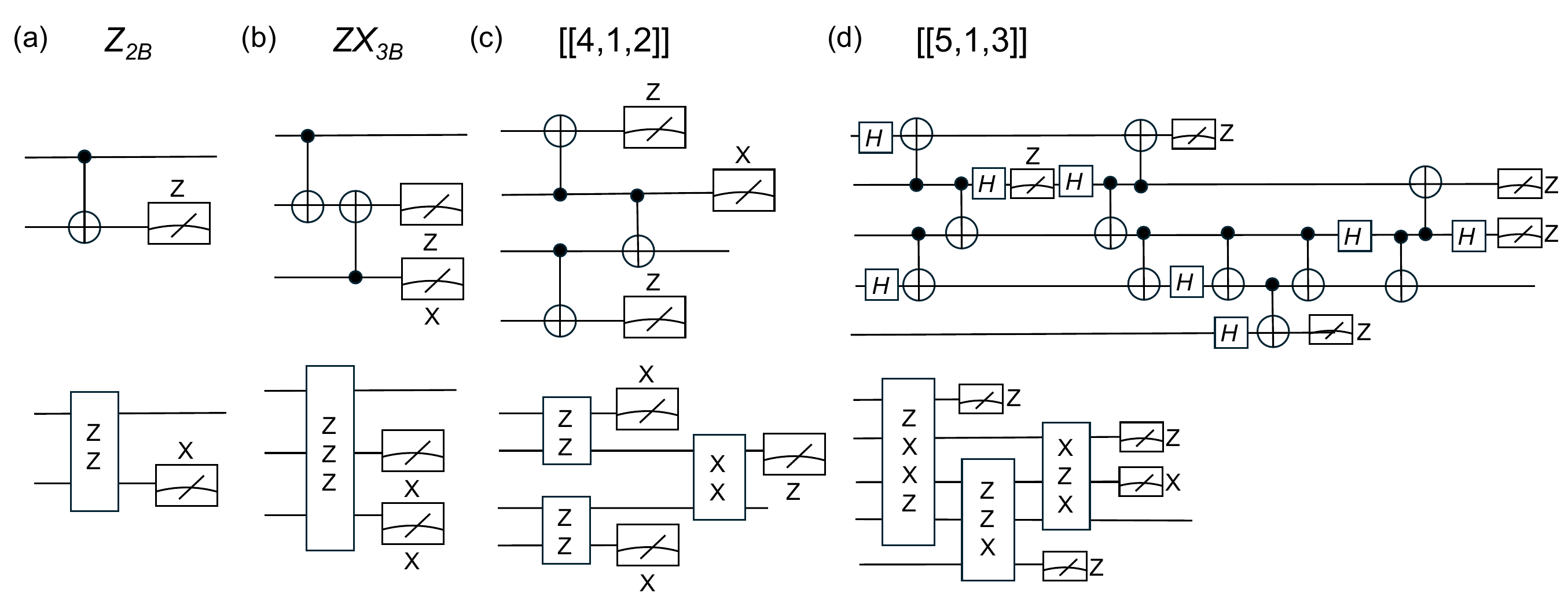}
		\caption
		{Implementation of distillation protocols, (a) $Z_\mathrm{2B}$, (b) $ZX_\mathrm{3B}$, (c) [[4,1,2]] and (d) [[5,1,3]] using a one-dimensional array of physical qubits (Top row) and multiple-Pauli measurement of logical qubits (Bottom row).
		}
		\label{fig:summary_distillation_circuit}
	\end{figure*}

	Figure \ref{fig:summary_distillation_circuit}, Tables \ref{tab:distillation_protocol_property} and \ref{tab:distillation_protocol_cost} summarize the distillation circuits considered in the main text. For physical Bell-pair distillation, a combination of CNOT between nearest-neighbor physical qubits, single-qubit operations, and single-qubit measurements can be used. For logical Bell-pair distillation, Pauli measurements for multiple logical qubits can be performed together with single logical qubit measurements. Although there may be room for further reduction of execution time, we use the quantum circuits shown here to calculate the distillation time in the subsequent analysis.
	
		\section{Expected number of Bell-pair generation  trials}
	The expected trial numbers, $\mathbb{E}[\mathcal{N}(p,\ n)]$, to generate $n$ Bell pairs under Bell-pair generation probability of $p$ can be derived under the following assumptions. 
	First, we have independent $n$ communication channels which generates Bell pair with finite probability, $0< p\le 1$. We repeat Bell-pair generation attempts until Bell pairs are shared in all $n$ communication channels. 
	\begin{equation}
		\mathbb{E}[\mathcal{N}(p,\ n)]=\sum_{i=1}^{\infty}i\times(P_i-P_{i-1}),	
	\end{equation}
	\begin{equation}
		P_i = [1-(1-p)^i]^n,
	\end{equation}
	where $P_i$ is the probability where $n$ Bell pair is generated until the $i$th trial. Thus, $P_i-P_{i-1}$ is the probability where $n$ Bell pair is generated at the $i$th trial. This is equivalent to
	\begin{equation}
		\mathbb{E}[\mathcal{N}(p,\ n)]
		=
		\sum_{j=1}^{n}
		\binom{n}{j}(-1)^{j+1}
		\frac{1}{1-(1-p)^j}.
	\end{equation}

\section{Bell-pair generation rate, latency, and throughput}
\label{app:latency_throughput}

In this work, we distinguish the physical Bell-pair source rate, a
latency-based effective rate, and a steady-state throughput.

The physical Bell-pair generation rate
$r_{\mathrm{Bell}}^{\mathrm{phys}}$ is defined as the average generation rate
of a single physical communication channel. For repeated independent
generation attempts with trial rate $R_{\mathrm{rep}}$ and success
probability $p_{\mathrm{success}}^{\mathrm{Bell}}$,
\begin{equation}
	r_{\mathrm{Bell}}^{\mathrm{phys}}
	=
	R_{\mathrm{rep}}
	p_{\mathrm{success}}^{\mathrm{Bell}}.
\end{equation}
We regard this quantity as a source-level parameter rather than exclusively
as a latency or throughput. For a single continuously operated channel,
$1/r_{\mathrm{Bell}}^{\mathrm{phys}}$ also gives the mean waiting time to the
next successful Bell pair, and the same rate describes the long-time average
generation rate. These interpretations generally differ when parallel
channels, buffering, or multistage distillation are introduced.

For a multistage process, we define the latency-based rate from the expected
time $T_{\mathrm{first}}$ required to obtain the first successful final Bell
pair from an initially empty pipeline,
\begin{equation}
	r_{\mathrm{lat}}
	=
	\frac{1}{\mathbb{E}[T_{\mathrm{first}}]}.
	\label{eq:latency_rate_definition}
\end{equation}
By contrast, the throughput is the long-time average output rate,
\begin{equation}
	r_{\mathrm{th}}
	=
	\lim_{T\rightarrow\infty}
	\frac{\mathbb{E}[N_{\mathrm{out}}(T)]}{T}.
	\label{eq:throughput_rate_definition}
\end{equation}
The two can differ because a continuously operated pipeline can reuse
partially prepared inputs and simultaneously operate different stages.

For the sequential logical-distillation model used in this work, the
latency-based description is convenient for analytical calculations.
If the $i$th distillation stage consumes $n_i$ Bell pairs, has success
probability $P_i$, and requires time $T_{\mathrm{distill},i}$, the
first-output latency is recursively evaluated as
\begin{equation}
	T_i
	=
	\frac{
		n_i T_{i-1}
		+
		T_{\mathrm{distill},i}
	}{
		P_i
	}.
	\label{eq:latency_recursive_definition}
\end{equation}
The corresponding average waiting time of the inputs is
\begin{equation}
	t_{\mathrm{wait},i}^{\mathrm{seq}}
	=
	\frac{n_i-1}{2}T_{i-1}.
\end{equation}
Here, the first prepared input waits for the remaining $n_i-1$ inputs,
whereas the last input does not wait, giving the average waiting time above.

When the required inputs are instead generated in parallel, their completion
times are generally different because Bell-pair generation is probabilistic.
An individual input requires $1/p$ attempts on average, whereas the complete
set becomes available after $\mathbb{E}[\mathcal{N}(p,n)]$ attempts on average.
Therefore, the average time for which an already generated input waits for
the remaining inputs is
\begin{equation}
	t_{\mathrm{wait},i}^{\mathrm{para}}
	=
	\left(
	\mathbb{E}[\mathcal{N}(p,n)]
	-
	\frac{1}{p}
	\right)
	\frac{p}{r},
	\label{eq:parallel_waiting_time}
\end{equation}
where $r=R_{\mathrm{rep}}p$ is the average Bell-pair generation rate of one
channel. The factor $p/r=1/R_{\mathrm{rep}}$ converts the number of generation
attempts into time. Thus, $t_{\mathrm{wait},i}^{\mathrm{para}}$ represents
the average time between the generation of an input Bell pair and the time
at which the last of the $n$ required inputs becomes available.

This formulation allows the waiting-induced memory error to be incorporated
without explicitly simulating the full time-dependent pipeline.

Monte Carlo simulation can evaluate either quantity: the first-output
latency can be obtained from repeated simulations starting from an empty
pipeline, whereas the throughput can be obtained from the long-time output
rate of a continuously operated pipeline.

In this work, the analytical treatment of physical Bell-pair distillation
uses the latency picture because it allows the multistage process to be
described by a simple recursive model. For the Monte Carlo analysis of
physical Bell-pair distillation, we instead use the throughput picture,
which more closely represents the intended implementation: physical Bell
pairs are continuously generated and distilled, and purified Bell pairs are
supplied successively to the logical communication layer. Thus, the
latency-based analytical model is used as a tractable approximation, whereas
the Monte Carlo simulation captures the continuous-operation throughput.

For logical Bell-pair distillation, we use the latency picture
in both the analytical and Monte Carlo analyses.
Logical distillation requires encoded code blocks and therefore
substantial physical-qubit resources.
For the sequential configuration, we assume one logical
distillation unit per stage to limit the required space.
For the parallel configuration, the full distillation tree
is provisioned as described below.

\section{Monte Carlo simulation for physical Bell-pair distillation}

Table~\ref{tab:Monte_calro_simulation_parameters} summarizes the parameters used for the analytical calculation and Monte Carlo simulation of physical Bell-pair distillation. The duration of each distillation protocol is taken from Table~\ref{tab:distillation_protocol_cost} and expressed as an integer number of simulation clocks.


The Monte Carlo simulation proceeds on a discrete time grid with clock period \(T_\mathrm{clock}=50~\mathrm{ns}\). Physical Bell-pair generation is attempted with repetition rate \(R_\mathrm{rep}=1~\mathrm{MHz}\), corresponding to one attempt every
$
	N_\mathrm{rep}
	=
	\frac{1}{R_\mathrm{rep}T_\mathrm{clock}}
	=
	20
$
simulation clocks. Each empty communication channel independently generates a Bell pair with probability \(p_\mathrm{success}^\mathrm{Bell}\) per attempt. A channel that already stores a Bell pair is not overwritten by a new generation attempt.

For a sequence of distillation protocols with input sizes \(n_i\), the simulation contains the minimum number of parallel communication channels required by the protocol tree. All protocols considered here have one output Bell pair per successful distillation, \(k_i=1\), and therefore
$
	N_\mathrm{comm}^\mathrm{phys}
	=
	\prod_i n_i .
$
At each distillation stage, the available Bell pairs are grouped into independent blocks, and different blocks can be distilled in parallel.

A distillation job starts when all required input Bell pairs are available. Bell pairs waiting in a buffer are discarded when their waiting time reaches \(T_\mathrm{threshold}\). Once a distillation job starts, its input Bell pairs are reserved until the operation finishes. To avoid producing an intermediate Bell pair before the corresponding higher-stage buffer becomes available, a lower-stage job is delayed when its designated output slot is occupied and the ongoing higher-stage job would finish later than the lower-stage job. Equivalently, the lower-stage job is postponed when
\begin{equation}
	N_\mathrm{remain}^{(k+1)}
	>
	N_\mathrm{dist}^{(k)},
\end{equation}
where \(N_\mathrm{remain}^{(k+1)}\) is the remaining number of clocks of the relevant higher-stage job and \(N_\mathrm{dist}^{(k)}\) is the duration of the lower-stage distillation.

Before each distillation operation, storage-induced relaxation is evaluated for each input Bell pair from its generation time to the start of the distillation job using the \(T_1\)- and \(T_2\)-based relaxation model. Because the analytical distillation formulas use a single input error rate, the largest relaxed infidelity among the input Bell pairs is used as the effective input infidelity for each distillation step. The effects of single-qubit gates, two-qubit gates, and measurements are then included using the operation error rates listed in Table~\ref{tab:Monte_calro_simulation_parameters}. The success or failure of each distillation step is sampled according to the corresponding success probability. Upon success, one output Bell pair is transferred to the next-stage buffer with the completion clock assigned as its new generation time; upon failure, all input Bell pairs used in that distillation job are discarded.

The simulation terminates after \(n_\mathrm{goal}\) successfully distilled Bell pairs have been obtained. If the \(n_\mathrm{goal}\)-th output is generated at time \(t_\mathrm{final}\), the total distilled Bell-pair generation rate is evaluated as
\begin{equation}
	r_\mathrm{Bell}^{\mathrm{phys(distill)}}
	=
	\frac{n_\mathrm{goal}}{t_\mathrm{final}} .
\end{equation}

The analytical calculation is performed over the full range of \(p_\mathrm{success}^\mathrm{Bell}\) considered in this work. Monte Carlo simulations are performed for the numerical points shown in the corresponding figure: \(p_\mathrm{success}^\mathrm{Bell}=1\) for all four input error rates and, in addition, \(p_\mathrm{success}^\mathrm{Bell}=10^{-1}\) for \(1-F_\mathrm{Bell}^\mathrm{phys}=0.1\). For lower generation probabilities, only the analytical results are shown because the finite buffer lifetime makes successful Monte Carlo output events increasingly rare.

\begin{table*}[t]
	\centering
	\caption{Parameters used in the analytical calculation and Monte Carlo simulation for physical Bell-pair distillation.}
	\label{tab:Monte_calro_simulation_parameters}
	\begin{tabular}{lll}
		\hline\hline
		Parameter & Symbol & Value \\
		\hline
		Input error rate of physical Bell pairs
		& \(1-F_\mathrm{Bell}^\mathrm{phys}\)
		& \(5,\ 10,\ 15,\ 20\%\) \\
		
		Physical Bell-pair generation attempt rate
		& \(R_\mathrm{rep}\)
		& \(1~\mathrm{MHz}\) \\
		
		Physical Bell-pair generation attempt time
		& \(T_\mathrm{rep}=1/R_\mathrm{rep}\)
		& \(1~\mu\mathrm{s}\) \\
		
		\begin{tabular}{l}
			Physical Bell-pair generation success probability\\
			per attempt (analytical calculation)
		\end{tabular}
		& \(p_\mathrm{success}^\mathrm{Bell}\)
		& \(1,\ 10^{-1},\ 10^{-2},\ 10^{-3}\) \\
		
		\begin{tabular}{l}
			Physical Bell-pair generation success probability\\
			per attempt (Monte Carlo)
		\end{tabular}
		& \(p_\mathrm{success}^\mathrm{Bell}\)
		& \(1,\ 10^{-1}\) \\
		
		Operation error rates for 1q, 2q, and measurement
		& \(p_{\mathrm{1q}},\,p_{\mathrm{2q}},\,p_{\mathrm{meas}}\)
		& \(10^{-3}\) \\
		
		Operation times for 1q, 2q, and measurement
		& \(T_{\mathrm{1q}},\,T_{\mathrm{2q}},\,T_{\mathrm{meas}}\)
		& \(50~\mathrm{ns},\ 250~\mathrm{ns},\ 100~\mathrm{ns}\) \\
		
		Simulation clock period
		& \(T_\mathrm{clock}\)
		& \(50~\mathrm{ns}\) \\
		
		Number of clocks per generation attempt
		& \(N_\mathrm{rep}\)
		& \(20\) \\
		
		Maximum buffer waiting time
		& \(T_\mathrm{threshold}\)
		& \(10~\mu\mathrm{s}\) \\
		
		Energy relaxation time
		& \(T_1\)
		& \(1~\mathrm{ms}\) \\
		
		Decoherence time
		& \(T_2=2T_1\)
		& \(2~\mathrm{ms}\) \\
		
		\begin{tabular}{l}
			Number of successful outputs required to\\
			terminate each Monte Carlo run
		\end{tabular}
		& \(n_\mathrm{goal}\)
		& \(1000\) \\
		\hline\hline
	\end{tabular}
\end{table*}

\section{Details of logical Bell-pair distillation}
\label{app:logical_distillation_details}

\subsection{Code-distance and protocol selection}
\label{app:logical_distill_code_distance}

The rate, fidelity, and waiting-time models for sequential and parallel
logical Bell-pair distillation are described in the main text. Here, we
summarize the procedure used to select the code distance and distillation
protocols.

At each distillation stage, the code distance must be sufficiently large to
suppress the memory error accumulated while the input Bell pairs are waiting.
At the same time, the resulting input infidelity must remain low enough for the
remaining distillation stages to reach the requested final fidelity.

At stage $i$, let $\epsilon_{\mathrm{in},i}$ be the Bell-pair infidelity before
including the waiting-induced error. The effective infidelity entering the
distillation operation is
\begin{equation}
	\epsilon_{\mathrm{eff},i}
	=
	\epsilon_{\mathrm{in},i}
	+
	\epsilon_{\mathrm{wait},i}(d_{\mathrm{code},i}),
\end{equation}
where $\epsilon_{\mathrm{wait},i}$ is the memory error accumulated during the
waiting time and depends on $d_{\mathrm{code},i}$.

For a fixed distillation sequence, the requested final infidelity is propagated
backward through the distillation maps to determine
$\epsilon_{\mathrm{eff},i}^{\max}$, the largest effective input infidelity at
stage $i$ that still allows the final target to be reached.

We then choose the smallest odd $d_{\mathrm{code},i}$ that satisfies
$
\epsilon_{\mathrm{wait},i}(d_{\mathrm{code},i})
\le
\eta\epsilon_{\mathrm{in},i},
$
and
$
\epsilon_{\mathrm{eff},i}(d_{\mathrm{code},i})
\le
\epsilon_{\mathrm{eff},i}^{\max}.
$
We use $\eta=1$. The first condition limits the additional memory error to no
more than the incoming Bell-pair infidelity, whereas the second imposes the
end-to-end fidelity requirement of the complete distillation sequence.

For the rate--infidelity trade-off calculation, the target infidelity is swept
from $10^{-2}$ to $10^{-15}$ in increments of 0.5 decades. For each target,
all ordered combinations of the $Z_{\mathrm{2B}}$, $ZX_{\mathrm{3B}}$,
$[[4,1,2]]$, and $[[5,1,3]]$ protocols containing up to six distillation
stages are evaluated. The stage code distances are optimized using the above
procedure, and the fastest feasible sequence is selected. No explicit code-distance cap is imposed. The resulting points are filtered to retain those
representing the best achievable trade-off between generation rate and infidelity.

For the representative resource comparison, physical-rate sweep, and Monte
Carlo validation, the target is fixed to
$1-F_{\mathrm{Bell}}^{\mathrm{logical(distill)}}\le10^{-12}$.
Candidate sequences are ranked using the communication-normalized rate
$r_{\mathrm{Bell}}^{\mathrm{logical(distill)}}/
N_{\mathrm{comm}}^{\mathrm{logical}}$.
For the sequential configuration,
$N_{\mathrm{comm}}^{\mathrm{logical}}=1$, so this metric is identical to the
unnormalized logical Bell-pair generation rate.

\subsection{Resource estimation}
\label{app:logical_distill_resource}

One logical surface-code block with code distance $d_{\mathrm{code}}$ is
assigned
\begin{equation}
	N_{\mathrm{phys}}^{\mathrm{block}}
	=
	2(d_{\mathrm{code}}+1)^2
\end{equation}
physical qubits. The $Z_{\mathrm{2B}}$, $ZX_{\mathrm{3B}}$, $[[4,1,2]]$,
and $[[5,1,3]]$ distillation units require 2, 5, 6, and 7 logical code
blocks, respectively.

For the sequential configuration, we consider a minimum-space implementation
consistent with the first-output latency model used in the main text.
Only one logical communication channel is used, and one distillation unit is
allocated to each distillation stage.
Logical Bell pairs are generated and accumulated until the number required for
the next distillation operation is available. Once a complete input set for a
stage has been prepared, further upstream preparation is paused until that set
has been consumed and storage space becomes available again. This prevents the
unnecessary accumulation of intermediate logical Bell pairs.
The physical-qubit requirement is evaluated by considering all possible
configurations encountered while producing one final logical Bell pair,
including logical Bell pairs temporarily stored between distillation stages and
the code blocks occupied by ongoing distillation operations. We define
\(N_{\mathrm{phys}}^{\mathrm{seq}}\) as the minimum physical-qubit capacity
required to realize this sequential first-output process, i.e., the maximum
instantaneous number of physical qubits occupied along the minimum-space
sequence of operations.
The final output Bell pair reuses a code block released by the final
distillation operation and is therefore not counted as an additional output
buffer.

For the parallel configuration, we provision the hardware required to operate
the full distillation tree without reusing distillation units between different
stages. We denote the corresponding provisioned physical-qubit requirement per
node by \(N_{\mathrm{phys}}^{\mathrm{para}}\). We label the first distillation
stage by $i=1$ and the final stage by $i=L$.
To supply the $n_L$ inputs required by the final stage, $n_L$ distillation
units are required at stage $L-1$. Each of these units in turn requires
$n_{L-1}$ outputs from stage $L-2$, and so forth. Consequently, the number
of distillation units required at stage $i$ is
\begin{equation}
	M_i
	=
	\prod_{j=i+1}^{L} n_j,
	\qquad
	M_L=1.
\end{equation}
For example, for a three-stage sequence with input sizes
$n_1$, $n_2$, and $n_3$, the numbers of units at the three stages are
$n_2n_3$, $n_3$, and $1$, respectively.

If one distillation unit at stage $i$ occupies $B_i$ logical code blocks, and
each code block with code distance $d_{\mathrm{code},i}$ contains
$2(d_{\mathrm{code},i}+1)^2$ physical qubits, the physical-qubit requirement
per node is
\begin{equation}
	N_{\mathrm{phys}}^{\mathrm{para}}
	=
	\sum_{i=1}^{L}
	M_i B_i
	2(d_{\mathrm{code},i}+1)^2.
\end{equation}
Because different stages of the full tree can operate at the same time, the
hardware assigned to all stages is counted simultaneously rather than being
reused between successive stages. In the main text, these peak physical-qubit requirements per node are
denoted by $N_{\mathrm{buffer}}^\mathrm{phys}$, corresponding to
$N_{\mathrm{phys}}^\mathrm{seq}$ and $N_{\mathrm{phys}}^\mathrm{para}$
for the sequential and parallel configurations, respectively.

The expected number of physical Bell pairs consumed through state injection
and logical Bell-pair distillation to obtain one final logical Bell pair is
\begin{equation}
	N_{\mathrm{consumed}}^{\mathrm{injection}}
	=
	\frac{
		\prod_i n_i
	}{
		p_{\mathrm{success}}^{\mathrm{injection}}
		\prod_i
		p_{\mathrm{success}}^{\mathrm{logical(distill},i)}
	}.
\end{equation}
When physical Bell-pair distillation is performed before state injection, the
effective number of raw physical Bell pairs consumed per physically distilled
Bell pair is
\begin{equation}
	N_{\mathrm{consumed}}^{\mathrm{phys(distill)}}
	=
	\frac{
		N_{\mathrm{comm}}^{\mathrm{phys}}
		r_{\mathrm{Bell}}^{\mathrm{phys}}
	}{
		r_{\mathrm{Bell}}^{\mathrm{phys(distill)}}
	}.
\end{equation}
Thus, the total number of raw physical Bell pairs consumed per final
logical Bell pair is
\begin{equation}
	N_{\mathrm{consumed}}^\mathrm{phys}
	=
	N_{\mathrm{consumed}}^{\mathrm{injection}}
	N_{\mathrm{consumed}}^{\mathrm{phys(distill)}}.
\end{equation}
When physical Bell-pair distillation is not used,
$N_{\mathrm{consumed}}^{\mathrm{phys(distill)}}=1$.
Additional movement and routing resources of logical code blocks are not
included.

\subsection{Monte Carlo validation}
\label{app:logical_distill_mc}

We perform discrete-event Monte Carlo simulations to validate the analytical
treatment of stochastic waiting and scheduling. The protocol sequence and
stage code distances are obtained from the target-aware analytical calculation
with
$1-F_{\mathrm{Bell}}^{\mathrm{logical(distill)}}\le10^{-12}$.
The Monte Carlo simulation samples Bell-pair generation, state-injection
success, distillation success, and the resulting waiting times; logical Pauli
errors are not sampled individually.

Physical Bell-pair generation trials are performed at a fixed rate of
$1~\mathrm{MHz}$, with
$p_{\mathrm{success}}^{\mathrm{Bell}}
=r_{\mathrm{Bell}}^{\mathrm{phys}}/(1~\mathrm{MHz})$.
Thus, the time required for successful Bell-pair generation is sampled from
the corresponding geometric distribution.

For the sequential configuration,
$N_{\mathrm{comm}}^{\mathrm{logical}}=1$, and one distillation unit is
assigned to each stage. Bell pairs are accumulated only until the input set
required for the next distillation operation is complete; further upstream
generation is then paused until storage space becomes available again.
In addition, Bell-pair generation and distillation at other stages are paused
while a distillation operation is active, consistent with the sequential
first-output latency model used in the analytical calculation.

\begin{figure*}[tb]
	\centering
	\includegraphics[width=140 mm]{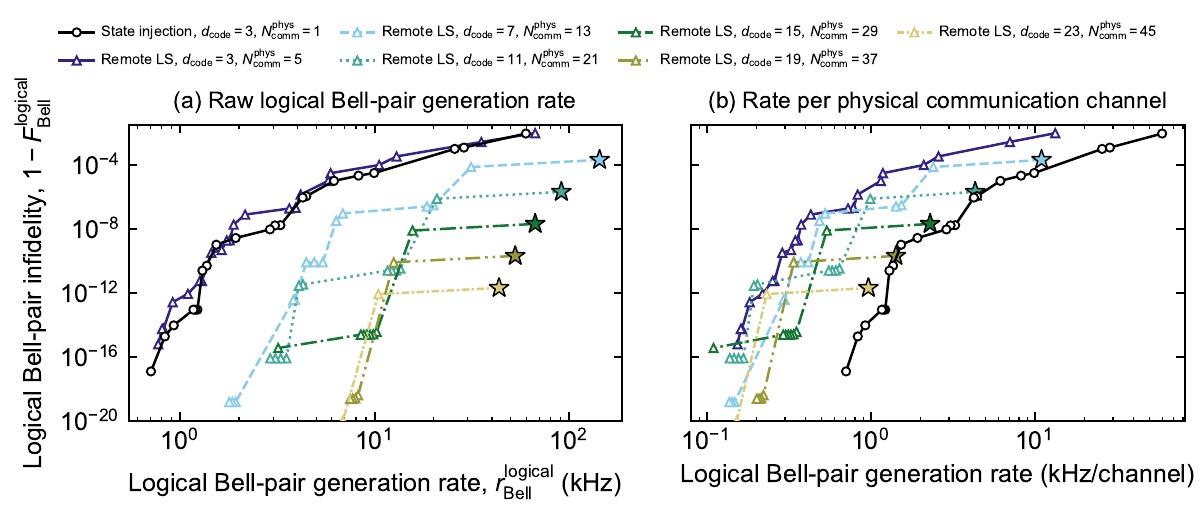}
	\caption{
		Comparison of state injection and remote lattice surgery followed by
		logical Bell-pair distillation.
		The physical Bell-pair fidelity and generation rate are fixed to
		$F_\mathrm{Bell}^\mathrm{phys}=0.99$ and
		$r_\mathrm{Bell}^\mathrm{phys}=1~\mathrm{MHz}$ per physical
		communication link, respectively.
		State injection uses one physical communication link with
		$d_\mathrm{code}=3$, whereas remote lattice surgery is evaluated for
		$d_\mathrm{code}=3,7,11,15,19,$ and $23$, using
		$2d_\mathrm{code}-1$ parallel communication links.
		For each requested final infidelity from $10^{-2}$ to $10^{-15}$ in
		0.5-decade steps, logical-distillation sequences containing up to four
		stages, including the case without logical distillation, are evaluated,
		and the fastest feasible sequence is selected.
		(a) Logical Bell-pair generation rate.
		(b) Logical Bell-pair generation rate normalized by the number of
		physical communication links.
		Star markers denote cases without logical distillation.
	}
	\label{fig:app_surgery_distillation}
\end{figure*}

For the parallel configuration, the full distillation tree is represented,
with
\begin{equation}
	N_{\mathrm{comm}}^{\mathrm{logical}}
	=
	\prod_i n_i
\end{equation}
independent communication channels and
$M_i=\prod_{j=i+1}^{L}n_j$ distillation units at stage $i$.
Independent branches are allowed to progress asynchronously.

The creation time of every stored logical Bell pair is recorded. For each
input Bell pair used at stage $i$, the waiting-induced contribution is
evaluated from its actual sampled waiting time as
\begin{equation}
	\epsilon_{\mathrm{wait}}
	=
	2p_{\mathrm{logical}}(d_{\mathrm{code},i})
	\frac{t_{\mathrm{wait}}}
	{\tau_{\mathrm{code}}^{\mathrm{cycle}}}.
\end{equation}
The resulting effective input infidelities are averaged over the $n_i$ input
Bell pairs before applying the same equal-input distillation map as in the
analytical calculation. The Monte Carlo simulation therefore validates the
effects of stochastic scheduling and asynchronous branch completion rather
than performing a circuit-level or Pauli-level error simulation.

Each trajectory starts from an empty distillation system and terminates when
the first successful logical Bell pair exits the final stage. The Monte Carlo
generation rate is evaluated using the same inverse-mean-first-output-latency
definition as in the main text. For the parallel configuration, a
communication-normalized rate can additionally be obtained by dividing the
result by $N_{\mathrm{comm}}^{\mathrm{logical}}$ after the simulation.

We use 300 independent trajectories for each Monte Carlo condition.

\subsection{Comparison of logical Bell-pair generation via state injection and remote lattice surgery}
\label{app:logical_distill_lattice_surgery}

We additionally compare state injection and remote lattice surgery as
methods for preparing the input logical Bell pairs for subsequent logical
Bell-pair distillation. Throughout this comparison, we assume
$F_\mathrm{Bell}^\mathrm{phys}=0.99$ and
$r_\mathrm{Bell}^\mathrm{phys}=1~\mathrm{MHz}$ per physical communication
link.

For state injection, the input logical Bell-pair rate and fidelity are
calculated using the model described in the main text with
$d_\mathrm{code}=3$ and one physical communication link. We restrict the
comparison to a single state-injection channel because parallel state
injection would require an additional scheduling and buffering model.

For remote lattice surgery, the input logical Bell-pair rate and fidelity
are calculated using the lattice-surgery model described in the main text.
We consider $d_\mathrm{code}=3,7,11,15,19,$ and $23$, with
$2d_\mathrm{code}-1$ parallel communication links, and use
$p_\mathrm{bulk}=10^{-3}$ and $p_\mathrm{seam}=10^{-2}$.

For both input-generation methods, the subsequent logical Bell-pair
distillation is evaluated using the same sequential first-output latency,
waiting-error, and code-distance-selection models described above.
For each requested final infidelity, all logical-distillation sequences
containing up to four stages, including the case without logical
distillation, are evaluated, and the sequence with the highest generation
rate satisfying the requested target is selected.

Because state injection and remote lattice surgery require different numbers
of physical communication links, Fig.~\ref{fig:app_surgery_distillation}(a)
shows the logical Bell-pair generation rate, whereas panel (b) shows the same
rate divided by the number of physical communication links. The divisor is
one for state injection and $2d_\mathrm{code}-1$ for remote lattice surgery.

\clearpage

%

\end{document}